%% file: main.tex
\documentclass[12pt]{article}
\usepackage{graphicx}
\usepackage{amsmath}
\usepackage{mathtools}
\usepackage{amssymb}
\usepackage{amsthm}

\usepackage{booktabs}
\usepackage{longtable}
\usepackage[dvipsnames,svgnames,table]{xcolor}
\usepackage{hyperref}
\usepackage{geometry}
\usepackage{parskip}
\usepackage{lmodern}
\usepackage[T1]{fontenc}
\usepackage{microtype}
\usepackage{natbib}
\usepackage{caption}
\usepackage{subcaption}
\usepackage{float}
\usepackage[section]{placeins}
\usepackage{multicol}
\usepackage{threeparttable}
\usepackage{siunitx}
\usepackage{pdflscape}
\usepackage{tikz}
\usetikzlibrary{arrows.meta, positioning, calc}
\usepackage[ruled,lined]{algorithm2e}
\usepackage[capitalise,noabbrev]{cleveref}  
\input{generated/macros}

\title{A Deep Learning Approach to Heterogeneous Consumer Aesthetics in Fast Fashion\thanks{I thank John Rust, Sanjog Misra, Nathan Miller, and Harry Paarsch for guidance and feedback.}}
\author{Pranjal Rawat\footnote{PhD Candidate, Georgetown University}}
\date{\today}

\begin{document}

\maketitle
\input{00_abstract}
\clearpage
\setcounter{tocdepth}{2}  
\tableofcontents
\clearpage

\input{01_introduction}       
\input{02_data}               
\input{03_embeddings}         
\input{04_theory}             
\input{05_demand}             
\input{08_hedonic}            
\input{09_event_study}        
\input{11_conclusion}         

%
%
%
\appendix
%
\input{B_robustness}          
%
%
\input{D_blp}                 
%
\input{E_cross_category}      
%
\input{H_embedding_validation} 
\input{I_glossary}             
%
%

\clearpage
\bibliographystyle{plainnat}
\bibliography{references}

\end{document}

%% file: generated/macros.tex
\input{generated/macros_data}

\input{generated/macros_supply}

\newcommand{\lcValNLLNoCF}{6.080}
\newcommand{\lcValRSqrNoCFPct}{6.40}
\newcommand{\lcAlphaOneShiftPct}{6.6}        
\newcommand{\lcAlphaTwoShiftPct}{1.7}        
\newcommand{\lcValNLLShift}{0.062}           
\newcommand{\lcValRSqrLiftPP}{0.96}          
\newcommand{\lcPiOneShift}{0.028}            

\input{generated/macros_cluster}

\newcommand{\lcEpsTauVarPct}{5}              

\newcommand{\sustReprice}{1.9}               
\newcommand{\sustNeutralLo}{30}              
\newcommand{\sustNeutralHi}{35}              
\newcommand{\sustPruneShallow}{2}            
\newcommand{\sustPruneDeep}{70}              

\newcommand{\aesthCosineOneBase}{0.322}          
\newcommand{\aesthCosineOneDelta}{0.134}         
\newcommand{\aesthCosineTwoBase}{0.361}          
\newcommand{\aesthCosineTwoDelta}{0.153}         

\input{generated/macros_rec}
\input{generated/macros_bnet}
\input{generated/macros_embeddings}

\newcommand{\datBootstrapB}{10}                     

\input{generated/macros_seas}

\newcommand{\hedFeatsBaseline}{8}
\newcommand{\hedFeatsTFIDF}{108}
\newcommand{\hedFeatsCLIP}{620}
\newcommand{\hedFeatsThreeTower}{72}
\newcommand{\hedRSqrInBaseline}{0.43}
\newcommand{\hedRSqrInTFIDF}{0.71}
\newcommand{\hedRSqrInCLIP}{0.73}
\newcommand{\hedRSqrInThreeTower}{0.65}
\newcommand{\hedRSqrInFull}{0.72}
\newcommand{\hedRSqrFwdBaseline}{0.25}
\newcommand{\hedRSqrFwdTFIDF}{0.43}
\newcommand{\hedRSqrFwdCLIP}{0.44}
\newcommand{\hedRSqrFwdThreeTower}{0.47}
\newcommand{\hedRSqrFwdFull}{0.45}
\newcommand{\hedRSqrLossBaseline}{43}
\newcommand{\hedRSqrLossTFIDF}{39}
\newcommand{\hedRSqrLossCLIP}{40}
\newcommand{\hedRSqrLossThreeTower}{27}
\newcommand{\hedRSqrLossFull}{39}

\input{generated/macros_hed}

\newcommand{\hedGapDress}{\hedGap}                  

\input{generated/macros_event}
\newcommand{\evtCellN}{132}                        
\newcommand{\evtClusterMinTxnsPerDay}{10}          

\newcommand{\evtUserRangePP}{28.4}                 
\newcommand{\evtUserVsDemo}{1.80}                  
\newcommand{\evtCellRangePP}{175.9}                

\newcommand{\evtPrelockStorePct}{30}
\newcommand{\evtPrelockOnlinePct}{70}
\newcommand{\evtPostlockStorePct}{34}
\newcommand{\evtPostlockOnlinePct}{66}

\newcommand{\evtCatDress}{\evtAggregatePct}        

\newcommand{\evtICzeroEffect}{38.0}
\newcommand{\evtIConeEffect}{1.9}
\newcommand{\evtICtwoEffect}{\evtItemICtwoMostNeg}
\newcommand{\evtICthreeEffect}{40.8}
\newcommand{\evtICfourEffect}{34.7}
\newcommand{\evtICfiveEffect}{27.3}
\newcommand{\evtICsixEffect}{\evtItemICsixMostPos}
\newcommand{\evtICsevenEffect}{33.2}
\newcommand{\evtICeightEffect}{45.0}
\newcommand{\evtICnineEffect}{28.8}
\newcommand{\evtICtenEffect}{43.1}
\newcommand{\evtICelevenEffect}{40.0}
\newcommand{\evtICzeroCILo}{52.8}
\newcommand{\evtICzeroCIHi}{18.6}
\newcommand{\evtIConeCILo}{50.6}
\newcommand{\evtIConeCIHi}{94.9}
\newcommand{\evtICtwoCILo}{62.7}
\newcommand{\evtICtwoCIHi}{40.7}
\newcommand{\evtICthreeCILo}{56.9}
\newcommand{\evtICthreeCIHi}{18.6}
\newcommand{\evtICfourCILo}{45.0}
\newcommand{\evtICfourCIHi}{22.5}
\newcommand{\evtICfiveCILo}{40.7}
\newcommand{\evtICfiveCIHi}{10.9}
\newcommand{\evtICsixCILo}{6.6}
\newcommand{\evtICsixCIHi}{157.1}
\newcommand{\evtICsevenCILo}{48.7}
\newcommand{\evtICsevenCIHi}{13.0}
\newcommand{\evtICeightCILo}{61.8}
\newcommand{\evtICeightCIHi}{20.9}
\newcommand{\evtICnineCILo}{40.5}
\newcommand{\evtICnineCIHi}{14.7}
\newcommand{\evtICtenCILo}{57.0}
\newcommand{\evtICtenCIHi}{24.6}
\newcommand{\evtICelevenCILo}{68.7}
\newcommand{\evtICelevenCIHi}{15.0}

\input{generated/macros_robtau}

\input{generated/macros_robwelfare}

\input{generated/macros_robieb}

\input{generated/macros_bnch}

\input{generated/macros_lcnocf}
\input{generated/macros_xcat}

\newcommand{\xcatHedRMin}{0.57}                     
\newcommand{\xcatHedRMax}{0.74}                     
\newcommand{\xcatItemRangeMin}{14.6}                
\newcommand{\xcatItemRangeMax}{144.4}               
\newcommand{\xcatUserRangeMin}{10.6}                
\newcommand{\xcatUserRangeMax}{55.9}                
\newcommand{\xcatUserDemoRatioMin}{0.5}             
\newcommand{\xcatUserDemoRatioMax}{1.8}             
\newcommand{\xcatARIEmbAgeTypical}{0.03}            

\newcommand{\xcatTrousersHedGap}{\hedGapTrousers}   
\newcommand{\xcatSweaterHedGap}{\hedGapSweater}     
\newcommand{\xcatTShirtHedGap}{\hedGapTShirt}       
\newcommand{\xcatShirtHedGap}{\hedGapShirt}         
\newcommand{\xcatShortsHedGap}{\hedGapShorts}       

\newcommand{\xcatDressAggregate}{\evtAggregatePct}
\newcommand{\xcatDressKItem}{\evtKItem}
\newcommand{\xcatDressRangeItem}{\evtItemRangePP}
\newcommand{\xcatDressKUser}{\evtKUser}
\newcommand{\xcatDressRangeUser}{\evtUserRangePP}
\newcommand{\xcatDressUserDemo}{1.8}                
\newcommand{\xcatTrousersAggregate}{\evtCatTrousers}  
\newcommand{\xcatSweaterAggregate}{\evtCatSweater}  
\newcommand{\xcatTShirtAggregate}{\evtCatTShirt}    
\newcommand{\xcatShirtAggregate}{\evtCatShirt}      
\newcommand{\xcatShortsAggregate}{\evtCatShorts}    



%% file: generated/macros_data.tex
\newcommand{\dataI}{\num{38918}}  
\newcommand{\dataJ}{\num{2034}}  
\newcommand{\dataT}{31}  

\newcommand{\dataTau}{0.04}  
\newcommand{\dataTauLo}{0.02}  
\newcommand{\dataTauHi}{0.30}  

\newcommand{\dataInsideShareLo}{3}  
\newcommand{\dataInsideShareHi}{5}  

\newcommand{\datFullPanelTxns}{\num{31788324}}  
\newcommand{\datFullPanelCustomers}{\num{1362281}}  
\newcommand{\datFullPanelArticles}{\num{104547}}  
\newcommand{\datGiniSalesCatalog}{0.772}  
\newcommand{\datRevTopTwentyPct}{80.6}  
\newcommand{\datPriceVarCatalogPct}{92.3}  
\newcommand{\datMeanPurchases}{23.3}  
\newcommand{\datPNinetyPurchases}{60}  
\newcommand{\datOneTimeBuyerPct}{9.7}  

\newcommand{\datDressRawArticles}{\num{10266}}  
\newcommand{\datDressRawTxnsM}{3.2}  
\newcommand{\datDressVolumePct}{10.2}  
\newcommand{\datDressRawCustomers}{\num{608936}}  
\newcommand{\datDressColorGroups}{47}  
\newcommand{\datDressPatternTypes}{28}  
\newcommand{\datDressPriceVarPct}{94.5}  
\newcommand{\datDressTextPct}{99.6}  
\newcommand{\datDressMedianDescWords}{30}  

%% file: generated/macros_supply.tex
\newcommand{\lcEpsBar}{2.13}  
\newcommand{\lcLernerPct}{60.6}  
\newcommand{\lcLernerNum}{0.606}  
\newcommand{\lcIIA}{1.043}  
\newcommand{\lcMcMean}{8.92}  
\newcommand{\lcMcPosCount}{\num{1897}}  
\newcommand{\lcMcPosPct}{93.3}  
\newcommand{\lcPiOne}{0.403}  
\newcommand{\lcAlphaOne}{-0.211}  
\newcommand{\lcAlphaTwo}{-0.120}  
\newcommand{\lcValNLL}{6.0181}  
\newcommand{\lcValRSqrPct}{7.36}  

\newcommand{\sustMethodATwoPct}{-0.01}  
\newcommand{\sustMethodASevenPct}{-10.94}  
\newcommand{\sustPriceTwoPct}{1.07}  
\newcommand{\sustPriceSevenPct}{0.64}  

\newcommand{\persGainOne}{0.224}  
\newcommand{\persGainTwo}{0.260}  
\newcommand{\persGainFour}{0.511}  
\newcommand{\persGainEight}{0.616}  
\newcommand{\persGainSixteen}{0.908}  
\newcommand{\persGainThirtyTwo}{0.993}  
\newcommand{\persGainSixtyFour}{1.011}  
\newcommand{\persGainOneTwentyEight}{1.234}  
\newcommand{\persPriceShiftOne}{1.04}  
\newcommand{\persPriceShiftTwo}{0.87}  
\newcommand{\persPriceShiftFour}{0.70}  
\newcommand{\persPriceShiftEight}{0.90}  
\newcommand{\persPriceShiftSixteen}{1.23}  
\newcommand{\persPriceShiftThirtyTwo}{1.37}  
\newcommand{\persPriceShiftSixtyFour}{1.56}  
\newcommand{\persPriceShiftOneTwentyEight}{1.75}  

\newcommand{\aesthCsChangePct}{-0.16}  
\newcommand{\aesthPriceChangePct}{1.99}  
\newcommand{\aesthProfitChangePct}{41.5}  
\newcommand{\aesthTotalWelfareChangePct}{0.35}  
\newcommand{\aesthBaselineProfit}{\num{14588}}  
\newcommand{\aesthBaselineCS}{101.85}  
\newcommand{\aesthBaselineTotalWelfare}{\num{3981733}}  
\newcommand{\aesthBaselineMeanPrice}{18.75}  

%% file: generated/macros_cluster.tex
\newcommand{\lcClusterAN}{\num{5859}}
\newcommand{\lcClusterBN}{\num{4928}}
\newcommand{\lcClusterAAge}{35.1}
\newcommand{\lcClusterBAge}{37.9}
\newcommand{\lcDiSeed}{7}
\newcommand{\lcDiKUsr}{8}
\newcommand{\lcDiFrobenius}{3.25}
\newcommand{\lcDiCorr}{+0.48}

%% file: generated/macros_rec.tex
\newcommand{\recWarmN}{47}  
\newcommand{\recColdN}{730}  

\newcommand{\recPopAll}{2.2}  
\newcommand{\recPopWarm}{6.4}  
\newcommand{\recPopCold}{1.9}  

\newcommand{\recCFDefAll}{8.0}  
\newcommand{\recCFDefWarm}{42.6}  
\newcommand{\recCFDefCold}{6.3}  

\newcommand{\recCFBestAll}{7.9}  
\newcommand{\recCFBestWarm}{34.0}  
\newcommand{\recCFBestCold}{6.5}  

\newcommand{\recBetaMixAll}{7.6}  
\newcommand{\recBetaMixWarm}{10.6}  
\newcommand{\recBetaMixCold}{7.3}  

\newcommand{\recBetaDomAll}{2.8}  
\newcommand{\recBetaDomWarm}{6.4}  
\newcommand{\recBetaDomCold}{2.7}  

%% file: generated/macros_embeddings.tex
\newcommand{\embTowerDim}{64}  
\newcommand{\embCLIPDim}{512}  
\newcommand{\embItemInputDim}{562}  
\newcommand{\embUserInputDim}{85}  
\newcommand{\embCategoricalDim}{50}  
\newcommand{\embTauInfoNCE}{0.1}  

\newcommand{\embHitTen}{0.1896}  
\newcommand{\embValPairsM}{1.12}  
\newcommand{\embNumCategories}{12}  
\newcommand{\embNumItemsApprox}{\num{39969}}  
\newcommand{\embPreCovidTxnsM}{8.8}  
\newcommand{\embFullPanelTxnsM}{31.8}  

\newcommand{\embDressN}{\num{5870}}  
\newcommand{\embClusterK}{4}  
\newcommand{\embARIUserAge}{0.047}  
\newcommand{\embCosineStart}{1.00}  
\newcommand{\embCosineEnd}{-0.74}  
\newcommand{\embClusterZeroItemPct}{20}  

\newcommand{\embClusterNZero}{\num{1169}}
\newcommand{\embClusterNOne}{\num{1740}}
\newcommand{\embClusterNTwo}{\num{1736}}
\newcommand{\embClusterNThree}{\num{1225}}
\newcommand{\embClusterPriceZero}{0.0358}
\newcommand{\embClusterPriceOne}{0.0219}
\newcommand{\embClusterPriceTwo}{0.0560}
\newcommand{\embClusterPriceThree}{0.0277}
\newcommand{\embClusterSalesZero}{53.0}
\newcommand{\embClusterSalesOne}{4.7}
\newcommand{\embClusterSalesTwo}{13.1}
\newcommand{\embClusterSalesThree}{29.1}

\newcommand{\embRSqrSharesCLIPOLS}{0.21}
\newcommand{\embRSqrSharesCLIPLGB}{0.39}
\newcommand{\embRSqrPricesCLIPOLS}{0.60}
\newcommand{\embRSqrPricesCLIPLGB}{0.71}
\newcommand{\embRSqrSharesTTOLS}{0.48}
\newcommand{\embRSqrSharesTTLGB}{0.48}
\newcommand{\embRSqrPricesTTOLS}{0.58}
\newcommand{\embRSqrPricesTTLGB}{0.65}

\newcommand{\embConcatDim}{576}  

%% file: generated/macros_seas.tex
\newcommand{\seasDeltaPeakPct}{57}  

%% file: generated/macros_hed.tex
\newcommand{\hedJevonsDecline}{60.5}  
\newcommand{\hedFisherDecline}{50.6}  
\newcommand{\hedGap}{9.9}  
\newcommand{\hedTotalFeats}{684}  
\newcommand{\hedMCDraws}{500}  

\newcommand{\xcatDressHedR}{0.72}  
\newcommand{\hedGapTrousers}{29.3}  
\newcommand{\xcatTrousersHedR}{0.68}  
\newcommand{\xcatTrousersJevons}{58.8}  
\newcommand{\xcatTrousersFisher}{29.5}  
\newcommand{\hedGapSweater}{16.0}  
\newcommand{\xcatSweaterHedR}{0.69}  
\newcommand{\xcatSweaterJevons}{51.6}  
\newcommand{\xcatSweaterFisher}{35.6}  
\newcommand{\hedGapTShirt}{22.3}  
\newcommand{\xcatTShirtHedR}{0.62}  
\newcommand{\xcatTShirtJevons}{46.8}  
\newcommand{\xcatTShirtFisher}{24.6}  
\newcommand{\hedGapShirt}{16.8}  
\newcommand{\xcatShirtHedR}{0.71}  
\newcommand{\xcatShirtJevons}{60.2}  
\newcommand{\xcatShirtFisher}{43.3}  
\newcommand{\hedGapShorts}{7.8}  
\newcommand{\xcatShortsHedR}{0.64}  
\newcommand{\xcatShortsJevons}{44.0}  
\newcommand{\xcatShortsFisher}{36.3}  
\newcommand{\hedGapTop}{17.0}  
\newcommand{\hedGapBlouse}{4.3}  
\newcommand{\hedGapSkirt}{8.1}  
\newcommand{\hedGapVestTop}{9.5}  
\newcommand{\hedGapJacket}{12.8}  

\newcommand{\obTotalPct}{13.9}  
\newcommand{\obCompPct}{3.0}  
\newcommand{\obValPct}{16.9}  
\newcommand{\obTotalCILo}{16.0}  
\newcommand{\obTotalCIHi}{11.4}  
\newcommand{\obCompCILo}{0.2}  
\newcommand{\obCompCIHi}{4.1}  
\newcommand{\obValCILo}{16.5}  
\newcommand{\obValCIHi}{14.8}  
\newcommand{\obConfLevel}{90}  
\newcommand{\obBootstrapN}{200}  
\newcommand{\obCompPosPct}{92}  
\newcommand{\obValNegPct}{100}  

\newcommand{\confConfidence}{90}  
\newcommand{\confBandWidthPP}{14}  
\newcommand{\confBandFigPP}{14}  
\newcommand{\confNewDressN}{39}  
\newcommand{\confInitialPct}{82.0}  
\newcommand{\confMonthThreePct}{91}  
\newcommand{\confMonthTwelvePct}{100}  

\newcommand{\hedPooledGap}{32.8}  
\newcommand{\hedPerPeriodGap}{28.4}  
\newcommand{\hedDeepFisherDecline}{58.3}  
\newcommand{\hedDeepGap}{2.2}  

\newcommand{\hedPredPriceLo}{14.6}  
\newcommand{\hedPredPriceHi}{37.8}  

%% file: generated/macros_event.tex
\newcommand{\evtAggregatePct}{28.6}  
\newcommand{\evtWHOPct}{31.5}  
\newcommand{\evtKItem}{12}  
\newcommand{\evtKUser}{10}  

\newcommand{\evtItemRangePP}{107.9}  
\newcommand{\evtItemICtwoMostNeg}{53.0}  
\newcommand{\evtItemICsixMostPos}{54.9}  

\newcommand{\evtICzeroItems}{600}  
\newcommand{\evtIConeItems}{608}  
\newcommand{\evtICtwoItems}{639}  
\newcommand{\evtICthreeItems}{572}  
\newcommand{\evtICfourItems}{532}  
\newcommand{\evtICfiveItems}{\num{1079}}  
\newcommand{\evtICsixItems}{820}  
\newcommand{\evtICsevenItems}{\num{1190}}  
\newcommand{\evtICeightItems}{984}  
\newcommand{\evtICnineItems}{\num{1012}}  
\newcommand{\evtICtenItems}{894}  
\newcommand{\evtICelevenItems}{\num{1309}}  
\newcommand{\evtICzeroTxns}{50}  
\newcommand{\evtIConeTxns}{10}  
\newcommand{\evtICtwoTxns}{384}  
\newcommand{\evtICthreeTxns}{19}  
\newcommand{\evtICfourTxns}{186}  
\newcommand{\evtICfiveTxns}{533}  
\newcommand{\evtICsixTxns}{235}  
\newcommand{\evtICsevenTxns}{307}  
\newcommand{\evtICeightTxns}{195}  
\newcommand{\evtICnineTxns}{803}  
\newcommand{\evtICtenTxns}{417}  
\newcommand{\evtICelevenTxns}{20}  

\newcommand{\evtARIEmbTFIDF}{0.086}  
\newcommand{\evtARIEmbHM}{0.479}  
\newcommand{\evtARITFIDFHM}{0.055}  
\newcommand{\evtNMIEmbTFIDF}{0.182}  
\newcommand{\evtNMIEmbHM}{0.656}  
\newcommand{\evtNMITFIDFHM}{0.168}  

\newcommand{\evtKmeansSeedsN}{20}  
\newcommand{\evtKmeansSeedMeanPP}{103.4}  
\newcommand{\evtKmeansSeedSDPP}{10.2}  
\newcommand{\evtKmeansSeedPctFifthLo}{90.4}  
\newcommand{\evtKmeansSeedPctFifthHi}{111.0}  
\newcommand{\evtKmeansSeedRangeLo}{91}  
\newcommand{\evtKmeansSeedRangeHi}{115}  
\newcommand{\evtKmeansSeedOutlierRange}{63}  

\newcommand{\evtCatSweater}{56.1}  
\newcommand{\evtCatBlouse}{46.2}  
\newcommand{\evtCatJacket}{40.2}  
\newcommand{\evtCatShirt}{37.5}  
\newcommand{\evtCatSkirt}{33.6}  
\newcommand{\evtCatTop}{24.1}  
\newcommand{\evtCatTrousers}{6.8}  
\newcommand{\evtCatTShirt}{14.2}  
\newcommand{\evtCatShorts}{14.6}  
\newcommand{\evtCatVestTop}{23.9}  

\newcommand{\xcatTrousersKItem}{15}  
\newcommand{\xcatTrousersRangeItem}{82.5}  
\newcommand{\xcatTrousersKUser}{6}  
\newcommand{\xcatTrousersRangeUser}{25.5}  
\newcommand{\xcatTrousersUserDemo}{0.9}  
\newcommand{\xcatSweaterKItem}{8}  
\newcommand{\xcatSweaterRangeItem}{55.3}  
\newcommand{\xcatSweaterKUser}{4}  
\newcommand{\xcatSweaterRangeUser}{10.5}  
\newcommand{\xcatSweaterUserDemo}{1.0}  
\newcommand{\xcatTShirtKItem}{15}  
\newcommand{\xcatTShirtRangeItem}{152.1}  
\newcommand{\xcatTShirtKUser}{10}  
\newcommand{\xcatTShirtRangeUser}{53.9}  
\newcommand{\xcatTShirtUserDemo}{1.1}  
\newcommand{\xcatShirtKItem}{10}  
\newcommand{\xcatShirtRangeItem}{201.8}  
\newcommand{\xcatShirtKUser}{6}  
\newcommand{\xcatShirtRangeUser}{32.1}  
\newcommand{\xcatShirtUserDemo}{1.7}  
\newcommand{\xcatShortsKItem}{4}  
\newcommand{\xcatShortsRangeItem}{147.0}  
\newcommand{\xcatShortsKUser}{4}  
\newcommand{\xcatShortsRangeUser}{47.3}  
\newcommand{\xcatShortsUserDemo}{0.5}  

%% file: generated/macros_robtau.tex
\newcommand{\robLernerSwingPP}{39}  
\newcommand{\robMcPosHi}{94.1}  
\newcommand{\robMcPosLo}{61.0}  

\newcommand{\robTauTwoVZero}{14.44}
\newcommand{\robTauTwoEps}{2.122}
\newcommand{\robTauTwoMc}{9.42}
\newcommand{\robTauTwoLerner}{0.579}
\newcommand{\robTauTwoMcPos}{\num{1913}}
\newcommand{\robTauFourVZero}{13.62}
\newcommand{\robTauFourEps}{2.128}
\newcommand{\robTauFourMc}{\lcMcMean}  
\newcommand{\robTauFourLerner}{\lcLernerNum}  
\newcommand{\robTauFourMcPos}{\lcMcPosCount}  
\newcommand{\robTauSixVZero}{13.11}
\newcommand{\robTauSixEps}{2.134}
\newcommand{\robTauSixMc}{8.46}
\newcommand{\robTauSixLerner}{0.632}
\newcommand{\robTauSixMcPos}{\num{1870}}
\newcommand{\robTauEightVZero}{12.72}
\newcommand{\robTauEightEps}{2.140}
\newcommand{\robTauEightMc}{8.01}
\newcommand{\robTauEightLerner}{0.658}
\newcommand{\robTauEightMcPos}{\num{1853}}
\newcommand{\robTauTenVZero}{12.41}
\newcommand{\robTauTenEps}{2.146}
\newcommand{\robTauTenMc}{7.56}
\newcommand{\robTauTenLerner}{0.683}
\newcommand{\robTauTenMcPos}{\num{1834}}
\newcommand{\robTauFifteenVZero}{11.80}
\newcommand{\robTauFifteenEps}{2.160}
\newcommand{\robTauFifteenMc}{6.43}
\newcommand{\robTauFifteenLerner}{0.748}
\newcommand{\robTauFifteenMcPos}{\num{1772}}
\newcommand{\robTauTwentyVZero}{11.32}
\newcommand{\robTauTwentyEps}{2.174}
\newcommand{\robTauTwentyMc}{5.24}
\newcommand{\robTauTwentyLerner}{0.817}
\newcommand{\robTauTwentyMcPos}{\num{1645}}
\newcommand{\robTauThirtyVZero}{10.55}
\newcommand{\robTauThirtyEps}{2.204}
\newcommand{\robTauThirtyMc}{2.59}
\newcommand{\robTauThirtyLerner}{0.972}
\newcommand{\robTauThirtyMcPos}{\num{1240}}

%% file: generated/macros_robwelfare.tex
\newcommand{\robProfitDropLo}{36}  
\newcommand{\robProfitDropHi}{44}  

\newcommand{\robTauTwoPS}{\num{8282}}
\newcommand{\robTauTwoProfitDelta}{43.94}  
\newcommand{\robTauTwoCSDelta}{0.08}  
\newcommand{\robTauFourPS}{\aesthBaselineProfit}  
\newcommand{\robTauFourProfitDelta}{41.46}  
\newcommand{\robTauFourCSDelta}{0.16}  
\newcommand{\robTauSixPS}{\num{28249}}
\newcommand{\robTauSixProfitDelta}{39.48}  
\newcommand{\robTauSixCSDelta}{0.25}  
\newcommand{\robTauEightPS}{\num{39548}}
\newcommand{\robTauEightProfitDelta}{37.80}  
\newcommand{\robTauEightCSDelta}{0.33}  
\newcommand{\robTauTenPS}{\num{51645}}
\newcommand{\robTauTenProfitDelta}{36.34}  
\newcommand{\robTauTenCSDelta}{0.42}  

%% file: generated/macros_robieb.tex
\newcommand{\robIEBlegacyValNLL}{6.1279}
\newcommand{\robIEBlegacyValRSqrPct}{5.67}
\newcommand{\robIEBlegacyItemParams}{\num{41474}}
\newcommand{\robIEBlegacyAlphaOne}{-0.164}
\newcommand{\robIEBlegacyAlphaTwo}{-0.119}
\newcommand{\robIEBlegacyPiOne}{0.363}
\newcommand{\robIEBlegacyEps}{2.09}
\newcommand{\robIEBlegacyLerner}{0.621}
\newcommand{\robIEBlegacyMcPos}{0.915}
\newcommand{\robIEBlegacyIIA}{0.683}
\newcommand{\robIEBlegacyItemRho}{1.000}
\newcommand{\robIEBlegacyUserRho}{1.000}
\newcommand{\robIEBlegacyItemRhoSpearman}{1.000}

\newcommand{\robIEBfcValNLL}{6.0408}
\newcommand{\robIEBfcValRSqrPct}{7.03}
\newcommand{\robIEBfcItemParams}{\num{156162}}
\newcommand{\robIEBfcAlphaOne}{-0.200}
\newcommand{\robIEBfcAlphaTwo}{-0.119}
\newcommand{\robIEBfcPiOne}{0.449}
\newcommand{\robIEBfcEps}{\lcEpsBar}  
\newcommand{\robIEBfcLerner}{\lcLernerNum}  
\newcommand{\robIEBfcMcPos}{0.933}  
\newcommand{\robIEBfcIIA}{\lcIIA}  
\newcommand{\robIEBfcItemRho}{0.743}
\newcommand{\robIEBfcUserRho}{0.760}
\newcommand{\robIEBfcItemRhoSpearman}{0.713}

\newcommand{\robIEBmainValNLL}{6.0181}
\newcommand{\robIEBmainValRSqrPct}{\lcValRSqrPct}  
\newcommand{\robIEBmainItemParams}{\num{172546}}
\newcommand{\robIEBmainAlphaOne}{\lcAlphaOne}  
\newcommand{\robIEBmainAlphaTwo}{\lcAlphaTwo}  
\newcommand{\robIEBmainPiOne}{\lcPiOne}  
\newcommand{\robIEBmainEps}{\lcEpsBar}  
\newcommand{\robIEBmainLerner}{\lcLernerNum}  
\newcommand{\robIEBmainMcPos}{0.933}  
\newcommand{\robIEBmainIIA}{\lcIIA}  
\newcommand{\robIEBmainItemRho}{0.791}
\newcommand{\robIEBmainUserRho}{0.895}
\newcommand{\robIEBmainItemRhoSpearman}{0.767}

%% file: generated/macros_bnch.tex
\newcommand{\bnchPCADim}{16}  
\newcommand{\bnchJ}{500}  

\newcommand{\bnchHedPooledR}{0.40}  
\newcommand{\bnchHedPooledDecline}{27.7}  
\newcommand{\bnchHedPerPeriodR}{0.39}  
\newcommand{\bnchHedPerPeriodDecline}{32.1}  
\newcommand{\bnchHedDeepR}{0.67}  
\newcommand{\bnchHedDeepStdPP}{2.9}  

%% file: generated/macros_lcnocf.tex
\newcommand{\lcAlphaOneNoCF}{-0.198}  
\newcommand{\lcAlphaTwoNoCF}{-0.118}  
\newcommand{\lcPiOneNoCF}{0.431}  


%% file: generated/macros_xcat.tex
\newcommand{\xcatTrousersI}{\num{68757}}  
\newcommand{\xcatTrousersJ}{\num{1818}}  
\newcommand{\xcatTrousersAlphaZero}{-0.142}  
\newcommand{\xcatTrousersAlphaOne}{-0.124}  
\newcommand{\xcatTrousersPiZero}{0.355}  
\newcommand{\xcatTrousersEps}{2.21}  
\newcommand{\xcatTrousersValNLLStatic}{11.18}  
\newcommand{\xcatTrousersValNLLTime}{10.65}  
\newcommand{\xcatTrousersRSqrP}{0.929}  
\newcommand{\xcatTrousersRSqrAes}{0.012}  
\newcommand{\xcatTrousersWT}{62.8}  

\newcommand{\xcatSweaterI}{\num{22808}}  
\newcommand{\xcatSweaterJ}{\num{1011}}  
\newcommand{\xcatSweaterAlphaZero}{-0.191}  
\newcommand{\xcatSweaterAlphaOne}{-0.139}  
\newcommand{\xcatSweaterPiZero}{0.539}  
\newcommand{\xcatSweaterEps}{1.85}  
\newcommand{\xcatSweaterValNLLStatic}{5.53}  
\newcommand{\xcatSweaterValNLLTime}{6.12}  
\newcommand{\xcatSweaterRSqrP}{0.850}  
\newcommand{\xcatSweaterRSqrAes}{0.141}  
\newcommand{\xcatSweaterWT}{47.9}  

\newcommand{\xcatTShirtI}{\num{26053}}  
\newcommand{\xcatTShirtJ}{\num{1367}}  
\newcommand{\xcatTShirtAlphaZero}{-0.210}  
\newcommand{\xcatTShirtAlphaOne}{-0.210}  
\newcommand{\xcatTShirtPiZero}{0.375}  
\newcommand{\xcatTShirtEps}{1.17}  
\newcommand{\xcatTShirtValNLLStatic}{5.58}  
\newcommand{\xcatTShirtValNLLTime}{5.69}  
\newcommand{\xcatTShirtRSqrP}{0.730}  
\newcommand{\xcatTShirtRSqrAes}{0.282}  
\newcommand{\xcatTShirtWT}{65.1}  

\newcommand{\xcatShirtI}{\num{4420}}  
\newcommand{\xcatShirtJ}{\num{496}}  
\newcommand{\xcatShirtAlphaZero}{-0.244}  
\newcommand{\xcatShirtAlphaOne}{-0.143}  
\newcommand{\xcatShirtPiZero}{0.374}  
\newcommand{\xcatShirtEps}{1.91}  
\newcommand{\xcatShirtValNLLStatic}{6.76}  
\newcommand{\xcatShirtValNLLTime}{6.10}  
\newcommand{\xcatShirtRSqrP}{0.930}  
\newcommand{\xcatShirtRSqrAes}{0.010}  
\newcommand{\xcatShirtWT}{56.6}  

\newcommand{\xcatShortsI}{\num{11693}}  
\newcommand{\xcatShortsJ}{\num{1029}}  
\newcommand{\xcatShortsAlphaZero}{-0.228}  
\newcommand{\xcatShortsAlphaOne}{-0.144}  
\newcommand{\xcatShortsPiZero}{0.335}  
\newcommand{\xcatShortsEps}{1.50}  
\newcommand{\xcatShortsValNLLStatic}{8.38}  
\newcommand{\xcatShortsValNLLTime}{8.35}  
\newcommand{\xcatShortsRSqrP}{0.801}  
\newcommand{\xcatShortsRSqrAes}{0.242}  
\newcommand{\xcatShortsWT}{55.9}  

\newcommand{\xcatDressRSqrP}{0.518}  
\newcommand{\xcatDressRSqrAes}{0.437}  

\newcommand{\xcatAlphaMinAbs}{0.24}  
\newcommand{\xcatAlphaMaxAbs}{0.12}  
\newcommand{\xcatGapMin}{0.00}  
\newcommand{\xcatGapMax}{0.10}  
\newcommand{\xcatEpsMin}{1.17}  
\newcommand{\xcatEpsMax}{2.21}  

%% file: 00_abstract.tex
\begin{abstract}
Aesthetics drives product differentiation in industries such as fashion, interior decor, luxury goods, real estate and hospitality. However, visual differentiation is hard to encode in formal economic analysis. This paper develops a methodology to parse, process and model the aesthetic dimension and applies it to fast fashion. It works with millions of purchase records from H\&M in the Netherlands, including product images, text descriptions, prices, and consumer demographics. I fine-tune Fashion CLIP embeddings with a three-tower approach that builds separate channels for product visuals and text, consumer history, and price, which makes downstream analysis tractable and scalable. The embeddings feed a latent-class deep demand system that captures price and taste sensitivities through deep nets, recovers rich substitution patterns, reveals meaningful heterogeneity, and performs much better than competing alternatives. Then, a supply-side inversion recovers sensible markups and costs and supports conduct tests and counterfactuals on sustainability practices. I also estimate machine learning hedonic pricing models that perform much better than competing alternatives. This model allows us to construct quality-adjusted price indices, make it possible to price completely new designs, and with an Oaxaca-Blinder decomposition reveal the underlying sources of price changes. Finally, a Poisson event study around the COVID-19 lockdown shows that the range of demand responses across embedding-based product and user clusters exceeds anything recoverable from simple text-based attributes or demographic labels alone. All these findings show that high-dimensional embeddings can be first class citizens in disciplined economic and statistical analysis. The methodology is portable to any market where products are differentiated along sensory dimensions that are hard to encode but meaningfully important for consumer choices.
\end{abstract}

%% file: 01_introduction.tex
\section{Introduction}
\label{sec:introduction}

Aesthetics drive product differentiation in fashion, furniture, automobiles, and consumer electronics. A dress's commercial success comes down to its silhouette, color palette, fabric, drape, proportion, texture, occasion fit, embellishment, print, and trend alignment. A human eye picks these up instantly. Encoding them into the structured variables that empirical economic analysis requires is the hard part. Demand systems need measurable dimensions of differentiation \citep{berry1995automobile, train2009discrete}. Hedonic pricing models need observable product characteristics \citep{rosen1974, court1939}. Event studies need meaningful product groupings. Whenever differentiation is primarily visual, all three frameworks hit the same obstacle, which is that the attributes that matter most are precisely the ones the researcher cannot measure easily. Fast fashion markets show this at scale, and therefore are an ideal ground to build models that can capture the great depth contained within ``taste".

Numerical product embeddings learned from data have entered economic analysis from multiple domains, including text \citep{gentzkow2019text, hansen2024text}, images \citep{dell2025deep, han2021using, zhang2022airbnb}, product design \citep{burnap2023aesthetic}, and financial portfolios \citep{gabaix2025asset}. In demand estimation, several recent papers incorporate embeddings from pre-trained models into structural choice frameworks \citep{quan2024extracting, bach2025adventures, magnolfi2023triplet, adam2024machine, gabel2022product, sifringer2020enhancing, wang2020deep, berry2024nonparametric}. \citet{compiani2025demand} is the closest specification to this paper, incorporating PCA-reduced pre-trained image and text embeddings as random-coefficient characteristics in mixed logit applied to Amazon data. \citet{donnelly2021counterfactual} learn joint product and consumer embeddings via variational inference\footnote{Variational inference is a technique for fitting probabilistic models with unobserved variables. The exact posterior distribution over those variables is usually intractable. Variational inference instead picks the closest member of a tractable family of distributions and optimizes that proxy. The technique scales to very large datasets where classical Bayesian methods do not.} for grocery basket prediction, producing representations that predict complementarity and substitution but are not deployed within a utility-maximizing demand model. \citet{bajari2025hedonic} use BERT and ResNet embeddings in a multi-task neural network for hedonic price indices. These approaches either treat pre-trained representations as fixed inputs to downstream estimation or learn embeddings without imposing economic structure during training. 

So what is new in this paper? First, the embeddings are fine-tuned on purchase behavior rather than borrowed off the shelf, and this gives us better consumer and product representation. Secondly, we not only have product embeddings but also user embeddings allowing us to capture a great range of sensitivities than before. Third, we do not reduce embeddings via PCA but show that it is possible to work fully with dense embeddings, and this allows a greater diversity of counterfactuals such as generative fashion design. Lastly, we also estimate hedonic and event study models to show that embeddings are not just a demand-side tool but can be used across the board. The analysis proceeds in two clearly separated stages, divided by a clean time break. Representation learning uses purchase records from September 2018 through June 2019. All three downstream applications run on a held-out window from July 2019 onwards.

The first stage is representation learning. I fine-tune a pre-trained vision-language model (OpenAI CLIP ViT-B/32, \citet{radford2021learning}) on roughly $\embPreCovidTxnsM$ million H\&M purchase records from a first-window sample (September 2018 through June 2019) using a contrastive purchase-prediction objective \citep{oord2018representation}. The architecture is a three-tower neural network trained jointly across $\embNumCategories$ product categories and $\embNumItemsApprox$ items. The item tower maps $\embItemInputDim$-dimensional features, $\embCLIPDim$ from the pre-trained CLIP encoder plus $\embCategoricalDim$ from ten categorical product attributes, to a $\embTowerDim$-dimensional aesthetic embedding. The consumer tower maps purchase history and demographics to a $\embTowerDim$-dimensional user embedding. The price tower maps log reference price through a small network to a separate $\embTowerDim$-dimensional price embedding. The contrastive objective pulls a buyer's embedding toward the purchased item and pushes it away from every other item in the same product category. This in-category negative sampling forces the item tower to encode aesthetic similarity from co-purchase patterns while the dedicated price tower absorbs product-level price variation. 

The fine-tune sits between two alternatives. Training a visual model directly on raw product images needs more data and compute than this study has. Plugging in an off-the-shelf representation is a reasonable fallback, and a pre-trained vision-language model is the natural choice because it has already learned to place product images and product descriptions into a shared vector space from a large external image-caption corpus. The approach taken here fine-tunes this pre-trained CLIP model once more on the purchase records in this catalogue, which distils the information in the pre-trained vector that matters for in-category choice. The dress demand master in \cref{sec:choice_model} then concatenates a separately extracted $\embCLIPDim$-dimensional FashionCLIP vector \citep{chia2022contrastive} alongside the three-tower item vector, so the master model gets the richest feasible visual signal. The same fine-tune also produces a $\embTowerDim$-number user embedding for every consumer, learned from first-window purchase behaviour, which carries richer information than any hand-coded set of demographic attributes. The two-stage split also gives fine-tuning access to all $\embNumCategories$ product categories and roughly $\embPreCovidTxnsM$ million H\&M transactions, while the downstream estimation runs on the dress sample alone, so representation learning gets the breadth it needs for generalisation and structural estimation gets the specific choice set it needs for identification.

The second stage deploys the learned representations. The embeddings are frozen and used as inputs to three distinct economic applications on a held-out second time window (July 2019 onwards), each asking a different question of the same representations. \Cref{sec:choice_model} estimates a latent-class deep logit demand model and inverts the estimated system through a supply-side markup recovery to recover implied marginal costs. \Cref{sec:counterfactuals} runs the counterfactual policy analysis on the estimated primitive. \Cref{sec:hedonic} uses the embeddings as regressors in a machine learning hedonic pricing model to construct quality-adjusted price indices and decompose price changes. \Cref{sec:event_study} clusters products and users in embedding space and tracks the dispersion of demand responses to the COVID-19 lockdown within a single retail category.

The demand model recovers meaningful consumer-specific price sensitivity through a small neural network, the $\alpha$-net, whose final layer forces the price coefficient to be negative for every consumer. A latent-class extension separates two consumer segments and delivers a realistic mean own-price elasticity. A supply-side inversion on the same estimated system yields positive implied marginal costs for the large majority of the $\dataJ$ dresses, with a mean Lerner index close to H\&M's reported gross margin (\cref{sec:choice_model}). The estimated demand-supply system then feeds a battery of policy-relevant counterfactuals (\cref{sec:counterfactuals}), including the profitability of sustainability practices, why visual differentiation in trend-chasing fast fashion is critical to profitability, a personalised pricing ladder over demographic segments, and a validation of the taste-matching network as a stand-alone recommender.

\begin{figure}[htbp]
  \centering
  \includegraphics[width=0.9\textwidth]{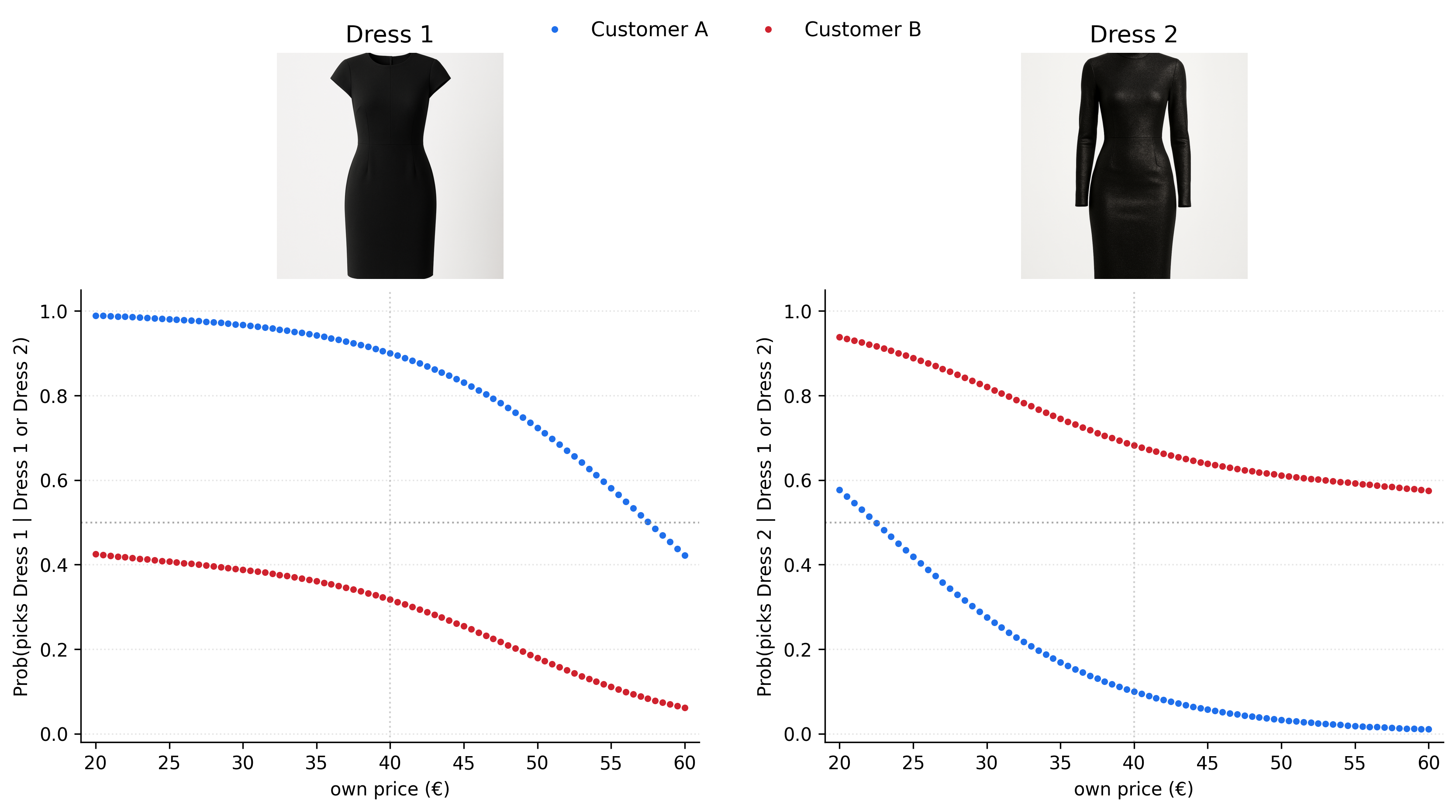}
  \caption{Two synthetic dresses produced by a generative image model. Each dot is the probability that a customer picks the labelled dress given they pick one of these two, under the two-type latent-class deep logit model of \Cref{sec:theory:logit}, evaluated at the named own price with the other dress held at a forty-euro reference. Blue: Customer A. Red: Customer B.}
  \label{fig:intro_teaser}
\end{figure}

\Cref{fig:intro_teaser} isolates aesthetic taste and price sensitivity in a single example. For two real H\&M customers drawn from the estimated demand system and two synthetic dresses that look almost identical at a glance, each panel sweeps one dress's own price across a realistic band while holding the other dress at a forty-euro reference price. Every dot is one direct evaluation of the two-type latent-class deep logit demand model for dresses. Two patterns emerge. First, every curve slopes downward, which is the price-sensitivity network responding to own-price increases. Second, the two customers disagree about which dress they prefer: in the Dress 1 panel, Customer A sits above Customer B at every price shown; in the Dress 2 panel, the pattern mirrors, and Customer B sits above Customer A. This is the ranking flip that the taste-matching network produces once it reads small design differences (sleeve length, fabric sheen, neckline) that a hand-entered attribute list (colour, length, silhouette, garment group) would collapse into a single identical label.

The hedonic price indices reveal a product-mix composition bias of roughly $\hedGap$ percentage points over two years relative to matched-item indices, meaning the average product sold in year two differs aesthetically from the average product sold in year one. An Oaxaca-Blinder decomposition points to quality shifts that categorical controls cannot capture, rather than aggregate price changes (\cref{sec:hedonic}). The lockdown event study surfaces wide dispersion in demand responses across embedding-based clusters within a single retail category. Both item and user clusters reveal heterogeneity substantially wider than anything the demographic labels alone can recover, and several embedding clusters cross-cut the retailer's own product taxonomy in ways no text-only or category-label partition can reproduce (\cref{sec:event_study}).

The methodology is portable to any market where products differ along sensory dimensions that matter for consumer choice but are hard to put into structured variables. The recipe is the same in each case. Start with a suitable pre-trained encoder, image models for visual products, text models for prose-heavy categories, audio models for music. Fine-tune it on observed purchase behavior using a contrastive loss with within-category negatives. Split the signal into dedicated channels whenever price and match quality must be identified independently. Freeze the learned representations and feed them to standard structural estimation. Luxury goods, interior decor, furniture, hospitality, fine dining, art markets, and high-end real estate all present the same measurement problem and the same solution path.

The framework offers a scalable way to relax the Independence of Irrelevant Alternatives restriction in applied demand estimation. Mapping visual similarity into substitution patterns produces cross-price elasticities that are non-proportional across product pairs at the consumer level and aggregate into realistic market-level substitution. Ignoring unobserved aesthetic differentiation can bias inflation measurement through quality-unadjusted price indices, misstate market power by reattributing differentiation rents to raw price sensitivity, and distort the welfare gain from product variety.

So where can practitioners apply this machinery? First, we can employ such models to get consumer-specific price and aesthetic sensitivities. These can be useful in downstream optimization tasks. Secondly, we can use the model to solve cold-start problems such as pricing new products or recommending products to new users. Thirdly, we can use the model to do ``generative fashion design", where design teams can leverage this model to run quick computer simulations to study both the price and demand impact of different designs. Fourthly, we can use the hedonic model to better understand a firm's (or the market's) pricing decisions and how they have evolved over time. This can be useful for both the firm itself and for policymakers who are interested in understanding the market dynamics. Lastly, the user and product embeddings themselves are very useful in causal machine learning applications, where one can use a two-step (cluster-then-estimate) or even a single-step (cluster-and-estimate) approach to estimating heterogeneous treatment effects across products and users.

Several limitations apply. The analysis uses data from a single retailer in one market segment over two years. The demand model is a deep neural network trained with stochastic gradient descent, which makes convergence hard to certify, standard errors hard to obtain, computation heavy, and estimates sensitive to architectural choices in the $\alpha$-net and the taste projection layers. The logit specification imposes IIA at the individual level, though heterogeneous price sensitivity generates market-level departures. The model abstracts from size and fit availability, browsing behavior, wardrobe complementarity, sequential purchasing dynamics, and competitor pricing or assortment responses.

The paper proceeds as follows. \Cref{sec:data} describes the data. \Cref{sec:embeddings} develops the three-tower representation learning framework. \Cref{sec:theory:logit} sets out the theoretical framework for the latent-class deep logit and the Bertrand-Nash markup recovery. \Cref{sec:choice_model} presents the demand and supply estimates and the seasonal extension. \Cref{sec:counterfactuals} runs the counterfactual policy analysis. \Cref{sec:hedonic} constructs quality-adjusted hedonic price indices. \Cref{sec:event_study} estimates heterogeneous lockdown effects. \Cref{sec:conclusion} concludes.

%% file: 02_data.tex
%
%
\section{Data and Context}
\label{sec:data}

The empirical analysis uses transaction data from H\&M, the world's second-largest fashion retailer by revenue, covering a single national market.\footnote{The data come from the Netherlands. Two independent checks confirm this. First, several hundred products in the catalog carry explicit ``COLLAB NL'' labels in their product names, corresponding to Netherlands-only H\&M designer collaborations sold during the sample period. Second, prices in the raw transaction file are in H\&M's internal units; multiplying by the community-derived rescaling factor of 590 produces EUR amounts that match H\&M Netherlands list prices for the same items verified against the retailer's website. The competition page that originally distributed the data notes that the currency was deliberately obfuscated to avoid revealing the market; cross-referencing product names and price levels identifies the market as the Netherlands.} The dataset contains $\datFullPanelTxns$ transactions from $\datFullPanelCustomers$ customers purchasing $\datFullPanelArticles$ unique articles between September 20, 2018, and September 22, 2020. \Cref{tab:data_structure} summarizes the four constituent files.

\begin{table}[H]
\centering
\small
\caption{Data Structure}
\label{tab:data_structure}
\begin{tabular}{lrp{10cm}}
\toprule
File & Observations & Key Variables \\
\midrule
Transactions & 31.8M & date (734 days: Sept 2018 to Sept 2020), customer, product, price, channel (online 70.4\%, store 29.6\%) \\
Articles & 105K & hierarchy (5 index groups, 21 garment groups, 131 types, 250 departments), aesthetic attributes (50 colors, 30 patterns), text descriptions \\
Customers & 1.36M & age (16 to 99), gender (inferred), postal code (hashed), club membership status, fashion news frequency \\
Images & 105K & product photographs \\
\bottomrule
\end{tabular}
\end{table}

\begin{figure}[H]
\centering
\includegraphics[width=0.7\textwidth]{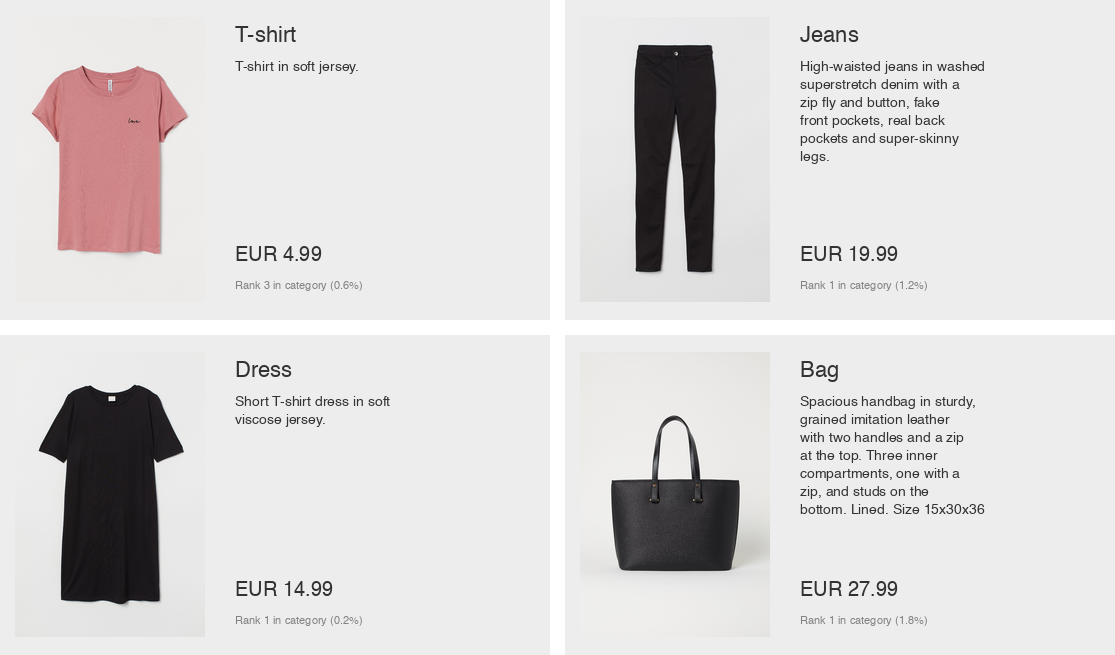}
\caption{Product assortment examples spanning price points and categories.}
\label{fig:product_examples}
\end{figure}

\Cref{fig:product_examples} shows representative products spanning price points and categories in the catalogue. The distribution of purchases exhibits positive skew across both products and customers. As detailed in \cref{tab:product_dists}, product revenues are highly concentrated; the SKU-level Gini coefficient equals $\datGiniSalesCatalog$, with the top 20\% of products generating $\datRevTopTwentyPct\%$ of total revenue. Pervasive promotional pricing yields temporal price variation for $\datPriceVarCatalogPct\%$ of articles, securing identification for the subsequent demand specifications.

On the consumer side (\cref{tab:product_dists}), the average purchase frequency rests at $\datMeanPurchases$ transactions, but the 90th percentile reaches $\datPNinetyPurchases$. This reflects a long tail of frequent buyers whose dense purchase histories provide strong signals for learning consumer embeddings, while the remaining infrequent buyers contribute primarily through demographic inputs.

Survey evidence from Dutch consumers provides a useful benchmark for the customer panel in this dataset. \citet{Leinenga2019} surveys 293 Dutch fashion consumers and finds wide variation in purchase frequency: 2\% buy clothing each week, 26\% each month, and 44\% once every six months. The average Dutch wardrobe holds 173 items, of which 28\% goes unworn in a given year, and approximately 46 new fashion items are purchased annually per person. This distribution is broadly consistent with the long right tail in \cref{tab:product_dists}, where purchase frequency spans from 2 to over $\datPNinetyPurchases$ transactions and $\datOneTimeBuyerPct\%$ of customers make only a single purchase over the two-year window. The same survey also documents that consumers who treat fashion as central to their identity buy more often and substitute less across product types, a split between consumer types that reappears in the latent-class demand model in \cref{sec:choice_model}.

Beyond static distributions, fast-fashion retail is defined by product turnover. \Cref{tab:fast_fashion_dynamics} summarises weekly new-article introductions, article lifespan, and the basic/fashion/fashion-basic tier split that governs markdown and clearance cycles.

\begin{table}[H]
\centering
\small
\caption{Market Distributions: Products and Customers}
\label{tab:product_dists}
\begin{tabular}{lrrr}
\toprule
 & \multicolumn{1}{c}{Products} & \multicolumn{2}{c}{Customers} \\
\cmidrule(lr){2-2}\cmidrule(lr){3-4}
Statistic & Price (EUR) & Purchases & Spending (EUR) \\
\midrule
10th percentile & 5.96 & 2 & 24.98 \\
25th percentile & 9.33 & 3 & 51.96 \\
Median & 14.99 & 9 & 144.90 \\
75th percentile & 19.99 & 27 & 413.70 \\
90th percentile & 29.99 & 60 & 957.51 \\
\midrule
Mean (SD) & 16.42 (11.32) & 23.3 (39.2) & 383.14 (707.00) \\
Range & 0.01--349.00 & -- & -- \\
\midrule
Gini coefficient & 0.772 & -- & -- \\
Top-20\% revenue share & 80.6\% & -- & -- \\
One-time rate & -- & 9.7\% & -- \\
\midrule
\multicolumn{4}{l}{\small\textit{Pricing dynamics (all 104{,}547 articles)}} \\
Articles with price variation & 92.3\% & -- & -- \\
Median distinct prices/article & 13 & -- & -- \\
Median max markdown depth & 60.0\% & -- & -- \\
Same-day markdowns & 43.5\% & -- & -- \\
\bottomrule
\end{tabular}
\end{table}

\begin{table}[H]
\centering
\small
\caption{Fast Fashion Product Dynamics}
\label{tab:fast_fashion_dynamics}
\begin{minipage}[t]{0.43\textwidth}
\centering
\textit{Panel A: Product Lifecycle (all 104{,}547 articles)}\\[4pt]
\begin{tabular}{lr}
\toprule
Statistic & Value \\
\midrule
New articles/week (median) & 706 \\
New articles/week (mean) & 986 \\
Median lifespan & 205 days \\
Lifespan $<$ 180 days & 45.3\% \\
Lifespan $<$ 90 days & 26.2\% \\
\bottomrule
\end{tabular}
\end{minipage}%
\hfill
\begin{minipage}[t]{0.53\textwidth}
\centering
\textit{Panel B: Three-Tier Supply Chain$^{\dagger}$}\\[4pt]
\setlength{\tabcolsep}{3pt}
\begin{tabular}{lrrrr}
\toprule
Tier & Articles & Revenue & p50 (EUR) & Lifespan \\
\midrule
Basic        & 15.3\% & 7.1\%  &  9.40 & 155d \\
Fashion      & 23.0\% & 29.4\% & 18.93 & 238d \\
Fashion-B.   & 61.7\% & 63.5\% & 13.03 & 202d \\
\bottomrule
\end{tabular}
\end{minipage}
\\[6pt]
\raggedright\footnotesize $^{\dagger}$Author-constructed classification from product metadata (garment group, price level, and lifespan). Basic = staples with low price CV; Fashion = trend-driven items; Fashion-Basic = hybrid.
\end{table}

\subsection{Primary Estimation Sample: Dresses}

The subsequent empirical analysis primarily uses the dress category as the estimation sample. The raw dress catalogue contains $\datDressRawArticles$ articles and $\datDressRawTxnsM$ million transactions ($\datDressVolumePct\%$ of total volume) from $\datDressRawCustomers$ customers. Dresses are suited for studying aesthetic differentiation for three reasons. The category carries high visual variance, with $\datDressColorGroups$ color groups and $\datDressPatternTypes$ pattern types producing far more visual combinations than standard categorical controls can capture. Seasonal demand amplitude is pronounced, peak summer volume exceeds the January trough by a factor of 3.2, which gives the seasonal demand model real temporal variation to fit. And $\datDressPriceVarPct\%$ of dress articles show within-product price variation, which is what the demand model needs to identify price elasticities. All $\embDressN$ dress articles in the embedding sample have product photographs, and $\datDressTextPct\%$ carry text descriptions (median $\datDressMedianDescWords$ words), confirming the three input streams for the joint embedding in \cref{sec:embeddings}.

Downstream sample filters narrow this raw set in two steps. The embedding-validation cuts in \cref{sec:embeddings} use the $\embDressN$ dress articles that remain after applying CLIP image-extraction filters (products must have a usable product photograph with non-trivial pixel coverage). The demand-estimation sample in \cref{sec:choice_model} narrows further to $\dataJ$ dress articles and $\dataI$ consumers over a $\dataT$-week pre-COVID window (July 2019 through January 2020), chosen to avoid contamination from the lockdown demand regime documented in \cref{sec:event_study}. Most of the article-count drop from $\embDressN$ to $\dataJ$ comes from two cuts inside that window: keeping only SKUs that fall inside the top 99 percent of category sales coverage, and requiring each SKU to match an entry in the three-tower embedding index. The consumer panel is trimmed in parallel by requiring at least two purchases per customer. Each of these filters is a mechanical data requirement rather than a modelling choice.

\subsection{Cross-Category Estimation Samples}
\label{sec:data:per_category}

The paper focuses on dresses for the main results, but the demand model runs identically on five additional categories. All six use the same pre-COVID estimation window ($T = \dataT$ weeks, July 2019 to January 2020, online channel only). \Cref{tab:per_category_samples} reports the per-category sample sizes, purchase counts, and median prices. The hedonic analysis in \cref{sec:hedonic} and the lockdown event study in \cref{sec:event_study} draw on the same product universe but extend to the full two-year panel.

\begin{table}[H]
\centering
\small
\caption{Per-category estimation samples}
\label{tab:per_category_samples}
\setlength{\tabcolsep}{4pt}
\begin{tabular}{lrrrrr}
\toprule
Category & Products $J$ & Consumers $I$ & Weeks $T$ & Purchases & Price p50 (EUR) \\
\midrule
Dress    & \num{2034}  & \num{38918}  & 31 & \num{162606}  & 19.21 \\
Trousers & \num{1818}  & \num{68757}  & 31 & \num{272949}  & 18.99 \\
T-shirt  & \num{1367}  & \num{26053}  & 31 & \num{87053}   &  9.55 \\
Sweater  & \num{1011}  & \num{22808}  & 31 & \num{69091}   & 14.72 \\
Shorts   & \num{1029}  & \num{11693}  & 31 & \num{42519}   & 12.61 \\
Shirt    & \num{496}   & \num{4420}   & 31 & \num{12293}   & 14.72 \\
\bottomrule
\multicolumn{6}{l}{\footnotesize All categories: pre-COVID window (Jul 2019 to Jan 2020, online channel). Price p50 from article-month panel.}
\end{tabular}
\end{table}

%% file: 03_embeddings.tex
%
%
%
\section{Representation Learning}
\label{sec:embeddings}

\subsection{Three-Tower Architecture}
\label{subsec:three_tower}

The first stage of the empirical pipeline compresses high-dimensional categorical features and raw product images into concise embeddings that feed the structural models in \cref{sec:choice_model,sec:hedonic,sec:event_study}. Throughout the paper I use the term embedding to mean a short numerical vector that summarizes a product, consumer, or price in a learned space. Vectors that sit close to each other in that space represent objects the model judges to be economically similar.

I fine-tune the embeddings once and then reuse them in everything that follows.\footnote{Fine-tuning starts with a neural network already trained on a large external dataset and continues training its parameters on task-specific data. In this paper, the pre-trained CLIP image-text encoder is used as a frozen 512-dimensional feature extractor. What is trained on H\&M purchase records is the three-tower projection on top of it, which compresses the CLIP vector together with ten product categoricals into a 64-dimensional task-specific embedding.} \Cref{fig:main_pipeline} summarizes the data flow from raw inputs to frozen embeddings to structural models.
\begin{figure}[H]
\centering
\begin{tikzpicture}[
  node distance=0.6cm and 1.2cm,
  arr/.style={-{Stealth[length=5pt]}, line width=0.8pt, color=gray},
  lbl/.style={font=\scriptsize\color{gray}, midway, above, yshift=1pt},
  box/.style={draw, rounded corners, minimum height=0.6cm, align=center, font=\footnotesize}
]
  \node[box] (img)   at (0, 0)     {Product\\Images};
  \node[box] (clip)  at (4.2, 0)   {CLIP\\$\mathbb{R}^{512}$};
  \node[box] (tower) at (8.4, 0)   {Three-Tower\\InfoNCE};
  \node[box, thick] (emb)   at (12.6, 0)  {Embeddings\\$\mathbf{v}_j,\mathbf{d}_i,\mathbf{p}_j$};
  \node[font=\scriptsize, below=0.06cm of emb, xshift=1cm] {frozen};

  \draw[arr] (img)   -- node[lbl]{encode} (clip);
  \draw[arr] (clip)  -- node[lbl]{fine-tune}  (tower);
  \draw[arr] (tower) -- node[lbl]{extract}(emb);

  \node[box]    (event)   at (5.6, -2.4)  {Event\\Study};
  \node[box]     (hedonic) at (9.4, -2.4)  {Hedonic\\Indices};
  \node[box] (demand)  at (13.2,-2.4)  {Deep Logit\\Demand};
  \node[box]    (supply)  at (13.2,-4.5)  {Supply-Side\\Markups};

  \draw[arr] (emb.south) -- ++(0,-0.5) -| (event.north);
  \draw[arr] (emb.south) -- ++(0,-0.5) -| (hedonic.north);
  \draw[arr] (emb.south) -- ++(0,-0.5) -| (demand.north);
  \draw[arr] (demand.south) -- (supply.north);
\end{tikzpicture}
\caption{The modular estimation pipeline. Embeddings are fine-tuned once and then frozen for all downstream identification strategies.}
\label{fig:main_pipeline}
\end{figure}

Two concerns motivate this fine-tuning step. First, the downstream demand, hedonic, and event-study pipelines are easier to run with a compact representation than with raw image features. A $\embTowerDim$-number vector per product is small enough to load, index, and pass through subsequent networks at reasonable speed and memory cost; the $\embCLIPDim$-dimensional CLIP output multiplied across roughly $\embNumItemsApprox$ products and every consumer-item pair in the panel would be slow and expensive to work with. Second, CLIP was trained on a different dataset and with a different objective. It is a rich encoder but rich about many things that do not matter for choice in this catalogue, such as photographic style, garment angle, lighting, and background texture. Fine-tuning uses actual purchases in this catalogue as the training target, which forces the $\embTowerDim$-number output to keep only the visual and categorical information that helps rank the product actually bought above the products that were not.

A critical challenge in learning aesthetic representations is avoiding \textit{price contamination}. If a single network ingests price alongside the visual inputs, the resulting product embedding merges price-positioning information with visual content. A typical two-tower recommendation model does exactly this by feeding log price alongside image and categorical features into a single item tower. That shortcut raises training-time purchase-prediction scores, but it hurts the downstream estimation. The item embedding ends up carrying part of the price variation. When this embedding is then passed to a logit demand model with price as a separate regressor, the estimated price coefficient is pulled toward zero, because some of the true price response has already been absorbed into the item vector. Price plays a special role in the downstream demand model as the scalar for elasticity estimation, so it has to sit in its own place in the feature pipeline. I therefore route user, item, and price inputs through three architecturally distinct towers that share no weights. The item tower requires a fixed-dimensional representation of each product's visual appearance, so I extract $\embCLIPDim$-dimensional embeddings from product photographs and product descriptions using a pre-trained vision-language model, alongside $\embCategoricalDim$ dimensions of categorical attributes (e.g., color group, garment type).\footnote{Visual and text features are extracted with the OpenAI CLIP ViT-B/32 model \citep{radford2021learning}, which was trained on a large external set of image and caption pairs using a contrastive objective that places matching pairs close together and non-matching pairs far apart in a shared $\embCLIPDim$-dimensional vector space. The image and text embeddings for each product are averaged and renormalised to give one $\embCLIPDim$-dimensional vector per product, which the three-tower fine-tune then compresses to the task-relevant directions.} The item tower thus receives a $\embItemInputDim$-dimensional input per product, the user tower receives an $\embUserInputDim$-dimensional input comprising a learnable ID embedding, demographics (e.g., age, postal code), and aggregate purchase history statistics, and the separate price tower maps the log reference price through a specialized feedforward network. The item tower maps its $\embItemInputDim$-dimensional input through a feedforward network to produce a $\embTowerDim$-dimensional, L2-normalized item embedding $\mathbf{v}_j$\footnote{L2 normalisation rescales each vector to unit length. After this step, the inner product between two embeddings is the cosine of the angle between them, so magnitudes no longer affect scoring.}; similarly, the user tower produces a $\embTowerDim$-dimensional user embedding $\mathbf{d}_i$, and the price tower produces $\mathbf{p}_j$. The predicted affinity between user $i$ and item $j$ is calculated as
\begin{equation}
\hat{y}_{ij} = \frac{\mathbf{d}_i^\top \mathbf{v}_j - \text{softplus}(\mathbf{d}_i^\top \mathbf{p}_j)}{\tau_{\mathrm{InfoNCE}}}
\label{eq:affinity}
\end{equation}
\noindent where $\tau_{\mathrm{InfoNCE}} = \embTauInfoNCE$ is a temperature parameter.\footnote{The temperature controls how sharply the loss penalizes wrong rankings. A lower value concentrates probability mass on the top-scoring item and makes training more aggressive; a higher value spreads mass more evenly and smooths the gradient signal. The role is analogous to the scale parameter in a logit model.} The softplus function,\footnote{Softplus is $\log(1+e^x)$, a smooth version of the positive-part function. It is always positive, so the price term in the score always subtracts from affinity.} ensures the price term always acts as a penalty against affinity. Because the item tower never sees price as an input, the learned item embedding $\mathbf{v}_j$ contains no direct price signal. Any residual correlation between $\mathbf{v}_j$ and price must go through genuine visual or categorical features. As \cref{tab:quantitative_validation} shows, the three-tower item embeddings still predict log prices with substantial fit, because visual style and categorical features are themselves correlated with price tier. The architectural claim is cleaner separation of price from visual content, not statistical orthogonality. A related line of work in recommender systems makes a similar argument, building price-aware models that keep price in its own channel rather than folding it into a single mixed item representation \citep{zheng2020price, zheng2023incorporating, cui2020disentangled}.

\subsection{Contrastive Training}
\label{subsec:contrastive_training}

The three towers are trained jointly by minimising an InfoNCE contrastive loss \citep{oord2018representation}\footnote{InfoNCE rewards the model for ranking the purchased item above a set of unpurchased distractors drawn from the same product category. When the distractor set covers the entire category, the loss coincides with the negative log-likelihood of a multinomial logit over that choice set.}, which pulls the buyer's representation toward the purchased item and pushes it away from the other items in the same product category. Formally, for each observed purchase $(i, j^+)$:
\begin{equation}
\mathcal{L}_{\text{InfoNCE}} = -\sum_{(i, j^+)} \log \frac{\exp\left(\hat{y}_{ij^+}\right)}{\sum_{k \in \mathcal{C}(j^+)} \exp\left(\hat{y}_{ik}\right)}
\label{eq:infonce_loss}
\end{equation}

In standard econometric terms, Equation~\ref{eq:infonce_loss} is the negative log-likelihood of a conditional logit in which each product $j$ carries a latent characteristic vector $\mathbf{v}_j$, each consumer $i$ carries a latent taste vector $\mathbf{d}_i$, and the systematic utility is $\mathbf{d}_i^\top \mathbf{v}_j / \tau_{\mathrm{InfoNCE}}$. The set $\mathcal{C}(j^+)$ is the choice set, the purchased item together with the competing alternatives in the category, so the denominator includes $\exp(\hat y_{ij^+})$ and the choice probabilities integrate to one. The absolute scale and orientation of the latent vectors are not point-identified by the loss alone, a known issue in the contrastive-learning literature \citep{rusak2024infonce}. The non-identification does not bind downstream because only inner products $\mathbf{d}_i^\top \mathbf{v}_j$ enter the demand utility, and inner products are invariant to joint orthogonal rotations and matched rescalings of the two vectors.

The two-stage design sits between two alternatives the paper does not take. Training a visual model from raw product images needs more data and deeper networks than this sample supports. A generic off-the-shelf image model is the next fallback but is weak here, since it is not trained on fashion style. OpenAI CLIP is the strongest available fallback, so the approach keeps the CLIP encoder frozen and trains only the three-tower projection head on its 512-dimensional output. The resulting 64-dimensional projection is faster to work with downstream and higher in retrieval accuracy than raw CLIP. The dress demand master in \cref{sec:choice_model} additionally concatenates a 512-dimensional FashionCLIP vector \citep{chia2022contrastive} alongside the three-tower item vector. The fine-tune uses the September 2018 through June 2019 window, and every downstream estimate uses the separate window from July 2019 onwards, so embedding-stage noise cannot leak into downstream coefficients through shared observations. The same fine-tune also produces a 64-dimensional user embedding $\mathbf{d}_i$ for every consumer, richer than any hand-coded demographic attribute set.

This training stage is used strictly to learn dense representations, not to recover structural behavioural parameters. The items competing against the purchased item in the denominator are drawn from the same product category, not from the entire catalogue, which forces the embeddings to capture fine-grained aesthetic differences within a valid choice set. When this denominator covers every other product in the category, \Cref{eq:infonce_loss} is algebraically identical to the log-likelihood of a standard multinomial choice model with utility $\mathbf{d}_i^\top \mathbf{v}_j / \tau_{\mathrm{InfoNCE}}$ and no price term. A recent line of work in operations research and transportation research develops this exact equivalence between contrastive representation learning and discrete choice estimation \citep{aouad_desir_2022,arkoudi_2023}. In practice the embedding stage uses a random subsample of competing items per batch\footnote{A batch is the small random slice of training examples used for one update to the model's parameters. Sampling negatives at the batch level keeps memory bounded as the catalogue grows.} rather than the full category, the same subsampling trick used in early methods for learning numerical embeddings from very large text datasets \citep{mikolov2013distributed}. The full structural demand system that uses these embeddings as fixed inputs to a price-sensitivity network and a taste-matching network is developed in \cref{sec:choice_model}.

Joint training that combines the contrastive loss with an explicit choice-model objective would pull the price response back into the item tower and defeat the separation just described. Other common contrastive objectives lose the clean mapping to discrete choice used above. A pairwise ranking loss contrasts one purchased item against a single unpurchased one rather than against a full choice set. A margin-based loss enforces a fixed gap between scores and is not a probability distribution. A batch-level objective treats every item in the training batch as a negative regardless of product category, which throws away the within-category choice set the downstream demand model relies on.\footnote{Pairwise ranking, triplet-margin, and SimCLR-style batch objectives are three well-known training losses for representation learning. None of the three delivers the exact log-likelihood of a choice model at convergence.} The trained three-tower model reaches Hit@10 retrieval accuracy of $\embHitTen$ on a held-out validation sample of $\embValPairsM$ million consumer-item pairs.\footnote{Hit@10 is the fraction of held-out purchases for which the item actually bought appears among the top ten items ranked by the model's affinity score within its own product category.}

The training dataset covers $\embNumCategories$ major product categories, roughly $\embNumItemsApprox$ items, and $\embPreCovidTxnsM$ million pre-COVID transactions drawn from the full $\embFullPanelTxnsM$ million transaction panel. Only pre-COVID records are used so the COVID shock does not contaminate the aesthetic space. \Cref{fig:main_timeline} illustrates this temporal separation. The resulting shared embedding space allows subsequent analyses to extract category-specific subsets.

\begin{figure}[H]
\centering
\begin{tikzpicture}[
    timeline/.style={draw=gray, line width=1.5pt, -{Stealth[length=6pt]}},
    tick/.style={draw=gray, line width=1.2pt}
]
    \draw[timeline] (0, 0) -- (12.5, 0);
    
    \draw[tick] (0.5, 0.15) -- (0.5, -0.15) node[below, font=\scriptsize, yshift=-2pt] {Sep 2018};
    \draw[tick] (7.5, 0.15) -- (7.5, -0.15) node[below, font=\scriptsize, yshift=-2pt] {Feb 2020};
    \draw[tick] (11.5, 0.15) -- (11.5, -0.15) node[below, font=\scriptsize, yshift=-2pt] {Sep 2020};
    
    \fill[gray!20, rounded corners=2pt] (0.5, 0.3) rectangle (7.4, 1.0);
    \node[font=\bfseries\footnotesize] at (3.95, 0.65) {Pre-COVID ($T_1$)};
    
    \fill[red!20, rounded corners=2pt] (7.6, 0.3) rectangle (11.5, 1.0);
    \node[font=\bfseries\footnotesize] at (9.55, 0.65) {COVID Shock ($T_2$)};

    \fill[red!55!white, rounded corners=1pt] (8.24, 2.55) rectangle (9.12, 2.85);
    \draw[red!75!black, line width=0.4pt, rounded corners=1pt] (8.24, 2.55) rectangle (9.12, 2.85);
    \node[font=\scriptsize] at (8.68, 3.05) {Lockdown};
    \draw[red!75!black, line width=0.4pt, dashed] (8.68, 2.55) -- (8.68, 1.02);
    
    \draw[draw=blue, fill=white, rounded corners=3pt, line width=0.6pt] (1.3, 1.5) rectangle (6.6, 2.3);
    \node[align=center, font=\scriptsize] at (3.95, 1.9) {\textbf{Embedding Fine-Tuning}\\(10 months)};
    \draw[blue, line width=0.8pt] (3.95, 1.5) -- (3.95, 1.0);

    \draw[draw=green!50!black, fill=white, rounded corners=3pt, line width=0.6pt] (3.5, -0.9) rectangle (11.0, -1.7);
    \node[align=center, font=\scriptsize] at (7.25, -1.3) {\textbf{Structural Estimation}\\Demand, Supply, \& Hedonics};
    
    \draw[green!50!black, line width=0.8pt] (4.0, -0.9) -- (4.0, -0.15);
    \draw[green!50!black, line width=0.8pt] (9.5, -0.9) -- (9.5, -0.15);
\end{tikzpicture}
\caption{Training timeline. Embeddings are learned strictly on the $T_1$ sequence to prevent demand shocks from altering visual characteristics.}
\label{fig:main_timeline}
\end{figure}

%% file: 04_theory.tex
\section{Theoretical Framework}
\label{sec:theory}

This section lays out the latent-class deep logit of \cref{sec:choice_model}, the Bertrand-Nash markup inversion that feeds off its demand Jacobian \citep{berry1995automobile, nevo2001measuring}, and the welfare formulas used in the counterfactuals of \cref{sec:choice_model}. Notation is introduced inline as each object appears.

\subsection{Latent Class Deep Logit}
\label{sec:theory:logit}

Consumer $i$ in class $c \in \{1, 2\}$ draws utility from product $j$ in week $t$,
\begin{equation}
U_{ijt \mid c} = u_{ijt \mid c} + \varepsilon_{ijt},
\label{eq:app_random_utility}
\end{equation}
with $\varepsilon_{ijt}$ i.i.d.\ Type~I Extreme Value \citep{mcfadden1974conditional}, giving the standard class-conditional logit
\begin{equation}
s_{ij \mid c} = \frac{\exp(u_{ijt \mid c})}{\exp(V_0) + \sum_{\ell \in \mathcal{J}_t} \exp(u_{i\ell t \mid c})},
\label{eq:app_class_cond_prob}
\end{equation}
where $V_0$ is a scalar outside-good utility shared across classes, calibrated by bisection below.

The within-class systematic utility decomposes into a price channel, a taste channel, an endogeneity correction, and a class-specific intercept,
\begin{equation}
u_{ijt \mid c} \;=\; \alpha_c(\mathbf{d}_i)\, p_{jt}
   \;+\; \bigl[\mathbf{r}_i^{c} + \boldsymbol{\delta}_m\bigr]^{\!\top}\! \mathbf{t}_j^{c}
   \;+\; \gamma_c\,\hat\lambda_j
   \;+\; b_c,
\label{eq:systematic_utility}
\end{equation}
where $\mathbf{d}_i \in \mathbb{R}^{64}$ is the pre-trained user embedding and $\mathbf{x}_j$ is the pre-trained item feature vector, both from the three-tower model in \cref{sec:embeddings}. The raw item feature $\mathbf{x}_j$ is the 64-dimensional three-tower item vector for the five non-Dress categories in \cref{sec:appendix_cross_category}, and the 576-dimensional concatenation of that same three-tower vector with the 512-dimensional raw FashionCLIP vector for the Dress master (\cref{sec:choice_model} footnote). The scalar intercept $b_c$ is a learnable class-specific bias estimated jointly with the networks; because the outside-good utility $V_0$ is shared across classes, $b_c$ is what lets each class have its own effective inside-vs-outside tilt.

Price sensitivity is a class-specific neural function of the user embedding,
\begin{equation}
\alpha_c(\mathbf{d}_i) = -\operatorname{softplus}\bigl(\operatorname{MLP}_\alpha^c(\mathbf{d}_i)\bigr) - 0.1.
\label{eq:app_alpha_net}
\end{equation}
$\operatorname{MLP}_\alpha^c$ is a small feedforward network \citep{goodfellow2016deep}\footnote{A feedforward network passes the input through a chain of linear transformations with nonlinear squashing functions in between, from input to output with no loops back. The ``small'' here is literal: one hidden layer of width 64.} with one hidden layer of width 64. The softplus-and-negate wrapper\footnote{$\operatorname{softplus}(x) = \log(1 + e^x) > 0$; negating enforces $\alpha_c < 0$ smoothly.} enforces the law of demand ($\alpha_c < 0$) without a hard cutoff that would break gradient-based training. The trailing $-0.1$ floor keeps $|\alpha_c(\mathbf{d}_i)| \geq 0.1$ for every consumer, which keeps the Bertrand-Nash markup inversion well conditioned even when the price signal is weak.

The class-$c$ user taste projection compresses the user embedding into a $K_{\mathrm{rank}} = 16$ taste bottleneck,
\begin{equation}
\mathbf{r}_i^{c} = \operatorname{MLP}_r^c(\mathbf{d}_i) \in \mathbb{R}^{16},
\label{eq:app_r_net}
\end{equation}
and the class-$c$ item taste projection lives in the same bottleneck, unit-normalised on the sphere,
\begin{equation}
\mathbf{t}_j^{c} = \frac{\operatorname{MLP}_t^c(\mathbf{x}_j)}{\lVert \operatorname{MLP}_t^c(\mathbf{x}_j) \rVert} \in \mathbb{R}^{16}.
\label{eq:app_t_net}
\end{equation}
The two MLPs compose into the $\beta$-net, whose class-$c$ scalar taste match between consumer $i$ and product $j$ is
\begin{equation}
\beta_c(\mathbf{d}_i, \mathbf{x}_j) \;:=\; \mathbf{r}_i^{c\,\top}\mathbf{t}_j^{c} \;=\; \operatorname{MLP}_r^c(\mathbf{d}_i)^{\top}\,\frac{\operatorname{MLP}_t^c(\mathbf{x}_j)}{\lVert \operatorname{MLP}_t^c(\mathbf{x}_j) \rVert},
\label{eq:app_beta_net}
\end{equation}
and this scalar is the taste channel that enters \cref{eq:systematic_utility}. The unit-norm constraint on $\mathbf{t}_j^{c}$ is a design choice that forces the price channel $\alpha_c(\mathbf{d}_i) p_{jt}$, not the taste channel, to carry price-correlated variation; without it, the taste inner product could absorb price information by scaling its own norm, which would attenuate $\alpha_c$ toward zero (see \citet{compiani_kitamura_2016} on the broader mixture-identification literature).

A shared monthly taste shift $\boldsymbol{\delta}_m \in \mathbb{R}^{16}$ rotates which latent taste directions attract demand in each calendar month, under a low-rank factorisation,
\begin{equation}
\boldsymbol{\delta}_m = W \mathbf{z}_m, \quad W \in \mathbb{R}^{16 \times 8}, \quad \mathbf{z}_m \in \mathbb{R}^{8}, \quad \sum_{m=1}^{12} \mathbf{z}_m = \mathbf{0}.
\label{eq:app_delta}
\end{equation}
The zero-sum constraint on $\mathbf{z}_m$ identifies the seasonal shift against the static user taste $\mathbf{r}_i^{c}$. The rank-8 basis is shared across months to avoid overfitting on the single year of pre-COVID data.

The term $\gamma_c \hat\lambda_j$ is the \citet{petrin_train_2010} control-function correction for price endogeneity. $\hat\lambda_j$ is a per-product pricing residual (the cross-sectional component of price the item embedding cannot explain, treated as a proxy for unobserved quality), loaded back into utility with a class-specific coefficient $\gamma_c$ so each latent type absorbs its own share of the price-correlated quality signal. See \cref{sec:appendix_robustness} for the construction and robustness to linear and nonparametric CF specifications.

Consumers are assigned to one of two classes with constant mixture weights $\pi_1$ and $\pi_2 = 1 - \pi_1$, estimated jointly. Demographic routing of class membership is not used; the gating is unconditional. The unconditional choice probability mixes the two class-conditional logits,
\begin{equation}
s_{ij} = \sum_{c=1}^{2} \pi_c \cdot s_{ij \mid c} = \sum_{c=1}^{2} \pi_c \cdot \frac{\exp\bigl(u_{ijt \mid c}\bigr)}{\exp(V_0) + \sum_{\ell \in \mathcal{J}_t} \exp\bigl(u_{i\ell t \mid c}\bigr)}.
\label{eq:app_choice_prob}
\end{equation}

Parameters are estimated by maximising the log-likelihood of observed purchases over each consumer's transaction history $\mathcal{T}_i$,
\begin{equation}
\mathcal{LL} = \sum_{i=1}^{I} \sum_{t \in \mathcal{T}_i} \log\bigl(s_{i, y_{it}, t}\bigr).
\label{eq:app_loglikelihood}
\end{equation}
Training uses the EM algorithm: the E step computes each consumer's posterior class probability given current parameters, and the M step takes one AdamW\footnote{AdamW is a gradient-based optimiser with decoupled weight decay, which separates the regularisation step from the adaptive-learning-rate step and typically generalises better than plain Adam.} update on the class-weighted log-likelihood. The outside-good utility $V_0$ is not part of the likelihood; it is calibrated post-estimation by bisection on
\[
\frac{1}{I}\sum_{i=1}^{I}\sum_{j=1}^{J} s_{ij}(V_0) = \hat\tau = \dataTau,
\]
consistent with H\&M's $\dataInsideShareLo\%$ to $\dataInsideShareHi\%$ single-category online share (\cref{sec:choice_model}). A single scalar $V_0$ is shared across both classes.

Standard errors for $\bar\alpha_c$, the coefficient gap $|\bar\alpha_1 - \bar\alpha_2|$, the mixture weight $\pi_1$, and the validation fit metrics come from a consumer-level bootstrap with $B = \datBootstrapB$ refits (\cref{tab:bootstrap_ses}), holding the first-stage embeddings fixed. The consumer-only bootstrap undercovers relative to a full two-stage correction that also propagates first-stage embedding noise; see \citet{karacamandic_train_2003, lennon_rubin_waddell_2025, battaglia_christensen_hansen_sacher_2024} for the relevant corrections when the second-stage regressors are variables generated by a first-stage machine-learning model. The contrastive InfoNCE loss of \cref{sec:embeddings} structurally mirrors this log-likelihood; see \citet{aouad_desir_2022, arkoudi_2023} for the formal link between contrastive representation learning and discrete choice estimation.

\subsection{Price Elasticities and IIA Departures}
\label{sec:theory:elasticities}

All derivatives are taken with respect to linear prices $p_j$. Differentiating the mixture choice probability $s_{ij}$ with respect to $p_j$ and $p_\ell$ gives the individual-level derivatives
\begin{align}
\frac{\partial s_{ij}}{\partial p_j} &= \sum_{c=1}^{2} \pi_c \cdot \alpha_c(\mathbf{d}_i) \cdot s_{ij \mid c} \cdot \bigl(1 - s_{ij \mid c}\bigr), \label{eq:app_own_deriv} \\
\frac{\partial s_{ij}}{\partial p_\ell} &= -\sum_{c=1}^{2} \pi_c \cdot \alpha_c(\mathbf{d}_i) \cdot s_{ij \mid c} \cdot s_{i\ell \mid c}, \quad j \neq \ell, \label{eq:app_cross_deriv}
\end{align}
where $s_{ij \mid c}$ is the choice probability conditional on class $c$. The own-price derivative is strictly negative and the cross-price derivative is strictly positive. Substitution between two products scales with $s_{ij \mid c} \cdot s_{i\ell \mid c}$, so products favoured by the same consumer class exhibit stronger cross-price effects. The aggregate market share derivative that fills the Jacobian matrix $\mathbf{J}$ averages these individual effects across all $I$ consumers,
\begin{equation}
\frac{\partial S_j}{\partial p_\ell} = \frac{1}{I} \sum_{i=1}^{I} \frac{\partial s_{ij}}{\partial p_\ell},
\label{eq:app_jacobian}
\end{equation}
yielding own-price elasticity $\varepsilon_{jj} = (p_j / S_j) \cdot \partial S_j / \partial p_j$ and cross-price elasticity $\varepsilon_{j\ell} = (p_\ell / S_j) \cdot \partial S_j / \partial p_\ell$. The standard IIA restriction of plain logit \citep{mcfadden1974conditional} breaks at the aggregate level through two channels. Heterogeneous tastes make two dresses favoured by the same consumer group closer substitutes than dresses favoured by disjoint groups. Heterogeneous price sensitivity across classes adds a second channel, in which highly price-sensitive consumers concentrate their probability mass on specific products and generate non-proportional substitution toward those products when prices change. Both channels have multiple sources in this model, including the two latent classes, the consumer-level variation within each class captured by the $\alpha$-net and $\beta$-net, and the seasonal taste shift $\boldsymbol{\delta}_m$.

\subsection{Supply-Side Markup Recovery}
\label{sec:theory:supply}

The markup inversion follows \citet{berry1995automobile} and \citet{nevo2001measuring} applied to a multi-product monopolist, which is the natural conduct for H\&M pricing its own dress catalog. The firm chooses prices to maximise static profit
\[
\pi \;=\; \sum_{j=1}^{J} (p_j - mc_j)\, M\, S_j(\mathbf{p}),
\]
with $mc_j$ a constant marginal cost and $M$ total market size. The first-order condition for product $j$ balances the direct revenue effect against cross-product cannibalisation across the full catalog,
\begin{equation}
S_j + \sum_{\ell = 1}^{J} (p_\ell - mc_\ell) \cdot \frac{\partial S_\ell}{\partial p_j} = 0.
\label{eq:app_foc_expanded}
\end{equation}
Defining the ownership matrix $\boldsymbol{\Omega}$ ($\Omega_{jk} = 1$ if the firm internalises the cross-effect between products $j$ and $k$; $\boldsymbol{\Omega} = \mathbf{1}\mathbf{1}^{\top}$ under the multi-product monopolist baseline) and the matrix $\boldsymbol{\Delta} = \boldsymbol{\Omega} \odot \mathbf{J}'$, provided $\boldsymbol{\Delta}$ is nonsingular the markup vector $\boldsymbol{\eta} = \mathbf{p} - \mathbf{mc}$ and marginal costs are recovered cleanly as
\begin{equation}
\boldsymbol{\eta} = -\boldsymbol{\Delta}^{-1} \mathbf{S}, \quad \widehat{\mathbf{mc}} = \mathbf{p} + \boldsymbol{\Delta}^{-1} \mathbf{S}.
\label{eq:app_marginal_cost}
\end{equation}
For a single-product firm ($\boldsymbol{\Delta}$ diagonal), the Lerner index, which measures the markup a firm charges above its own marginal cost as a fraction of the final price, reduces to $L_j = \eta_j / p_j = 1/|\varepsilon_{jj}|$. Given H\&M's small individual item shares, cross-tier cannibalization is weak, rendering multi-product monopolist pricing nearly indistinguishable from independent single-product pricing. \Cref{fig:markup_inversion_flow} sketches the inputs that feed the markup inversion.

\begin{figure}[H]
\centering
\begin{tikzpicture}[
  box/.style={draw, rounded corners, minimum height=1cm, minimum width=2.2cm,
              align=center, font=\small},
  arr/.style={-{Stealth[length=3mm]}, thick},
  node distance=1.2cm and 0.8cm
]
\node[box] (prices) {Observed\\prices $\mathbf{p}$};
\node[box, right=of prices] (demand) {Estimated\\demand $\mathbf{J}$};
\node[box, right=of demand] (omega) {Ownership\\$\boldsymbol{\Omega}$};

\node[box, below=1.2cm of demand, fill=gray!10] (invert) {Invert:\\$\boldsymbol{\eta} = -\boldsymbol{\Delta}^{-1}\mathbf{S}$};

\node[box, below=1.2cm of invert, fill=gray!10] (mc) {Recovered $\widehat{\mathbf{mc}}$\\$= \mathbf{p} + \boldsymbol{\Delta}^{-1}\mathbf{S}$};

\node[box, below left=1cm and 0.3cm of mc] (lerner) {Lerner\\index};
\node[box, below right=1cm and 0.3cm of mc] (welfare) {Welfare\\analysis};

\draw[arr] (prices) -- (invert);
\draw[arr] (demand) -- (invert);
\draw[arr] (omega) -- (invert);
\draw[arr] (invert) -- (mc);
\draw[arr] (mc) -- (lerner);
\draw[arr] (mc) -- (welfare);
\end{tikzpicture}
\caption{Markup inversion pipeline. Observed prices, estimated demand (the Jacobian $\mathbf{J}$ from \cref{sec:choice_model}), and the ownership matrix $\boldsymbol{\Omega}$ are combined to recover unobserved marginal costs.}
\label{fig:markup_inversion_flow}
\end{figure}

\subsection{Consumer Surplus}
\label{sec:theory:surplus}

Applying the log-sum formula \citep{train2009discrete}, individual consumer surplus is the probability-weighted sum of conditional expected maximum utilities,
\begin{equation}
CS_i = \sum_{c=1}^{2} \frac{\pi_c}{|\alpha_c(\mathbf{d}_i)|} \cdot \log\!\left(\sum_{j \in \mathcal{J}} \exp\bigl(u_{ij \mid c}\bigr) + \exp(V_0)\right),
\label{eq:app_consumer_surplus}
\end{equation}
where $V_0$ is the outside-good utility from \cref{eq:app_class_cond_prob}, shared across classes and calibrated once by bisection. The compensating variation from moving between market states $0$ and $1$ is $CV_i = CS_i^{1} - CS_i^{0}$; because $V_0$ is held fixed across counterfactuals, the $V_0$-induced additive term in $CS_i$ cancels in $CV_i$. Aggregate policy analysis evaluates the sum of changes in consumer surplus and producer profit, $\Delta TW = M \cdot \overline{CV} + \Delta PS$.

%% file: 05_demand.tex
\section{Demand and Supply Estimation}
\label{sec:choice_model}

I estimate the two-class latent deep logit from \cref{sec:theory:logit} on the dress category, restricting to the pre-COVID window (July 2019 through January 2020) to avoid contamination from the lockdown demand regime documented in \cref{sec:event_study}. The estimation sample has $I = \dataI$ online consumers choosing among $J = \dataJ$ products over $\dataT$ weeks. The outside good utility $V_0$ is calibrated by bisection to a target inside share $\hat\tau = \dataTau$, the midpoint of the $\dataInsideShareLo\%$ to $\dataInsideShareHi\%$ single-category online share that is a reasonable estimate for H\&M dresses in this window. Elasticity estimates are invariant to the precise choice of $\hat\tau$ when individual item shares remain small (\cref{tab:supply-sensitivity} in \cref{sec:appendix_tau_robustness}).

The two-class latent deep logit gives mean absolute own-price elasticity $|\bar\varepsilon| = \lcEpsBar$ and an implied Lerner of $\lcLernerPct\%$ under multi-product monopolist conduct. Validation McFadden $R^2$ is strongly positive on held-out weeks; \cref{tab:cf_robustness} reports the point estimate alongside the control-function robustness run.\footnote{The $\beta$-net item input for the Dress master concatenates the three-tower item vector with the raw FashionCLIP vector. The concatenation lifts validation $R^2$ over three-tower only; see \cref{subsec:item_embedding_robustness} for the full comparison.} The remainder of this section traces where that fit comes from. A benchmark against plain aggregate logit, random-coefficient logit, BLP with GMM \citep{berry1995automobile}, and individual mixed logit is reported in \cref{sec:demand_comparison}.

\subsection{Latent Class Estimates: Taste vs Price}
\label{sec:choice_results}

The model recovers two classes of roughly comparable size, with the minority slightly smaller than the majority. Mean price sensitivity differs meaningfully across the two classes, with the minority running nearly twice as price-elastic as the majority. Mixture weights and class-mean price coefficients are in \cref{tab:cf_robustness}. The class split combines price and taste: the price-sensitive minority shows a broader taste distribution, while the majority has a tighter one. Within each class, individual-level $\alpha$-net outputs still vary smoothly across consumers.

\begin{figure}[H]
\centering
\includegraphics[width=0.9\textwidth]{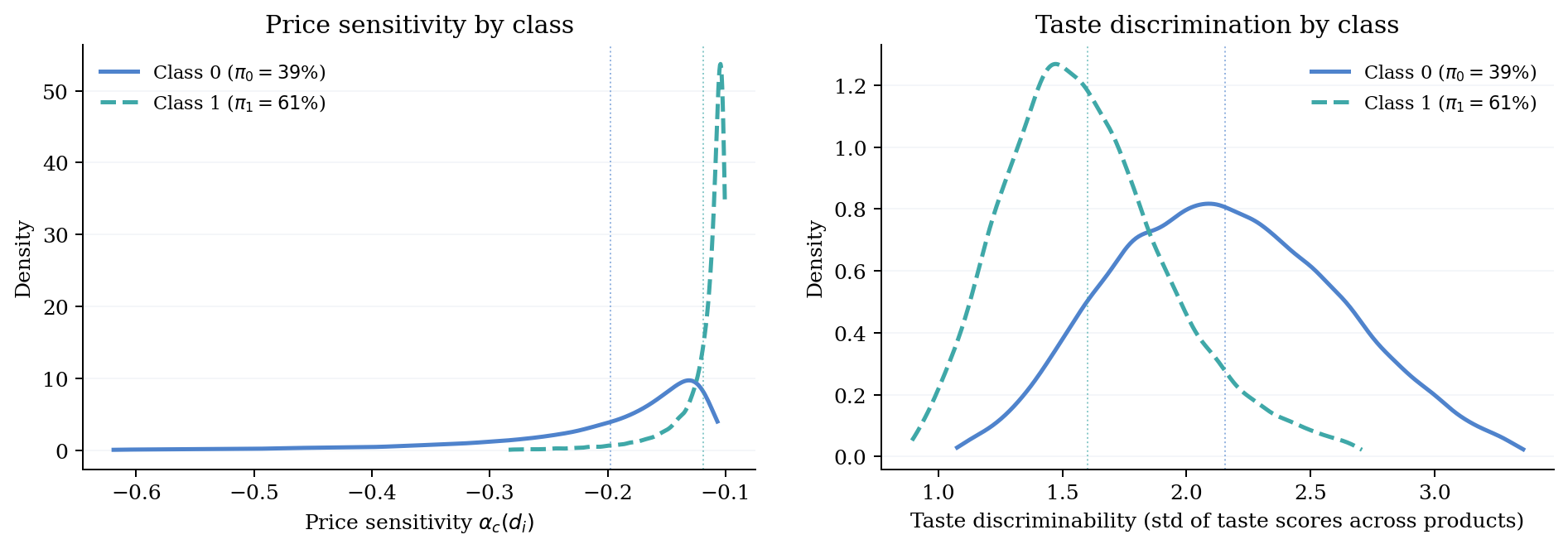}
\caption{Individual-level price sensitivity (left) and the dispersion of taste scores across all products per consumer (right), split by latent class. Mean $\alpha$ separates the classes; taste dispersion separates them further.}
\label{fig:alpha_taste_distribution}
\end{figure}

\begin{figure}[H]
\centering
\includegraphics[width=0.9\textwidth]{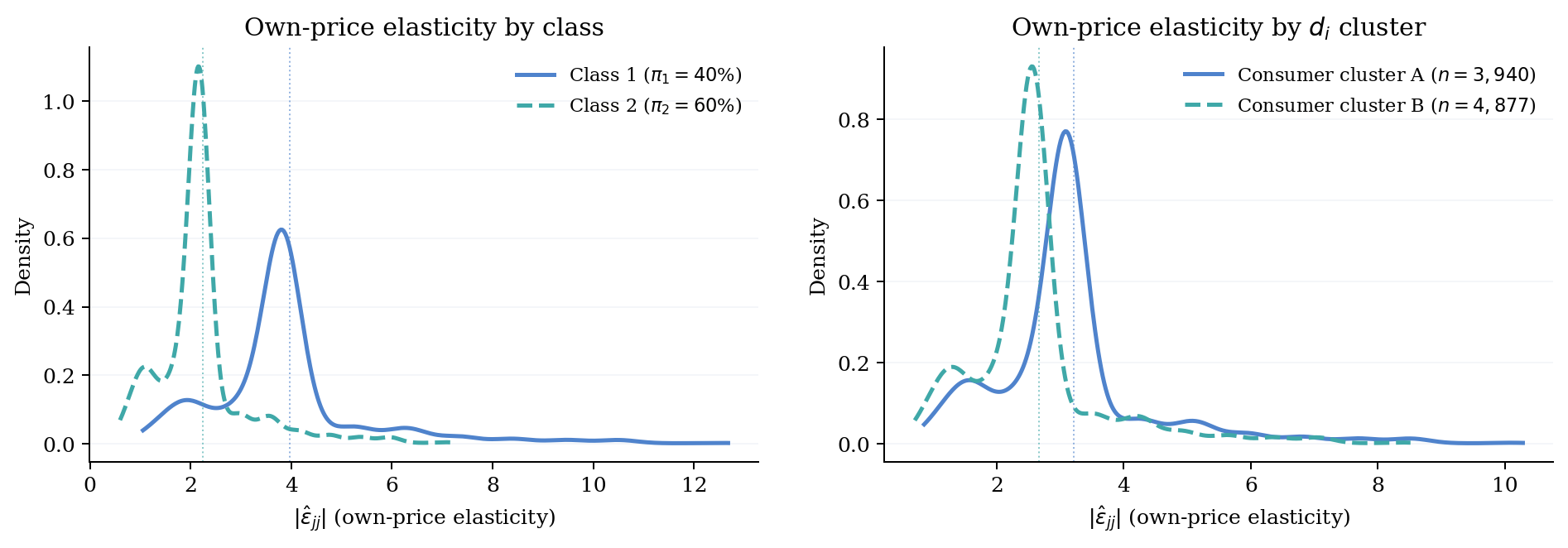}
\caption{Own-price elasticity heterogeneity. Left, the distribution of $|\hat\varepsilon_{jj}|$ across the $J = 2{,}034$ dresses computed under each latent class separately. Right, the same distribution computed within the youngest age decile (age $\leq 23$) and the oldest age decile (age $\geq 55$).}
\label{fig:own_elast_by_class_segment}
\end{figure}

\Cref{fig:alpha_taste_distribution} shows the two classes separating more sharply on taste dispersion than on price sensitivity, so the split is primarily taste-driven. \Cref{fig:own_elast_by_class_segment} maps the $\alpha$ gap into product-level elasticities, where the minority class runs substantially more elastic than the majority. The same gap appears when consumers are grouped directly on the user embedding $\mathbf{d}_i$ into clusters of near-identical mean age (see \cref{fig:user_cluster_substitution} below), so the pattern reflects taste, not age.

\Cref{fig:demand_carousel_dress} shows the visual pattern behind the taste split on the dress estimation sample itself. Each panel shows the dresses the $\beta$-net picks as top matches at the mean consumer embedding of one age and spending segment. The separation by age and by spending is visible by eye.

\begin{figure}[H]
\centering
\begin{minipage}{0.48\textwidth}
\centering\small\textbf{Low spend}
\end{minipage}\hfill
\begin{minipage}{0.48\textwidth}
\centering\small\textbf{High spend}
\end{minipage}
\vspace{0.15cm}
\begin{minipage}{0.48\textwidth}
\centering\includegraphics[width=\textwidth]{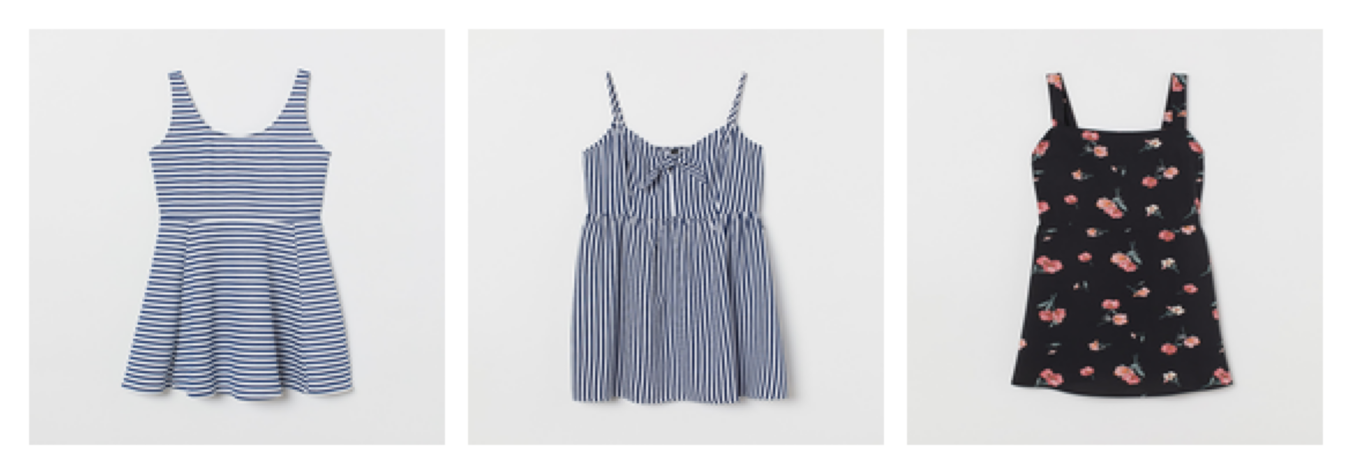}
\end{minipage}\hfill
\begin{minipage}{0.48\textwidth}
\centering\includegraphics[width=\textwidth]{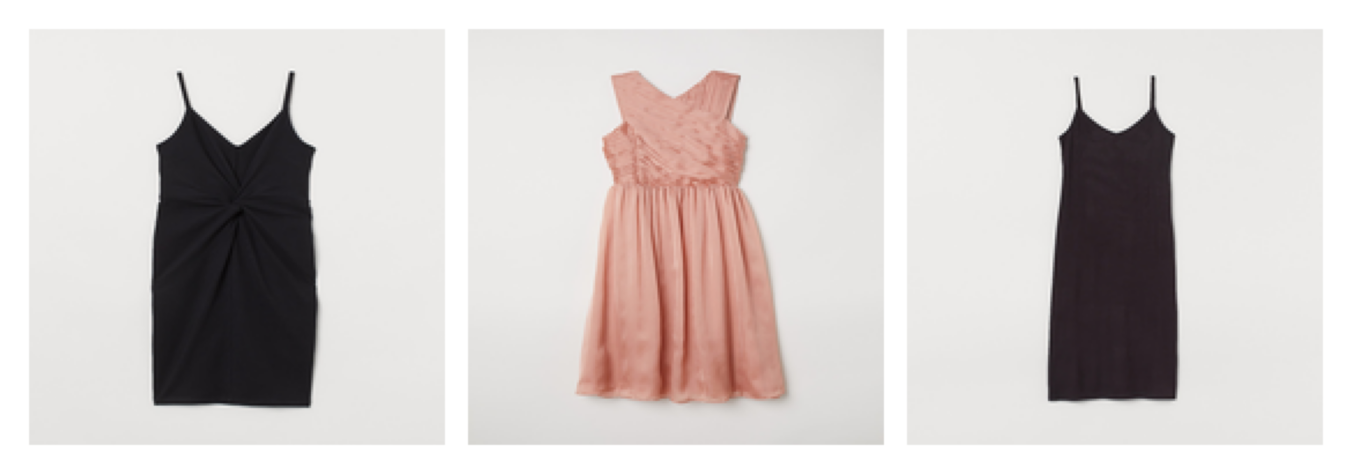}
\end{minipage}
\begin{minipage}{0.48\textwidth}
\centering\includegraphics[width=\textwidth]{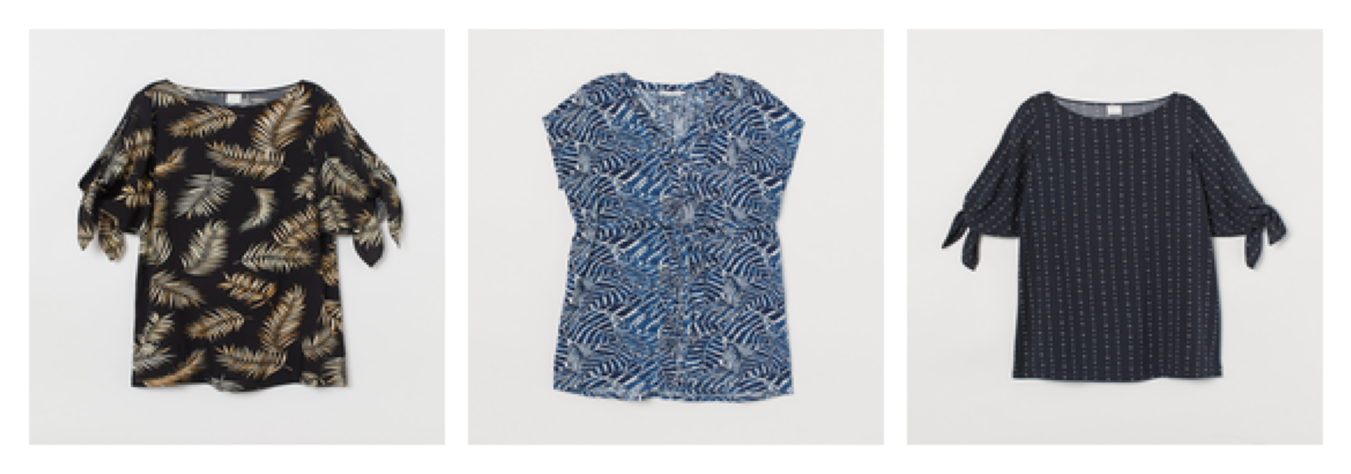}
\end{minipage}\hfill
\begin{minipage}{0.48\textwidth}
\centering\includegraphics[width=\textwidth]{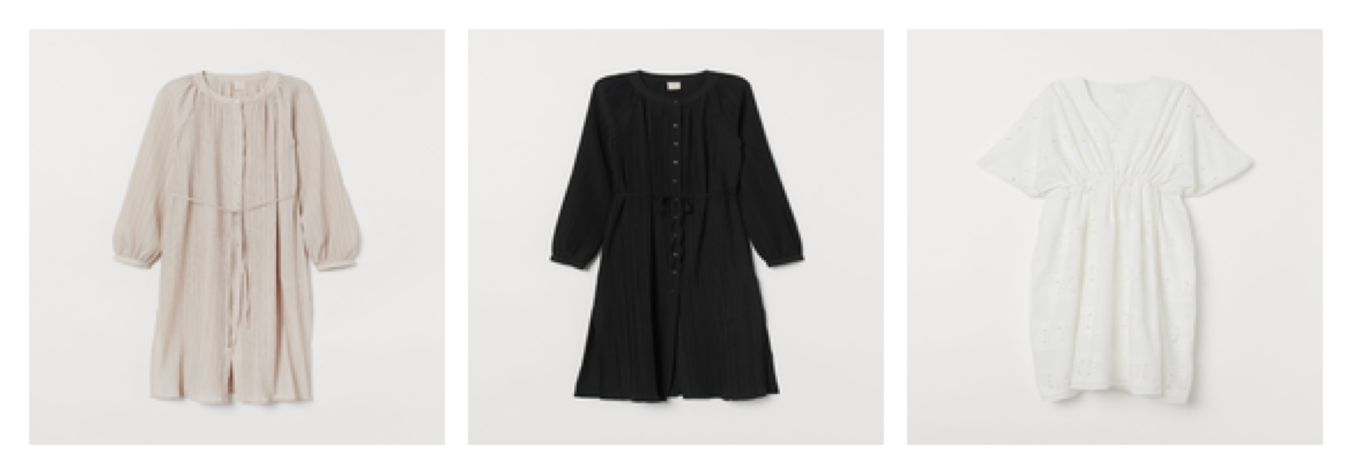}
\end{minipage}
\caption{Younger and higher-spending shoppers prefer visibly different dresses. Columns group shoppers by total spending (low spend left, high spend right). Rows group shoppers by age (younger above, older below). Each panel shows the top dress matches under the $\beta$-net at that segment's mean consumer embedding.}
\label{fig:demand_carousel_dress}
\end{figure}

\begin{figure}[!tbp]
\centering
\begin{minipage}{0.8\textwidth}
\centering
\begin{minipage}{0.48\textwidth}
\centering\small\textbf{Shirts}
\end{minipage}\hfill
\begin{minipage}{0.48\textwidth}
\centering\small\textbf{Shorts}
\end{minipage}
\vspace{0.15cm}
\begin{minipage}{0.48\textwidth}
\centering\includegraphics[width=\textwidth]{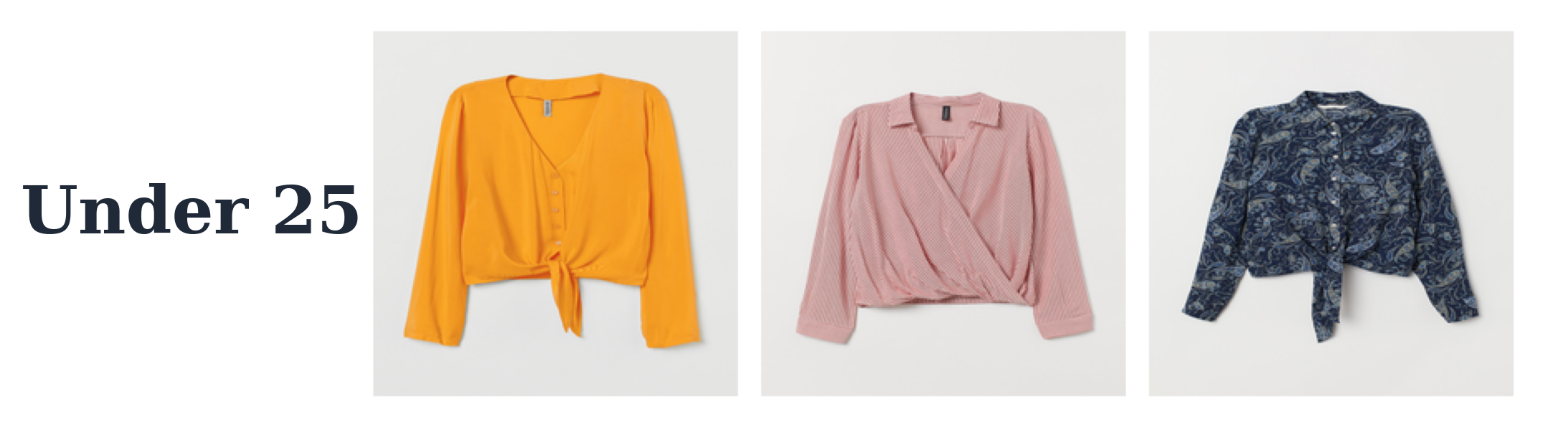}
\end{minipage}\hfill
\begin{minipage}{0.48\textwidth}
\centering\includegraphics[width=\textwidth]{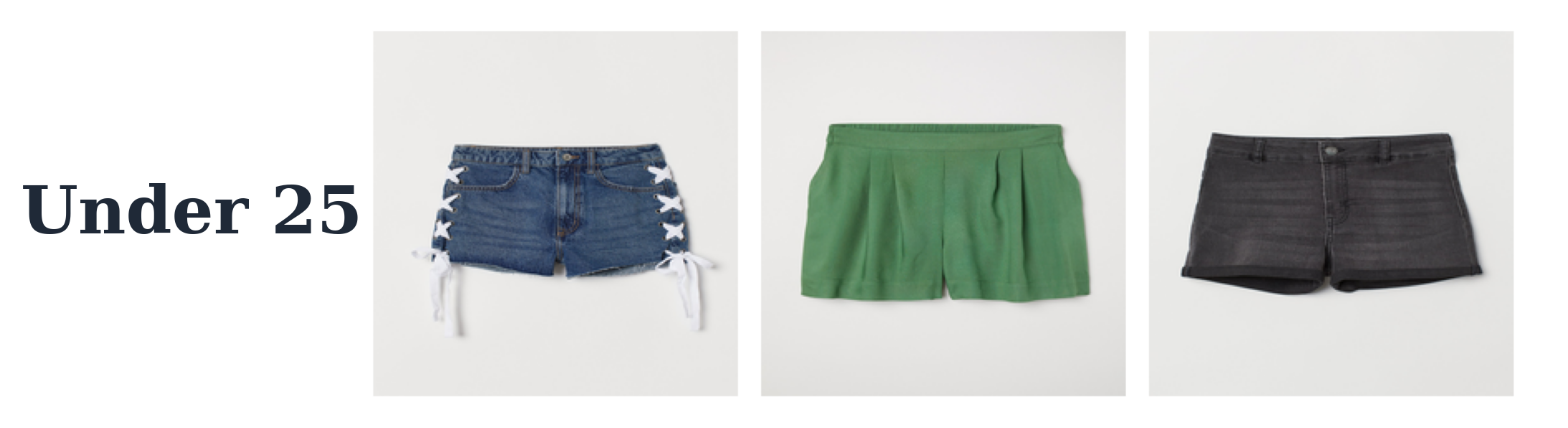}
\end{minipage}
\begin{minipage}{0.48\textwidth}
\centering\includegraphics[width=\textwidth]{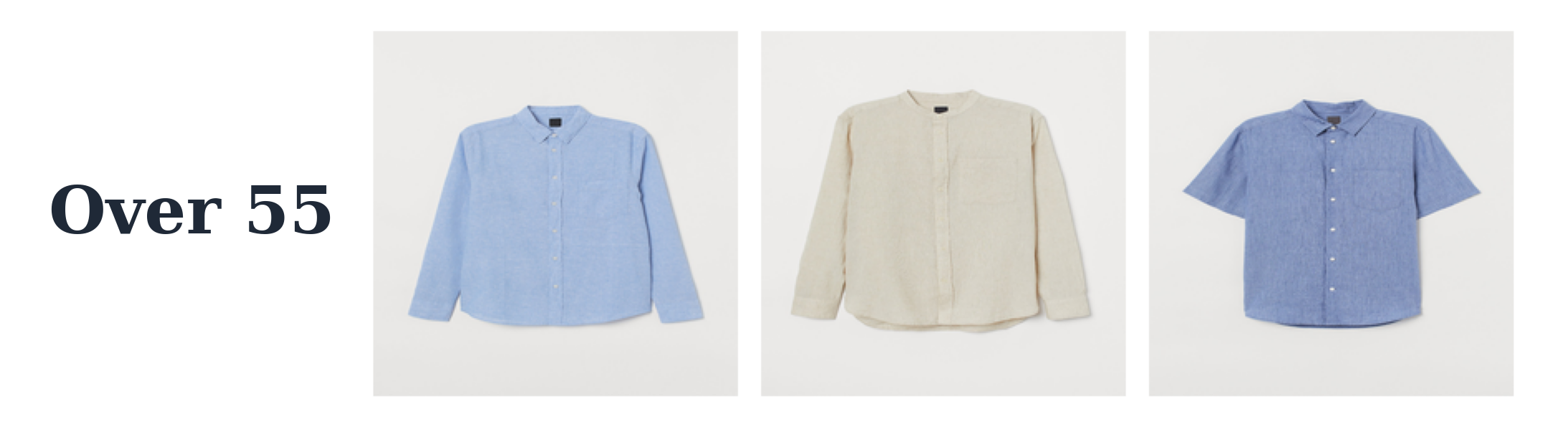}
\end{minipage}\hfill
\begin{minipage}{0.48\textwidth}
\centering\includegraphics[width=\textwidth]{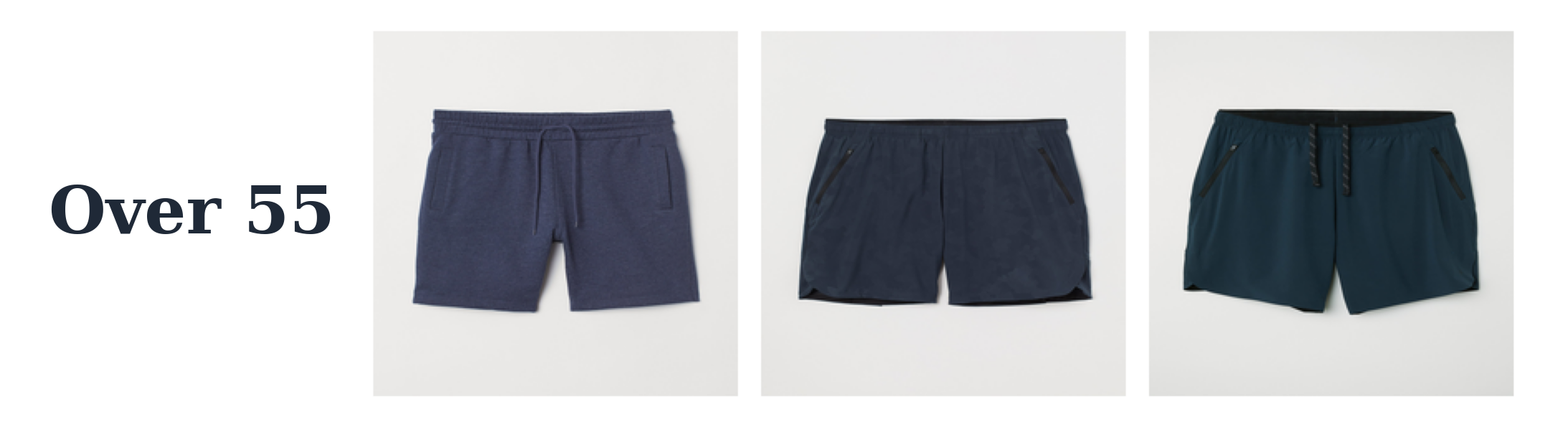}
\end{minipage}
\begin{minipage}{0.48\textwidth}
\centering\includegraphics[width=\textwidth]{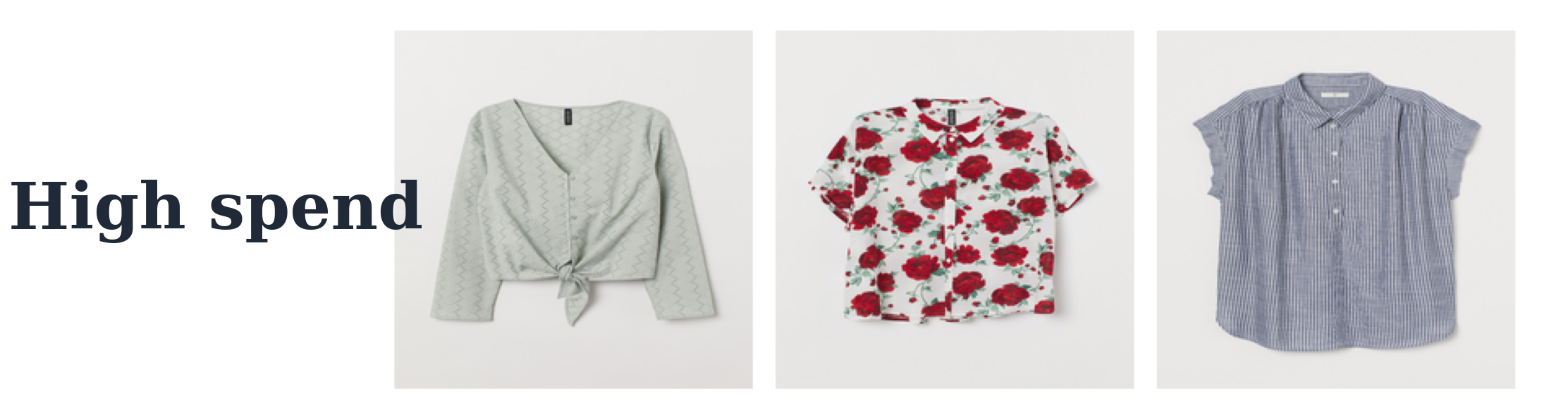}
\end{minipage}\hfill
\begin{minipage}{0.48\textwidth}
\centering\includegraphics[width=\textwidth]{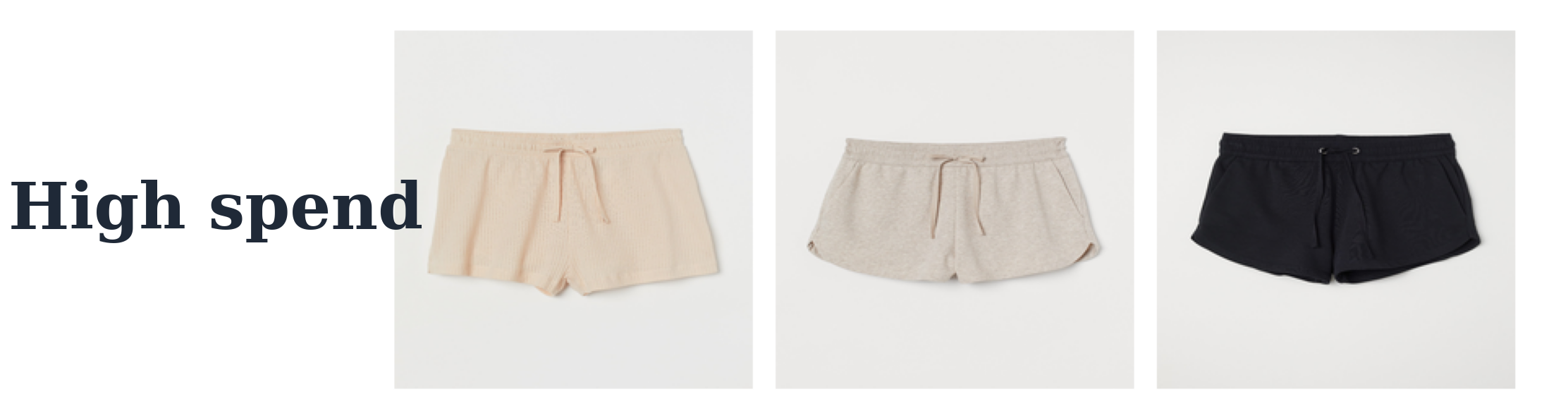}
\end{minipage}
\begin{minipage}{0.48\textwidth}
\centering\includegraphics[width=\textwidth]{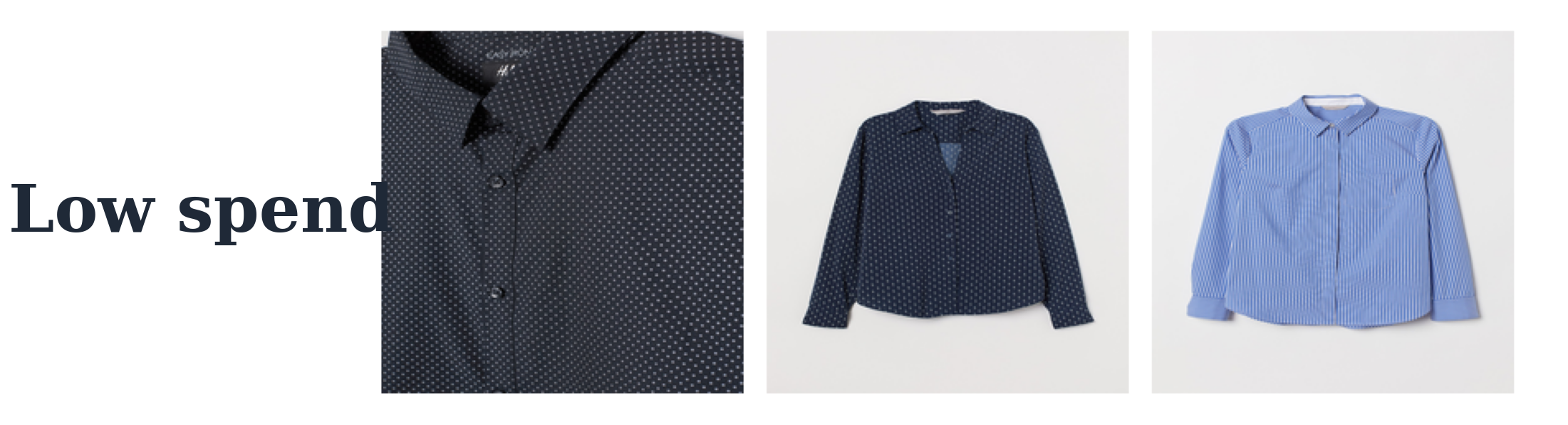}
\end{minipage}\hfill
\begin{minipage}{0.48\textwidth}
\centering\includegraphics[width=\textwidth]{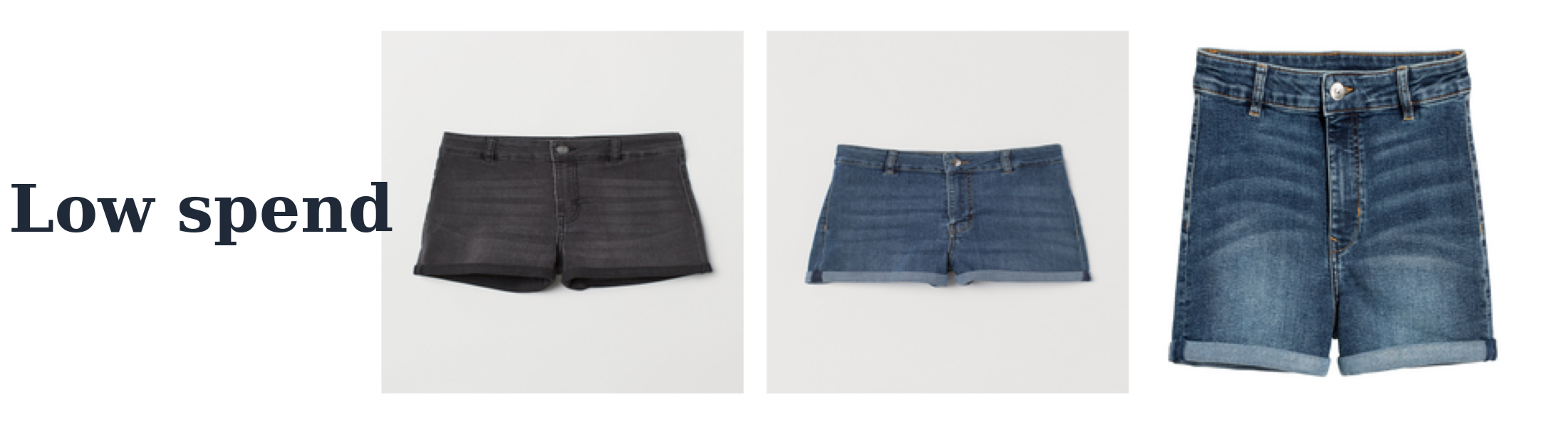}
\end{minipage}
\end{minipage}
\caption{Younger and higher-spending shoppers prefer visibly different shirts and shorts, and the pattern holds across categories. Rows from top to bottom: young, older, high-spending, low-spending segments. Each row shows the top taste picks under the $\beta$-net at that segment's mean user embedding.}
\label{fig:demand_carousel}
\end{figure}

\Cref{fig:demand_carousel} repeats the same picture on shirts and shorts. Both categories sit outside the dress estimation sample, so they serve as an external generalization check. The same age and spending separation shows up in both. The taste dimensions are general rather than dress-specific.

\subsubsection{Control Function Robustness}
\label{sec:cf_robustness}

I re-fit the same specification without the hedonic residual entering utility. \Cref{tab:cf_robustness} reports the comparison. The control function makes both classes more price-sensitive, with the minority class shift larger than the majority class shift. Signs, class ordering, the mixture split, and the aggregate own-price elasticity ordering are preserved. Validation log-likelihood improves modestly with the control function, so the qualitative picture does not depend on the correction.

\begin{table}[H]
\centering
\caption{Control function robustness under the $\embConcatDim$D Dress master. Same specification, seed, and data; only the hedonic residual is dropped from per-class utility. Both columns train for the same epoch budget.}
\label{tab:cf_robustness}
\begin{tabular}{l ccc}
\toprule
 & With CF & Without CF & Change \\
\midrule
$\bar\alpha_1$ (minority, more price-sensitive) & $\lcAlphaOne$ & $\lcAlphaOneNoCF$ & $+\lcAlphaOneShiftPct\%$ with CF \\
$\bar\alpha_2$ (majority, less price-sensitive) & $\lcAlphaTwo$ & $\lcAlphaTwoNoCF$ & $+\lcAlphaTwoShiftPct\%$ with CF \\
Val NLL                    & $\phantom{-}\lcValNLL$  & $\phantom{-}\lcValNLLNoCF$  & $-\lcValNLLShift$ with CF \\
Val R\textsuperscript{2}   & $+\lcValRSqrPct\%$ & $+\lcValRSqrNoCFPct\%$ & $+\lcValRSqrLiftPP$ pp with CF \\
$\pi_1$ (minority mixture weight) & $\phantom{-}\lcPiOne$ & $\phantom{-}\lcPiOneNoCF$ & $-\lcPiOneShift$ with CF \\
\bottomrule
\end{tabular}
\end{table}

\subsection{Heterogeneity Decomposition}
\label{sec:choice_decomposition}

To see what the two structural utility components do separately, I regress model-predicted log market shares on the price channel and the taste channel. Each channel alone carries a substantial share of the variance, and together they capture essentially all of it, so nothing outside these two channels matters for the model's aggregate predictions. Inside each price tier, the aesthetic channel is exactly what sorts products from most- to least-preferred: among products with identical posted prices, the actual chosen product's aesthetic utility ranks well above the 50\% random baseline on average and is highly significant.

\begin{figure}[H]
\centering
\includegraphics[width=\textwidth]{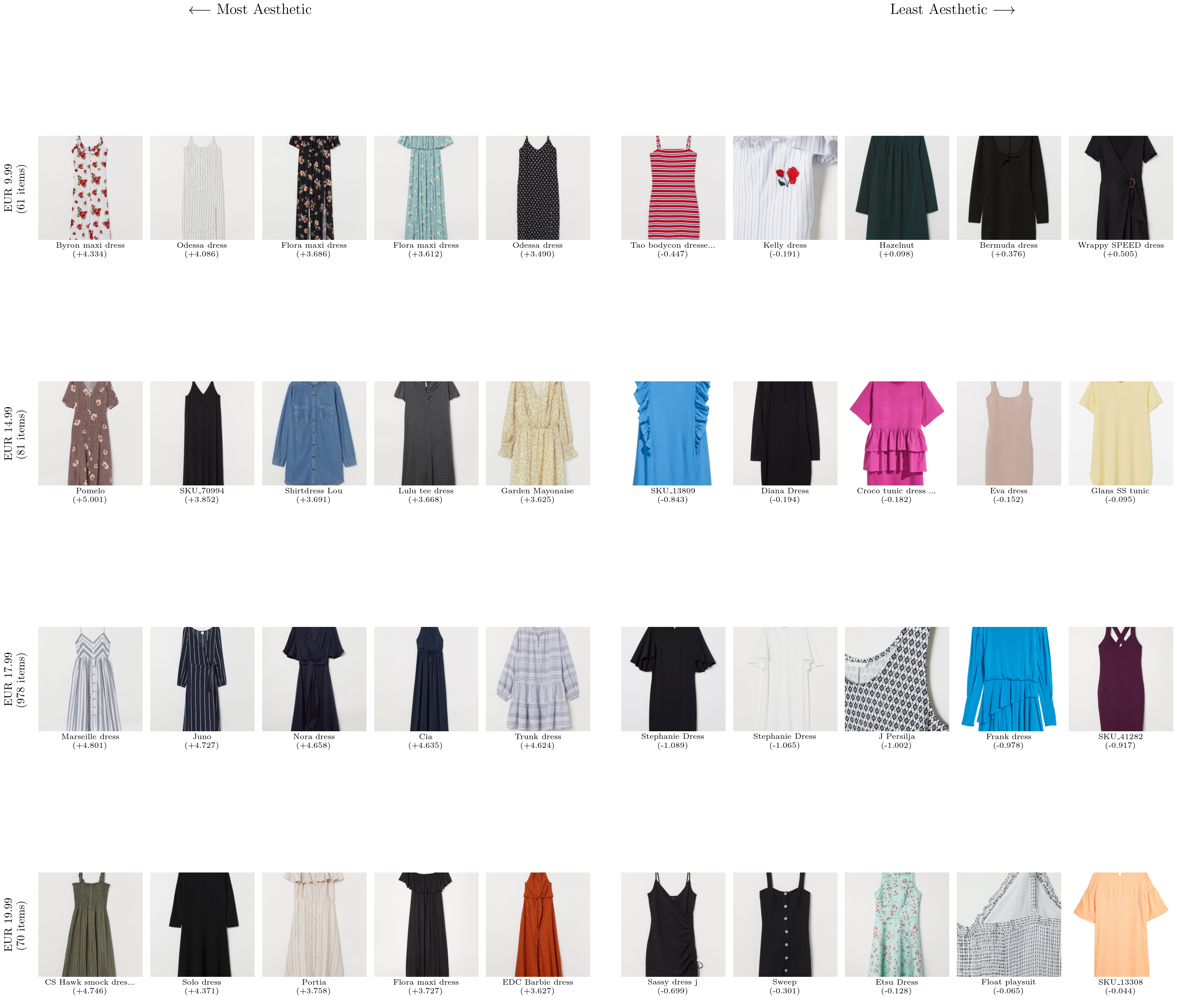}
\caption{Within-tier aesthetic ranking. Each row shows products that share the same posted price, ordered by the $\beta$-net's predicted taste utility. Price determines which tier a product lives in, and the taste channel decides which product wins inside the tier.}
\label{fig:within_tier_grid}
\end{figure}

\Cref{fig:within_tier_grid} shows that the pre-trained embeddings carry enough resolution to rank products inside a price tier even though horizontal characteristics like price are held fixed. Price sorts products vertically across tiers, and the $\beta$-net sorts products horizontally inside each tier.

\subsection{Substitution Structure}
\label{sec:choice_substitution}

\begin{figure}[H]
\centering
\includegraphics[width=0.9\textwidth]{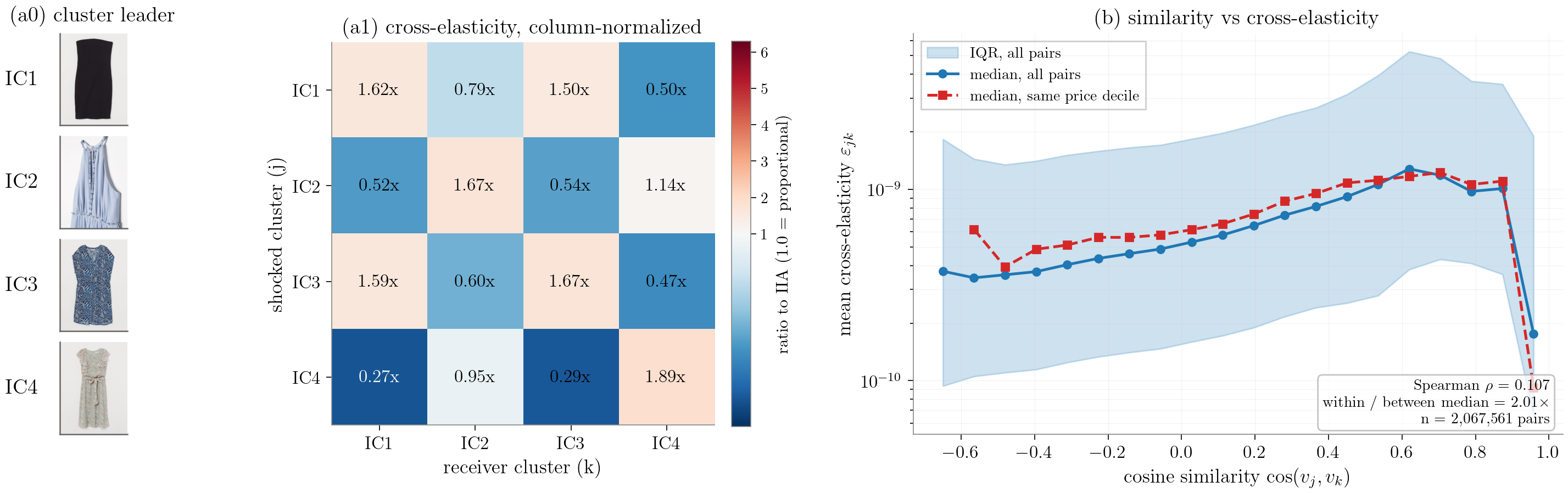}
\caption{Substitution structure at the cluster level. Left, the column-normalized four-by-four cross-elasticity matrix from the two-class latent deep logit. Middle, the similarity-to-elasticity scatter on the same-price-decile pair sample. Right, the Spearman rank correlation by similarity bin.}
\label{fig:substitution_structure}
\end{figure}

Aggregating the product-level cross-elasticity matrix over four $K$-means clusters on the three-tower aesthetic features gives \cref{fig:substitution_structure}: an average dress loses meaningfully more to other dresses in its own visual cluster than to dresses in other clusters. Substitution follows look, not price tier. Most of the response to a price change still leaks out of the category entirely; the cluster pattern describes the structure inside the residual that stays in the category.

\subsection{Consumer-Cluster Substitution}
\label{sec:user_cluster_substitution}

\begin{figure}[H]
\centering
\includegraphics[width=\textwidth]{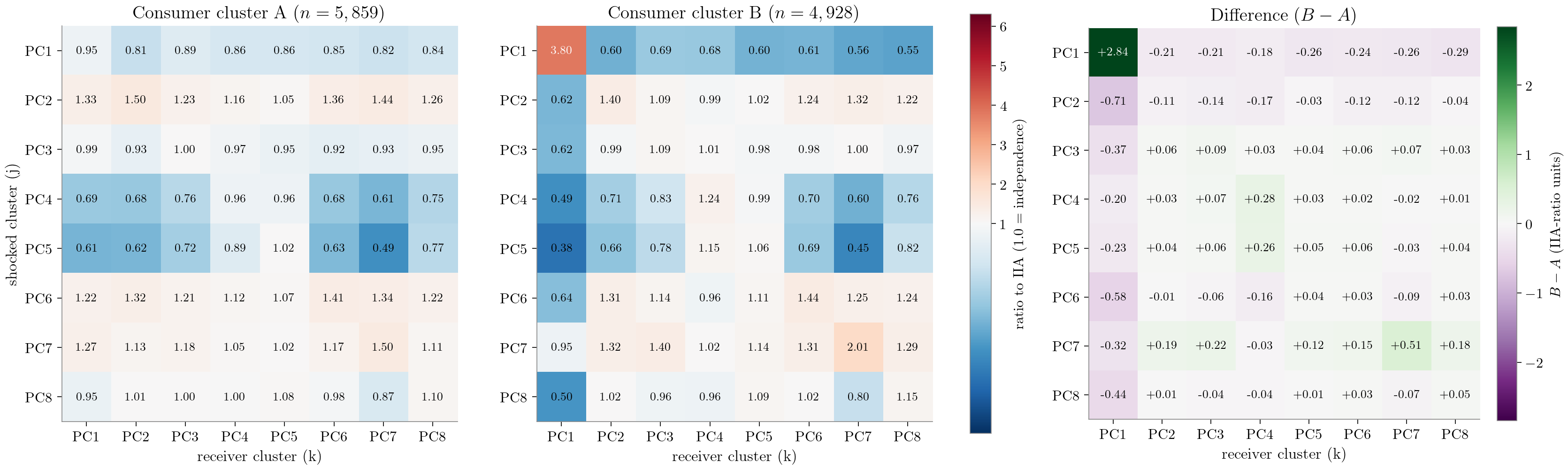}
\caption{Cross-elasticity matrices for two consumer clusters obtained by $K$-means (K=$\lcDiKUsr$, seed=$\lcDiSeed$) on the user embedding $\mathbf{d}_i$. Left panel: cluster A ($n = \lcClusterAN$, mean age $\lcClusterAAge$). Middle panel: cluster B ($n = \lcClusterBN$, mean age $\lcClusterBAge$). Right panel: the cell-by-cell difference $B - A$. The pair was chosen to maximise the Frobenius distance between the two matrices across a sweep over $K$ and random seeds (selected pair: $\lcDiFrobenius$ Frobenius, Pearson $\lcDiCorr$). The left two panels are aggregated to eight aesthetic product clusters from $K$-means on item embeddings and column-normalized so $1.0$ is the independence baseline; the difference panel shares row and column clusters and is centered at $0$ with a separate divergent scale.}
\label{fig:user_cluster_substitution}
\end{figure}

The matrix above pools every consumer in the sample. To see whether different consumers substitute differently, I group consumers using the user embedding $\mathbf{d}_i$ and pick the two groups whose substitution matrices look the most different.\footnote{Method: $K$-means on $\mathbf{d}_i$ with $K=\lcDiKUsr$, seed $\lcDiSeed$; among the eight-cluster pairs, the pair reported here maximises the Frobenius distance between cluster-aggregated cross-elasticity matrices across a sweep over $K\in\{2,4,6,8\}$ and three random seeds. The pair's Frobenius distance is $\lcDiFrobenius$ with cell-by-cell Pearson correlation $\lcDiCorr$.} \Cref{fig:user_cluster_substitution} shows their cross-elasticity matrices alongside the cell-by-cell difference. The difference panel makes the divergence explicit: several product clusters attract substitution from cluster A but not from cluster B, and one (a Divided line of cheap youth-targeted dresses) flips diagonal dominance by a factor of four across the two consumer clusters. The model therefore picks up real taste-driven differences across consumers that an age-only split would miss.

\input{06_seasonal}

\subsection{Markup Recovery}
\label{sec:demand:supply}

Applying the multi-product monopolist inversion from \cref{sec:theory:supply} to the latent-class demand Jacobian gives mean Lerner index $\bar{L} = \lcLernerNum$ with mean marginal cost \texteuro{\lcMcMean} among positive-cost items. Of \dataJ{} dresses, \lcMcPosCount{} ($\lcMcPosPct\%$) have $\widehat{mc}_j > 0$; the remainder are cheap jersey basics where the model estimates inelastic demand ($|\varepsilon| < 1$) and the pre-trained embeddings do not provide enough differentiation to generate elastic substitution. The result is robust to the choice of outside share: \cref{tab:supply-sensitivity} in \cref{sec:appendix_tau_robustness} shows mean $|\varepsilon|$ essentially unchanged across $\hat\tau$ values spanning the plausible single-category online share band. The recovered Lerner sits modestly above H\&M's reported gross margin. The static Bertrand-Nash model overstates realized margins because it assumes prices are always set at the instantaneous profit maximum, while in practice markdown calendars and seasonal clearance pricing push realized margins below that benchmark \citep{soysal2012demand, caro2020future}. Cross-category substitution beyond the dresses in the estimation sample would further moderate margins.

\begin{figure}[H]
\centering
\begin{minipage}{0.52\textwidth}
\centering
\includegraphics[width=\textwidth]{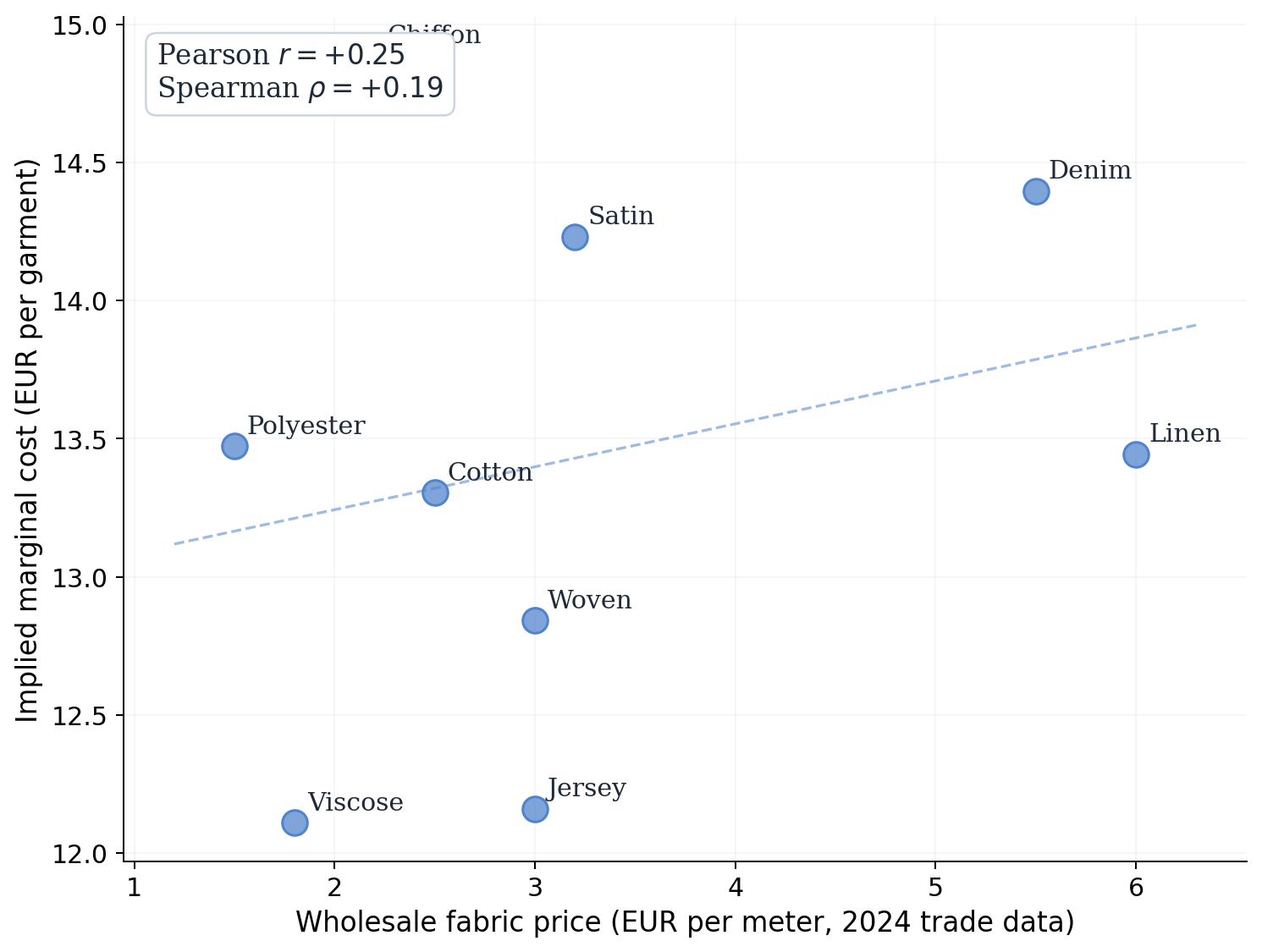}
\end{minipage}\hfill
\begin{minipage}{0.44\textwidth}
\centering
\includegraphics[width=\textwidth]{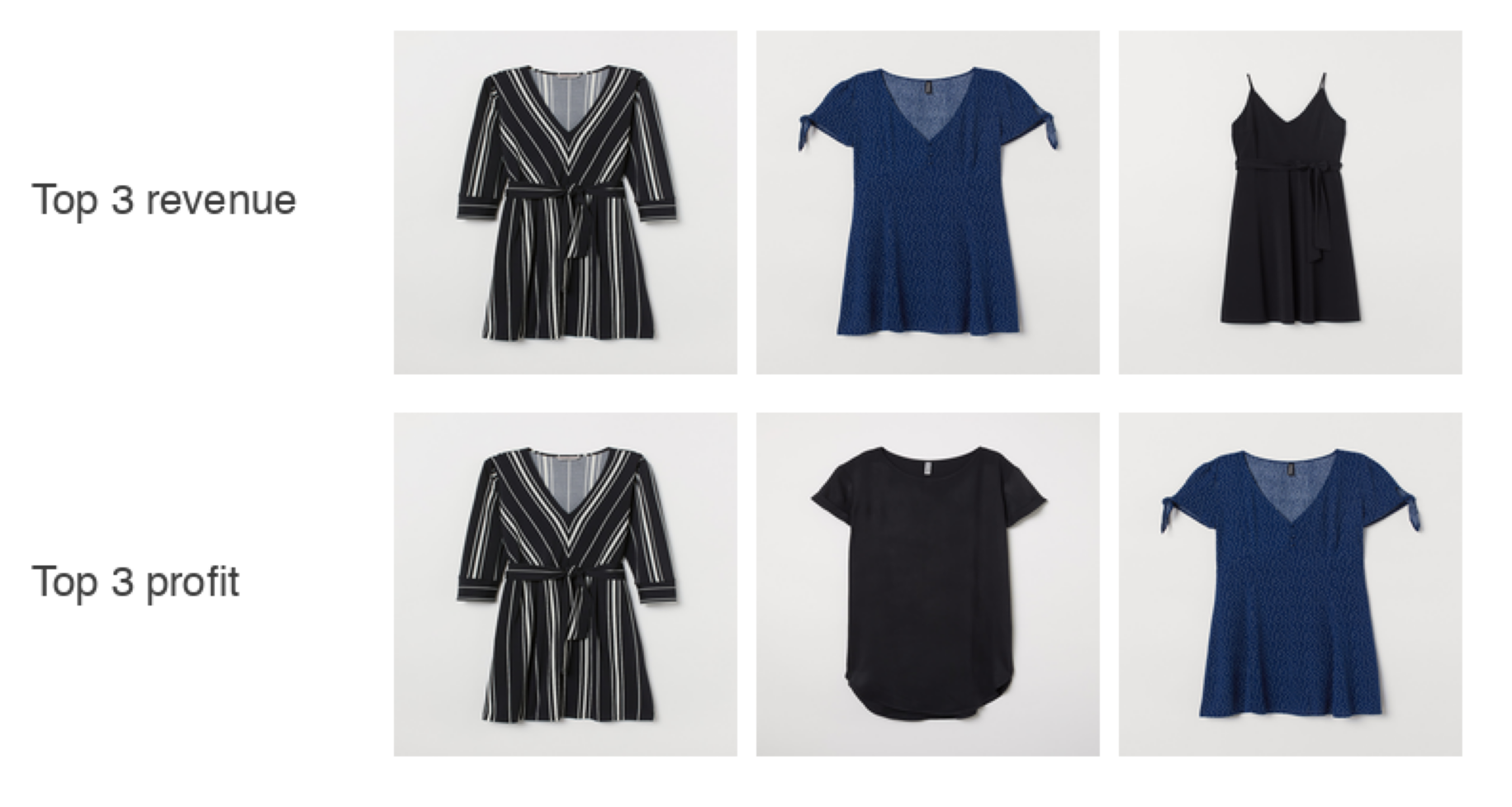}
\end{minipage}
\caption{Implied costs and the revenue-versus-profit ranking. Left, mean implied marginal cost per garment (blue) grouped by the primary fabric keyword in the product description, plotted next to the median wholesale fabric price (grey) from industry sources, each min-max rescaled. Right, the top three dresses ranked by total revenue (upper row) and by total profit (lower row). The two lists barely overlap.}
\label{fig:supply_validation}
\end{figure}

\Cref{fig:supply_validation} supports the inversion in two ways. The material breakdown on the left shows that structured fabrics such as woven and satin have higher implied costs than jersey and viscose, which are stretch materials that are cheaper to cut and sew. The model never saw any information about materials, so the ordering is a qualitative out-of-sample check that the implied costs line up with production economics. The revenue-versus-profit panel on the right shows that the top sellers by revenue are not the top earners by profit, which is what the Lerner formula implies when low-priced products have the highest elasticity and therefore the thinnest margins.

\section{Counterfactuals}
\label{sec:counterfactuals}

\subsection{Sustainability Practices}

\begin{figure}[H]
\centering
\includegraphics[width=0.8\textwidth]{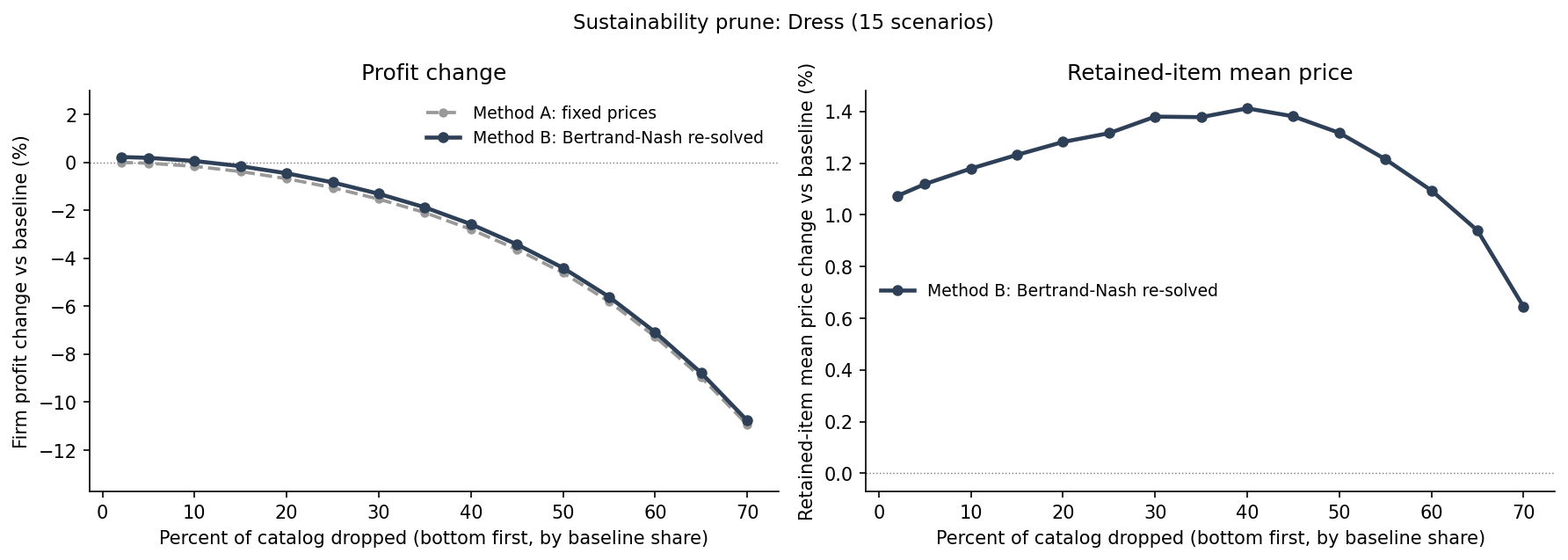}
\caption{Sustainability prune for dresses, ordered by baseline inside share (smallest first). Left panel: firm profit change against prune depth under two methods. The dashed gray line holds prices at observed values (Method A), the solid dark line re-solves the multi-product Bertrand-Nash equilibrium on the retained subset (Method B). Method A loses profit monotonically, from $\sustMethodATwoPct\%$ at a $\sustPruneShallow\%$ prune to $\sustMethodASevenPct\%$ at a $\sustPruneDeep\%$ prune. Method B earns a re-pricing premium of roughly $+\sustReprice$ percentage points over Method A at every depth, which keeps profit positive up to about a $\sustNeutralLo\%$ cut and crosses the profit-neutral line between $\sustNeutralLo\%$ and $\sustNeutralHi\%$. Right panel: retained-item mean price under Method B, rising from $+\sustPriceTwoPct\%$ at a $\sustPruneShallow\%$ prune to $+\sustPriceSevenPct\%$ at a $\sustPruneDeep\%$ prune. Cross-category curves are in \cref{sec:appendix_cross_category}.}
\label{fig:sustainability_prune}
\end{figure}

The first counterfactual tests a sustainability practice. Fast-fashion retailers face pressure to reduce SKU counts, so I order dresses by their baseline market share and drop the bottom quantile. I compute the profit consequences two ways. Method A holds prices at the observed values and lets shares redistribute across the surviving items, which is the conventional accounting used in the existing SKU-pruning literature. Method B re-solves the multi-product Bertrand-Nash equilibrium on the retained subset using the latent-class mixture Jacobian, so the firm is allowed to re-optimize prices on the products it keeps. \Cref{fig:sustainability_prune} compares the two profit curves. A cross-category view covering the other five categories is in \cref{sec:appendix_cross_category}.

Under the latent-class demand model, dropping the lowest-share dresses costs the firm profit on the observed-price path and recovers part of that loss on the re-pricing path. Method A profit falls monotonically and convexly as the prune deepens. Method B earns a positive re-pricing premium over Method A at every prune depth, because removing close aesthetic substitutes lets the firm lift markups on what it keeps. The premium keeps Method B profit positive up to a moderate prune depth and crosses the profit-neutral line near a one-third catalogue cut. The mean retained-item price rises with deeper prunes. \Cref{fig:sustainability_prune} reports the profit curves and retained-item price path.

\subsection{Aesthetic Differentiation Collapse}

The second counterfactual collapses aesthetic differentiation instead of catalogue size. I hold the catalogue fixed and push each class's item projections one standard deviation toward the class-specific centroid, then re-solve the multi-product Bertrand-Nash equilibrium on the flattened taste space. Consumer surplus barely moves and mean price drops slightly, but firm profit collapses (a drop of $\aesthProfitChangePct\%$). \Cref{tab:aesthetic_collapse} reports the full decomposition. The profit collapse is driven by the Bertrand-Nash markup formula. When products become more similar in taste space, substitution inside the catalogue intensifies, and in the re-solved system that stronger cannibalisation lowers equilibrium markups across the board. In this model, most dress profit comes from product differences in the learned demand space, which is the economic value of the three-tower representation in euros.

\begin{table}[H]
\centering
\caption{Effect of collapsing per-class aesthetic differentiation by one standard deviation toward the class-specific centroid. Dress master, $J = 2{,}034$, $I = 38{,}918$, $K = 2$. The Bertrand-Nash equilibrium is re-solved on the flattened taste space. The mean within-class item cosine similarity to the class centroid is reported separately for each latent class; both go up because items move closer to their class centroid.}
\label{tab:aesthetic_collapse}
\begin{tabular}{lrr}
\toprule
 & Baseline & $\Delta$ vs baseline \\
\midrule
Firm profit (\texteuro{} per week)                    & $\aesthBaselineProfit$   & $-\aesthProfitChangePct\%$ \\
Consumer surplus (\texteuro{} per consumer)           & $\aesthBaselineCS$       & $\aesthCsChangePct\%$ \\
Total welfare (\texteuro{} per week)                  & \phantom{1}$\aesthBaselineTotalWelfare$ & $-\aesthTotalWelfareChangePct\%$ \\
Mean price                                             & $\aesthBaselineMeanPrice$ & $-\aesthPriceChangePct\%$ \\
Mean within-class item cosine to centroid (class 1)    & $\aesthCosineOneBase$    & $+\aesthCosineOneDelta$ \\
Mean within-class item cosine to centroid (class 2)    & $\aesthCosineTwoBase$    & $+\aesthCosineTwoDelta$ \\
\bottomrule
\end{tabular}
\end{table}

\subsection{Personalized Pricing Ladder}

The firm clusters consumers into $N_s$ segments by K-means on the 64-dimensional user embedding $\mathbf{d}_i$.\footnote{$N_s$ is the pricing-segment count, distinct from the $K=2$ latent classes inside the demand model. The segment index $g$ below is locally scoped and does not collide with the latent-class index $c$ of \cref{sec:theory}.} Within each segment $g$ with consumer set $\mathcal{I}_g$ of size $I_g$, the firm solves the multi-product Bertrand-Nash problem
\begin{equation}
  \max_{\mathbf{p}_g \,\in\, \mathbb{R}^J_+} \; \sum_{j=1}^J (p_{gj} - \widehat{mc}_j)\, S_{gj}(\mathbf{p}_g),
  \qquad
  S_{gj}(\mathbf{p}_g) \;=\; \frac{1}{I_g} \sum_{i \in \mathcal{I}_g} \sum_{c=1}^{K} \pi_{c}\, s_{ij \mid c}(\mathbf{p}_g),
  \label{eq:segment_bn}
\end{equation}
under multi-product monopolist conduct ($\boldsymbol{\Omega}=\mathbf{1}\mathbf{1}^{\top}$). The first-order condition $\mathbf{p}_g - \widehat{\mathbf{mc}} = -\bigl[\boldsymbol{\Omega} \odot \partial\mathbf{S}_g/\partial\mathbf{p}_g\bigr]^{-1} \mathbf{S}_g$ is solved numerically, and total firm profit is $\Pi(N_s) = \sum_g I_g\,(\mathbf{p}_g^\star - \widehat{\mathbf{mc}})\cdot \mathbf{S}_g(\mathbf{p}_g^\star)$. $N_s=1$ is uniform pricing and the finest segmentation I run is $N_s=128$ (roughly $300$ consumers per cluster at $I=\num{38918}$).

The $N_s=1$ rung lands essentially at the static FOC baseline, confirming that the FOC-inverted optimal uniform price is close to the observed mean. Segmentation then builds monotonically up to the finest rung at $N_s=128$, which caps at a profit gain of $+\persGainOneTwentyEight\%$ over uniform pricing. Most of the gain accrues in the middle rungs; the final doublings add little. That saturation is the signature of bounded heterogeneity in $\alpha(\mathbf{d}_i)$ within the two latent classes: the user embedding carries a real pricing signal, but one that does not get reliably bigger by cutting the consumer base more finely. \Cref{tab:personalized_pricing_ladder} reports the full ladder.

\begin{table}[H]
\centering
\caption{Personalization ladder: Dress, under the canonical $K=2$ latent-class deep logit with the multi-product monopolist first-order condition solved separately inside each of $N_s$ K-means segments on $\mathbf{d}_i$. $N_s$ is the segmentation count, distinct from the $K=2$ latent classes inside the demand model.}
\label{tab:personalized_pricing_ladder}
\begin{tabular}{rrrr}
\toprule
$N_s$ & Profit gain (\%) & Mean price shift (\%) & Converged \\
\midrule
\phantom{0}\phantom{0}\phantom{0}1 & $+\persGainOne$             & $+\persPriceShiftOne$             & $1/1$     \\
\phantom{0}\phantom{0}\phantom{0}2 & $+\persGainTwo$             & $+\persPriceShiftTwo$             & $2/2$     \\
\phantom{0}\phantom{0}\phantom{0}4 & $+\persGainFour$            & $+\persPriceShiftFour$            & $4/4$     \\
\phantom{0}\phantom{0}\phantom{0}8 & $+\persGainEight$           & $+\persPriceShiftEight$           & $8/8$     \\
\phantom{0}\phantom{0}16           & $+\persGainSixteen$         & $+\persPriceShiftSixteen$         & $16/16$   \\
\phantom{0}\phantom{0}32           & $+\persGainThirtyTwo$       & $+\persPriceShiftThirtyTwo$       & $32/32$   \\
\phantom{0}\phantom{0}64           & $+\persGainSixtyFour$       & $+\persPriceShiftSixtyFour$       & $64/64$   \\
\phantom{0}128                     & $+\persGainOneTwentyEight$  & $+\persPriceShiftOneTwentyEight$  & $128/128$ \\
\bottomrule
\end{tabular}
\end{table}

\subsection{Recommendation Validation: \texorpdfstring{$\beta$}{β}-net vs Collaborative Filtering}

The fourth counterfactual asks whether the $\beta$-net is useful as a recommender in its own right. Holding the price band fixed and training both models on the same earlier weeks, the $\beta$-net beats a tuned collaborative-filtering baseline on items the user has never bought before. The Recall@10\footnote{Recall@10 is the share of a user's held-out purchases that the model ranks among its top ten dresses for that user, averaged across users.} lead is reported in \cref{tab:recval} and survives every one-at-a-time perturbation of the active-user threshold, the price-tile granularity, the held-out horizon, and all three hyperparameters of the collaborative filter. The collaborative filter does win on items the user already bought during training, by simply recognising past purchases, but that is a memorisation gap, not a taste-prediction gap. The $\beta$-net's user-side projections $\mathbf{r}_i^{c}$ are a function of the pre-trained user embedding $\mathbf{d}_i$ alone, so it is the pre-trained taste signal by itself that beats collaborative filtering on the new-purchase task.

\begin{table}[H]
\centering
\caption{Recall@10 on the held-out weeks; both models train on the same earlier weeks. \emph{All}: every held-out purchase. \emph{Warm}: items the user bought during training (memorisation slice, $N{=}\recWarmN$). \emph{Cold}: items new to this user ($N{=}\recColdN$). \emph{Best of grid}: the highest-cold-recall cell of an eighteen-point tuning grid for the collaborative filter. The cold-slice lead is robust to all sensitivity sweeps reported in the validation script's output file.}
\label{tab:recval}
\begin{tabular}{lrrr}
\toprule
Method & all & warm & cold \\
\midrule
Popularity                          & $\recPopAll\%$    & $\recPopWarm\%$    & $\recPopCold\%$ \\
Collaborative filter, default       & $\recCFDefAll\%$  & $\recCFDefWarm\%$  & $\recCFDefCold\%$ \\
Collaborative filter, best of grid  & $\recCFBestAll\%$ & $\recCFBestWarm\%$ & $\recCFBestCold\%$ \\
$\beta$-net, mixture-weighted       & $\recBetaMixAll\%$ & $\recBetaMixWarm\%$ & $\mathbf{\recBetaMixCold\%}$ \\
$\beta$-net, dominant-class         & $\recBetaDomAll\%$ & $\recBetaDomWarm\%$ & $\recBetaDomCold\%$ \\
\bottomrule
\end{tabular}
\end{table}

\subsection{Class-Level Taste Heterogeneity in the \texorpdfstring{$\beta$}{β}-net}
\label{sec:counterfactuals:betanet_heterogeneity}

Fifty synthetic dresses are generated through a vision-language design pipeline, projected through the frozen item tower, and scored separately under each of the two latent classes.\footnote{A vision-capable language model writes a design brief and ten categorical attribute fields; an image model renders the design; the item tower from \cref{sec:embeddings} maps the result to a sixty-four-dimensional unit-norm embedding in the same space as $\mathbf{v}_j$. The class-$c$ score is the inner product of that embedding with the grand-mean $\beta$-net output within class $c$.} \Cref{fig:betanet_class_polarising} shows the six designs on which the two classes disagree most. \Cref{tab:betanet_class_polarising} reports per-design class ranks and the rank-correlation summary.

Two findings. The demand model extends cleanly to designs outside the estimation sample, which supports generative counterfactuals on the catalogue without re-estimation. The $\beta$-net recovers consumer-side aesthetic heterogeneity that the aggregate share model averages away, so the same new product is more or less attractive depending on which latent class looks at it.\footnote{The disagreement is a property of the latent-class split, not of user-embedding geometry. Within any $K$-means cluster on $\mathbf{d}_i$, ranking agreement across clusters is high on both synthetic and held-out real dresses, while cross-class agreement is substantially lower. Full Kendall-$\tau$ values appear in the validation script output.}

\input{betanet_class_polarising_fragment}
\input{betanet_class_polarising_table}

%% file: 06_seasonal.tex
\subsection{Seasonal Demand Dynamics}
\label{sec:time_varying}

A seasonal extension of the demand model lets the taste vector depend on calendar month through the rank-eight shift $\boldsymbol{\delta}_m$ of \cref{eq:app_delta}. Price sensitivity is not given a seasonal component; the channel that moves with the calendar is aesthetic taste, not the price coefficient.

\subsubsection{Taste Rotation}

\begin{figure}[H]
\centering
\includegraphics[width=0.85\textwidth]{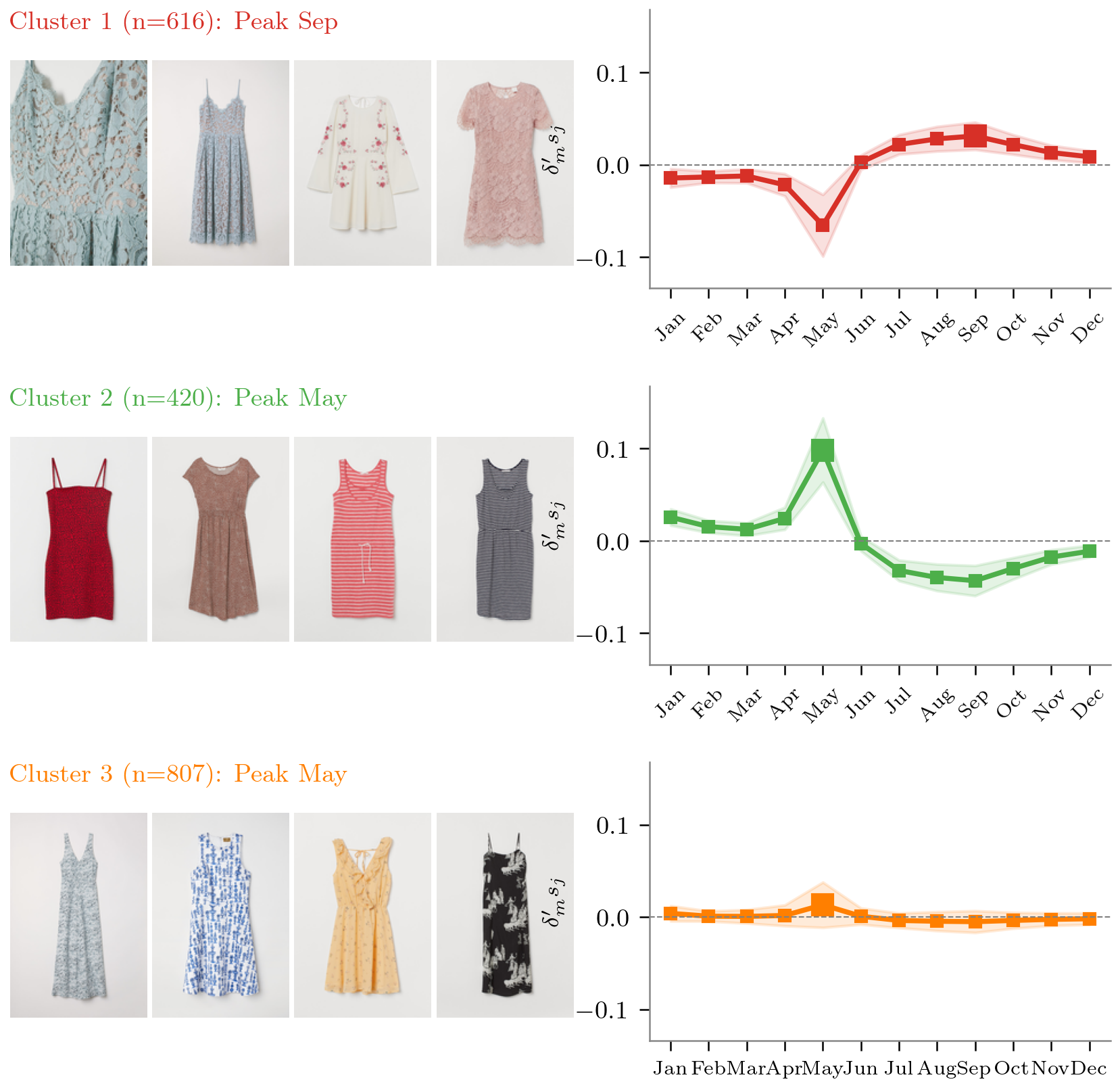}
\caption{Monthly aesthetic cluster rotation driven by the low-rank taste shift $\boldsymbol{\delta}_m = W \mathbf{z}_m$. Each row shows, for one month, the item cluster that gains the most utility next to the cluster that loses the most.}
\label{fig:seasonal_cluster_images}
\end{figure}

The two-class latent deep logit identifies a monthly taste shift inside the pre-COVID estimation window. The peak occurs in June at roughly $\seasDeltaPeakPct\%$ of the grand-mean taste norm, and the rotation persists through summer and fall. \Cref{fig:seasonal_cluster_images} visualises the month-by-month shift in the item clusters that gain and lose utility. Implied own-price elasticities vary across months because $\boldsymbol{\delta}_m$ rotates which clusters consumers attend to, and the cross-cluster spread exceeds the cross-month spread; seasonal markdown decisions should therefore be anchored to the taste rotation, not to a calendar of price moves.

%% file: betanet_class_polarising_fragment.tex
\begin{center}
\begin{minipage}[b]{0.14\textwidth}
  \centering
  \includegraphics[width=\linewidth]{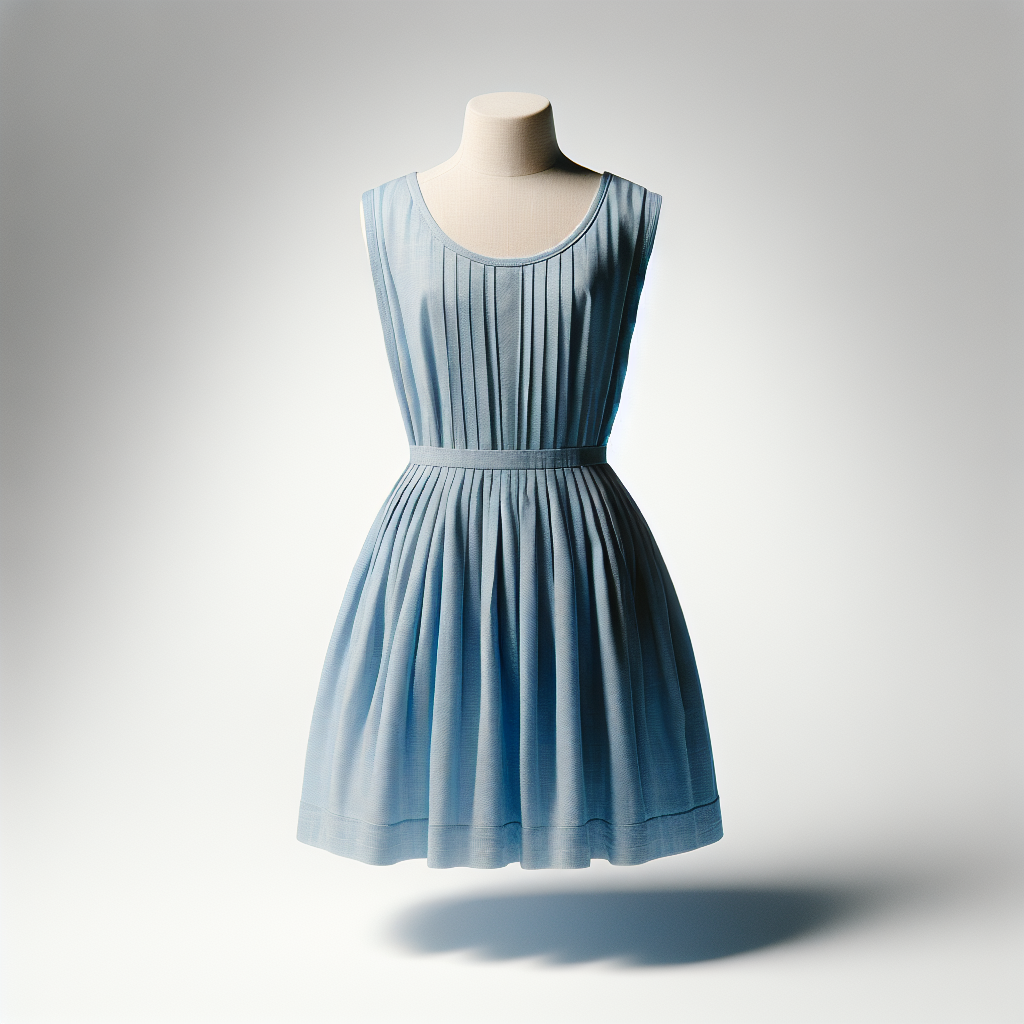}
  \par\smallskip {\tiny Design 1}
\end{minipage}
\hfill
\begin{minipage}[b]{0.14\textwidth}
  \centering
  \includegraphics[width=\linewidth]{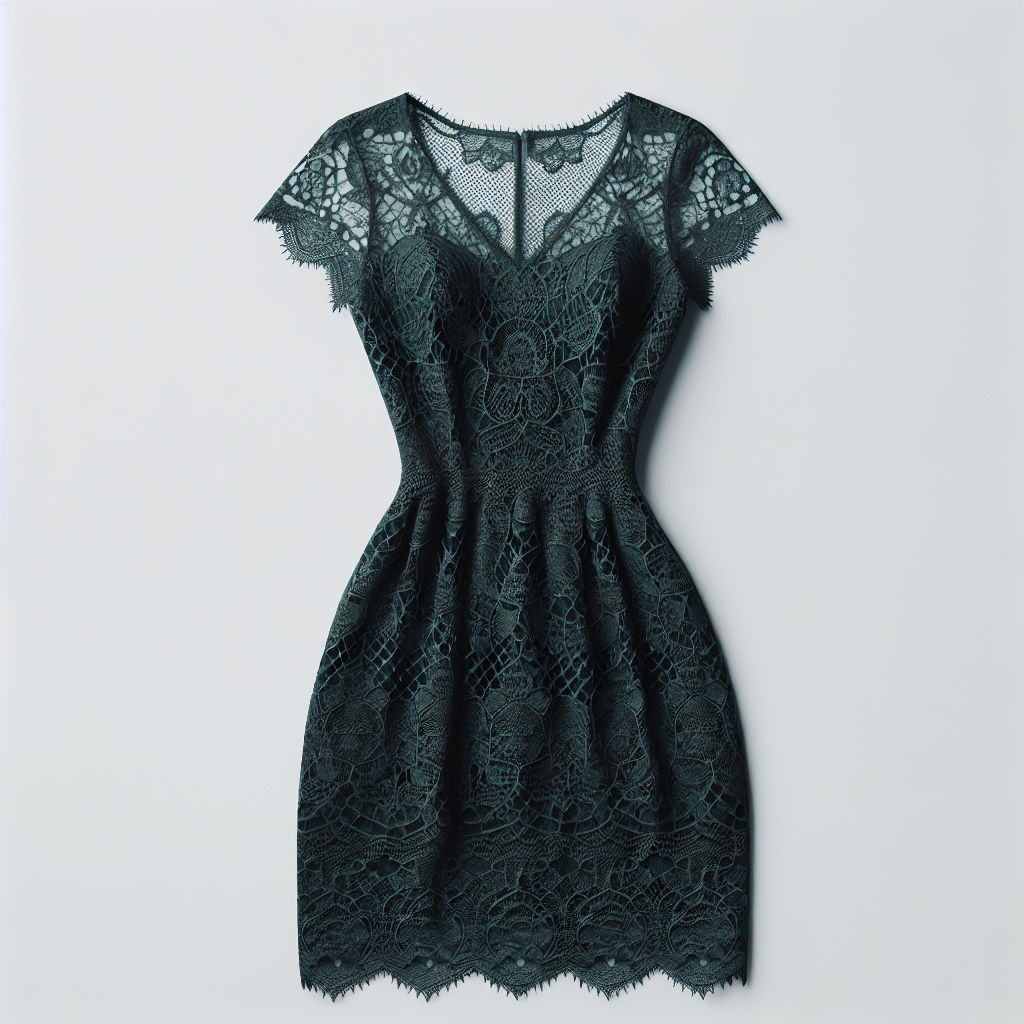}
  \par\smallskip {\tiny Design 2}
\end{minipage}
\hfill
\begin{minipage}[b]{0.14\textwidth}
  \centering
  \includegraphics[width=\linewidth]{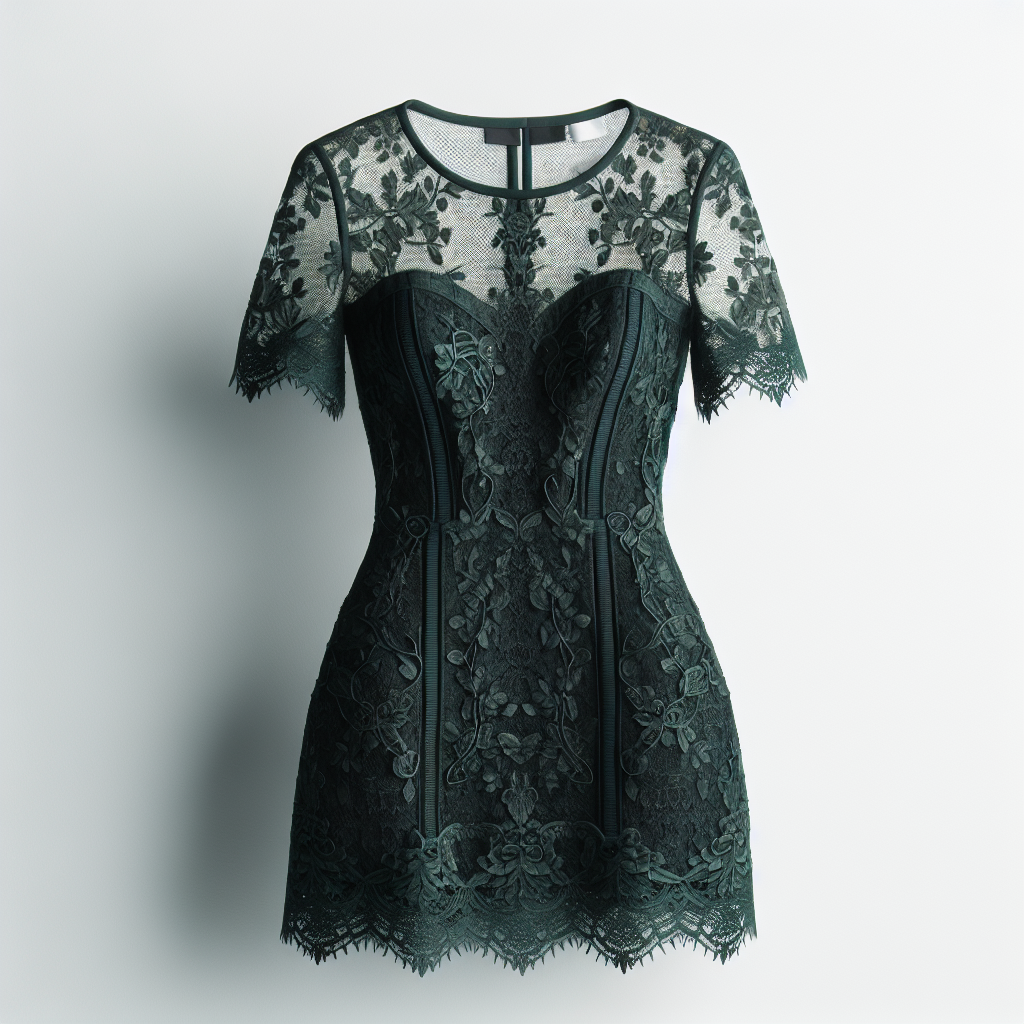}
  \par\smallskip {\tiny Design 3}
\end{minipage}
\hfill
\begin{minipage}[b]{0.14\textwidth}
  \centering
  \includegraphics[width=\linewidth]{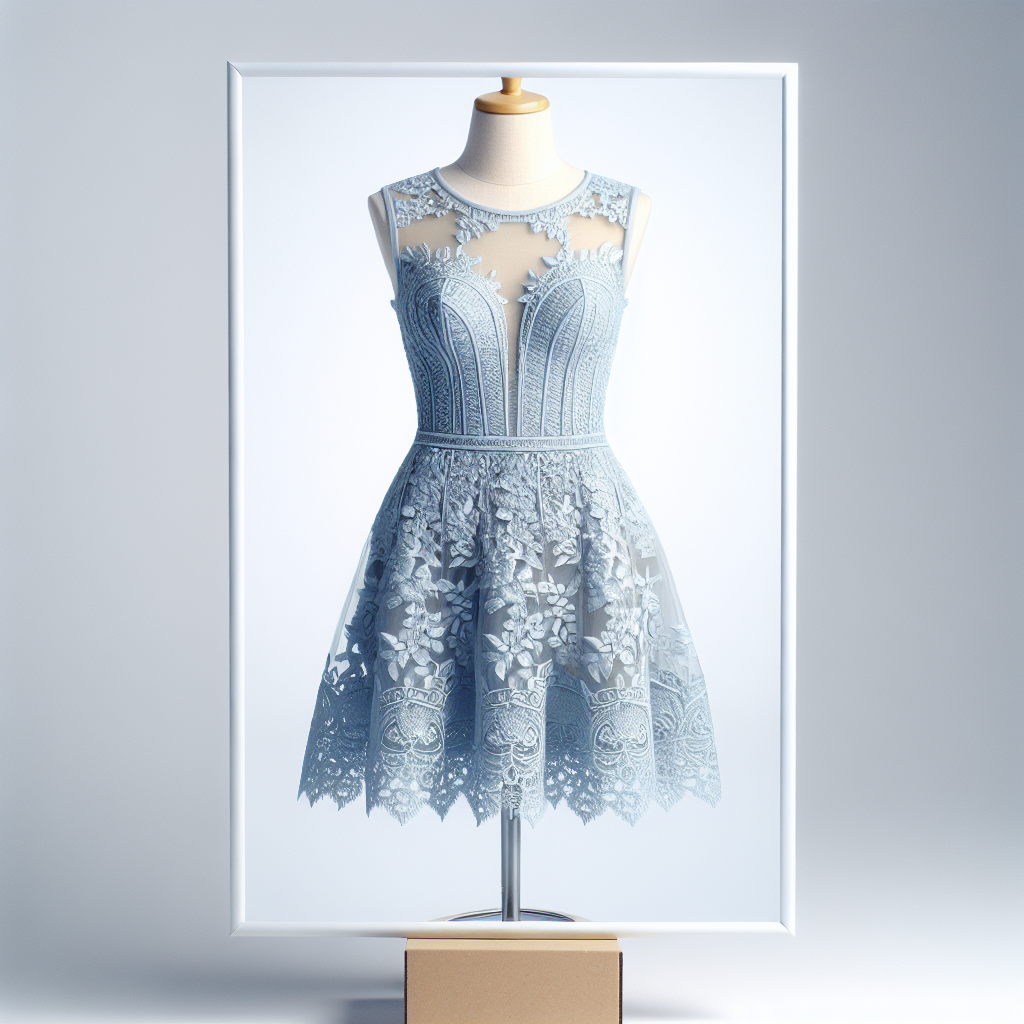}
  \par\smallskip {\tiny Design 4}
\end{minipage}
\hfill
\begin{minipage}[b]{0.14\textwidth}
  \centering
  \includegraphics[width=\linewidth]{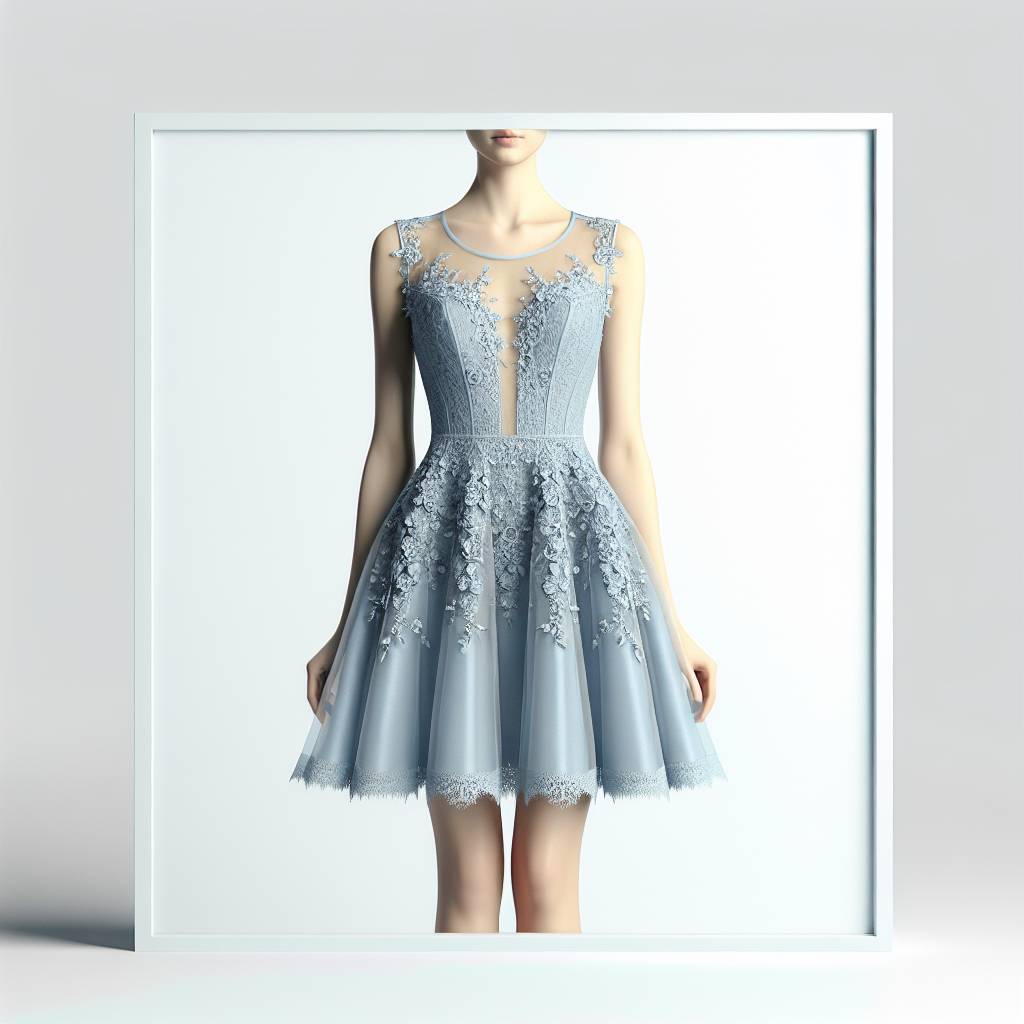}
  \par\smallskip {\tiny Design 5}
\end{minipage}
\hfill
\begin{minipage}[b]{0.14\textwidth}
  \centering
  \includegraphics[width=\linewidth]{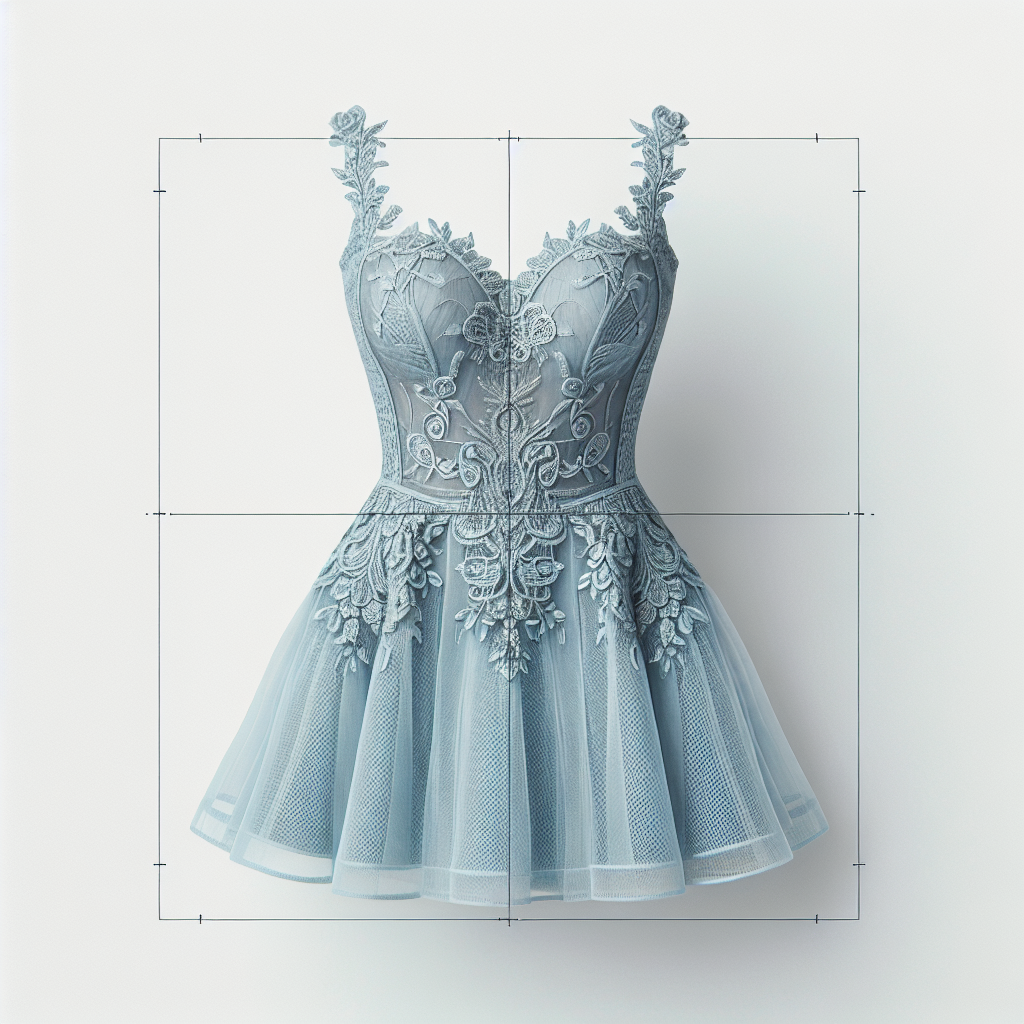}
  \par\smallskip {\tiny Design 6}
\end{minipage}
\captionof{figure}{The six synthetic dresses on which the two latent consumer classes disagree most in ranking.  Per-design ranks and summary rank-agreement statistics appear in Table~\ref{tab:betanet_class_polarising}.}
\label{fig:betanet_class_polarising}
\end{center}

%% file: betanet_class_polarising_table.tex
\begin{center}
\captionof{table}{Class-level ranking disagreement on the six most-polarising synthetic dresses.  Rank under class 1 is each design's position among the fifty synthetic designs under the minority, more price-sensitive class; rank under class 2 is the position under the majority, less price-sensitive class; lower is more preferred.  The lower panel reports summary rank-agreement statistics across all fifty designs.}
\label{tab:betanet_class_polarising}
\begin{tabular}{lccc}
\toprule
Design & Rank under class 1 & Rank under class 2 & Rank gap \\
\midrule
Design 1 (Light Blue Solid) & 6 & 50 & 44 \\
Design 2 (Dark Green Lace) & 47 & 6 & 41 \\
Design 3 (Dark Green Lace) & 48 & 9 & 39 \\
Design 4 (Light Blue Lace) & 4 & 38 & 34 \\
Design 5 (Light Blue Lace) & 14 & 46 & 32 \\
Design 6 (Light Blue Lace) & 9 & 41 & 32 \\
\midrule
\multicolumn{4}{l}{\emph{Summary across all $N=50$ synthetic designs}} \\
Kendall-$\tau$ (class~1 vs class~2) & \multicolumn{3}{c}{$+0.058$ ($p < 10^{-3}$)} \\
Mean absolute rank gap & \multicolumn{3}{c}{$16.0$ positions} \\
Maximum absolute rank gap & \multicolumn{3}{c}{$44$ positions} \\
\bottomrule
\end{tabular}
\end{center}

%% file: 08_hedonic.tex
\section{Hedonic Price Indices}
\label{sec:hedonic}

\subsection{Model}
\label{subsec:hedonic_model}

Matched-model indices track price changes for surviving products but fail when assortment turnover is rapid. Hedonic regressions \citep{rosen1974} instead treat listed price as a function of observable product characteristics. Listed prices respond to many forces at once, including assortment turnover, seasonality, discounts, underlying quality and construction, input costs, and competition, and the hedonic surface summarises how those forces are reflected in the implicit price of each attribute. I specify the hedonic surface as
\begin{equation}
\log p_{jt} = f_t(\mathbf{v}_j, \mathbf{h}_j) + \varepsilon_{jt},
\label{eq:hedonic_model}
\end{equation}
where $f_t$ denotes the evaluation at month $t$ of a single gradient-boosted tree, fitted jointly on all months in the sample using LightGBM \citep{ke2017lightgbm}. LightGBM is an open-source gradient-boosted tree library optimized for large datasets. The month and year enter as features of $\mathbf{h}_j$, so the same fitted surface returns a period-specific prediction. The dress category averages about three article-month observations per product per period, too few to fit the 684-feature surface period by period, so pooling is a sample-size compromise rather than a methodological preference. A gradient-boosted tree is a flexible nonlinear regressor that builds many small decision trees in sequence, with each new tree fitted to the residual errors of the trees before it. The feature space includes the $\embTowerDim$-dimensional pre-trained item embeddings $\mathbf{v}_j$, $\embCLIPDim$ raw CLIP dimensions (the same pre-trained CLIP image-text encoder that enters the three-tower model), 100 TF-IDF text dimensions from product descriptions\footnote{TF-IDF (term frequency-inverse document frequency) turns a text description into a numerical vector by counting how often each word appears in that product's description, then down-weighting words that appear in nearly every product. Common words like ``dress'' get low weight; distinctive words like ``lace'' or ``ruffled'' get high weight.}, and a small set of temporal and categorical controls, for a total of $\hedTotalFeats$ features.

\Cref{tab:hedonic_results} reports the ablation. Text features carry most of the within-time $R^2$ gain. However, the three-tower embeddings encode the most time-stable product properties, losing the least predictive power in out-of-time evaluation compared to text.

\begin{table}[H]
\centering
\caption{Hedonic model ablation and temporal stability}
\label{tab:hedonic_results}
\begin{tabular}{lrccc}
\toprule
Specification & Feat & $R^2$ (in-time) & $R^2$ (forward) & $R^2$ loss \\
\midrule
Baseline (cat + time) & $\hedFeatsBaseline$  & $\hedRSqrInBaseline$  & $\hedRSqrFwdBaseline$  & $\hedRSqrLossBaseline\%$ \\
+ TF-IDF text         & $\hedFeatsTFIDF$     & $\hedRSqrInTFIDF$     & $\hedRSqrFwdTFIDF$     & $\hedRSqrLossTFIDF\%$ \\
+ CLIP                & $\hedFeatsCLIP$      & $\hedRSqrInCLIP$      & $\hedRSqrFwdCLIP$      & $\hedRSqrLossCLIP\%$ \\
+ Three-tower         & $\hedFeatsThreeTower$ & $\hedRSqrInThreeTower$ & $\hedRSqrFwdThreeTower$ & $\mathbf{\hedRSqrLossThreeTower\%}$ \\
Full (all combined)   & $\hedTotalFeats$     & $\mathbf{\hedRSqrInFull}$ & $\hedRSqrFwdFull$   & $\hedRSqrLossFull\%$ \\
\bottomrule
\end{tabular}
\end{table}

\begin{figure}[H]
\centering
\includegraphics[width=0.7\textwidth]{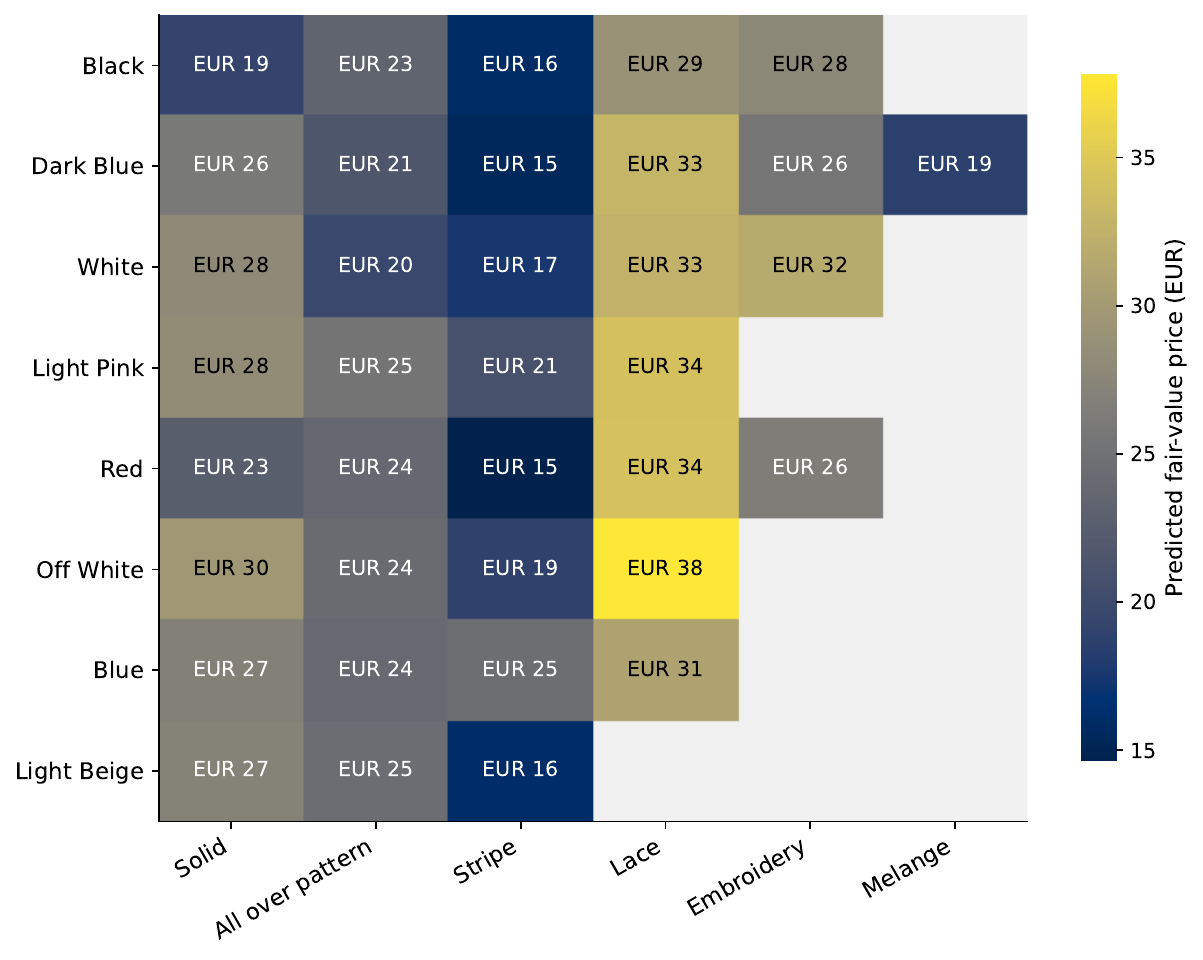}
\caption{Predicted fair-value dress prices by color and pattern. Cells with fewer than five dresses are masked.}
\label{fig:colour_pattern_grid}
\end{figure}

\Cref{fig:colour_pattern_grid} visualizes the fitted surface at the attribute level. For each color and pattern I average the product feature vectors across every dress in that cell, feed the averaged vector through the fitted model, and read off the predicted fair-value price. Across the eight most common colors and six most common patterns, predicted prices span EUR $\hedPredPriceLo$ to EUR $\hedPredPriceHi$. Lace dresses carry a consistent premium across every color, while stripe dresses sit at the low end regardless of color. The gradients are smooth and directionally consistent with merchandising intuition.

\begin{figure}[H]
\centering
\includegraphics[width=0.9\textwidth]{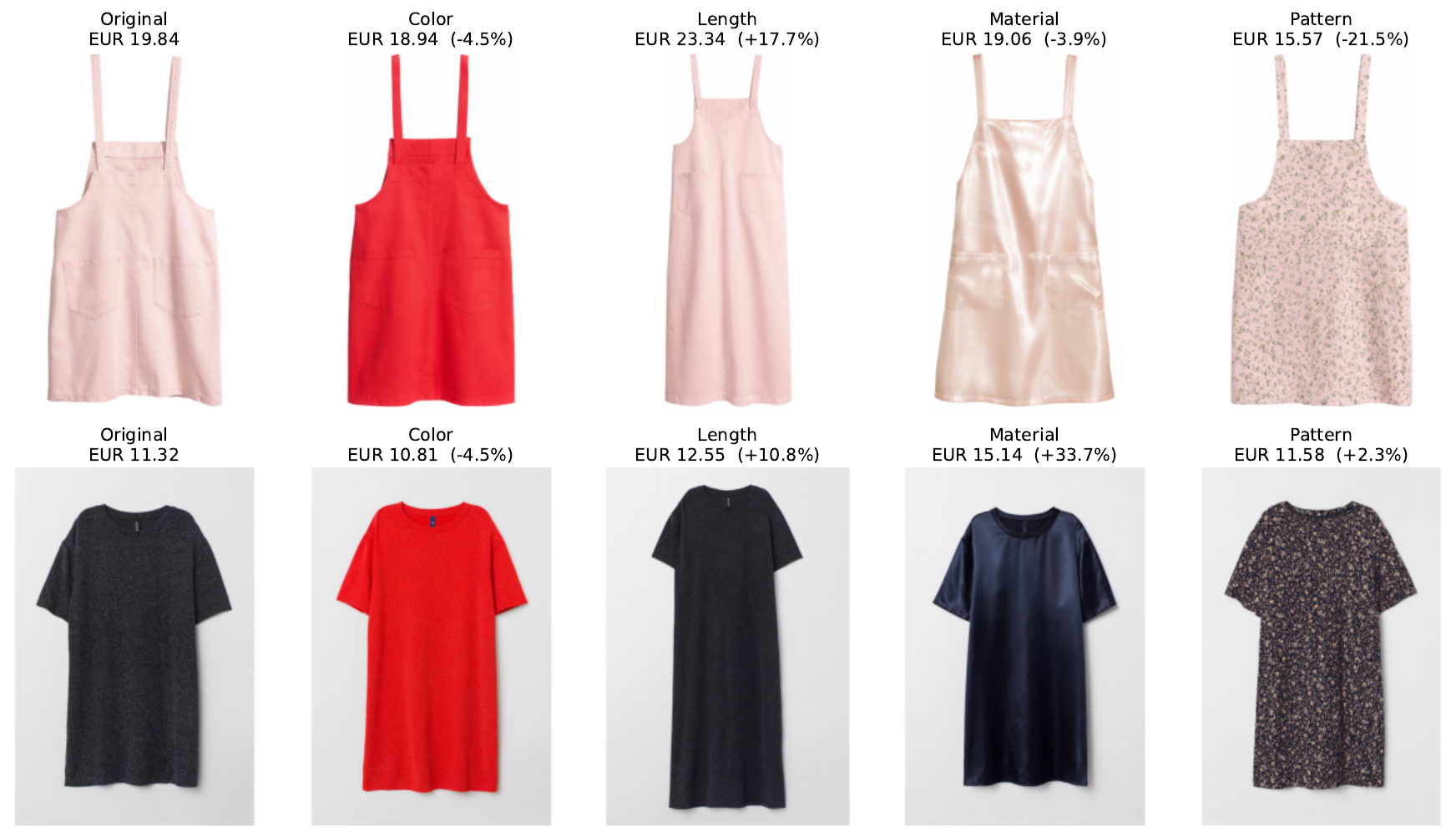}
\caption{Image-edit counterfactuals on two real H\&M dresses. Each column applies one standardized attribute change and reports the predicted fair-value price delta relative to the original. Two edits (length, color) move in the same direction for both dresses. The fabric and print edits reverse: switching to silk satin and adding a floral print both lower the dungaree's predicted price while raising the T-shirt dress's, showing the hedonic surface captures attribute interactions rather than fixed additive premia.}
\label{fig:dress_counterfactual}
\end{figure}

The color and pattern gradients average over all dresses of a given type. A complementary test asks whether the same attribute edit produces the same price change for every individual dress. \Cref{fig:dress_counterfactual} applies four standardized edits, changing color to red, lengthening to mid-calf, swapping fabric to glossy silk satin, and adding a small floral print, to two real H\&M dresses: a pink twill dungaree and a grey jersey T-shirt dress. Two of the four edits move in the same direction for both dresses: length adds roughly eleven to eighteen percent, and changing to red takes a small amount away. The other two edits reverse across the two dresses. Switching to glossy silk satin lowers the dungaree's predicted price by about four percent while raising the T-shirt dress's by about thirty-four percent, and adding a small floral print lowers the dungaree's predicted price by about twenty-two percent while raising the T-shirt dress's by about two percent. A model that treated satin or print as a fixed additive premium would miss both reversals. The hedonic surface captures that the value of a material or pattern depends on the rest of the dress. Satin on a dungaree looks out of place, while satin on a dressier jersey dress looks like an upgrade, and the same logic applies to florals.

\subsection{Interpretation}
\label{subsec:hedonic_interpretation}

The literature gives three main ways to read a hedonic regression. The first, due to \citet{rosen1974} and sharpened by \citet{ekeland2004identification}, reads the slope of price on a feature as the additional amount a buyer would pay for a small increase in that feature, an equilibrium outcome of many buyers with different tastes meeting many sellers with different costs. The standard example is the gradient of house price in floor area, which recovers what the marginal buyer pays for an extra square metre of living space. The second tradition, running from \citet{court1939} and \citet{griliches1961} to the handbook of \citet{triplett2006}, treats the regression as an accounting device. Its purpose is to hold measured features fixed so the price change of a like-for-like bundle can be separated from shifts in the mix of products on sale. This is the reading the United States Bureau of Labor Statistics uses when it imputes the current price of an exiting laptop model from its features and removes quality change from the consumer price index. A third tradition extends the second. \citet{bajari2025hedonic} replace the hand-picked list of characteristics with learned image and text representations for categories with rich variety, without changing the accounting interpretation. In their application to Amazon apparel listings, features from a pre-trained image network and a pre-trained text network feed a hedonic Fisher price index that is then compared directly with the official consumer price index for the same product category.

Our setup sits in the second and third traditions. The fitted surface is used only through its predictions, which feed a quality-adjusted index and a decomposition in the spirit of \citet{oaxaca1973male} and \citet{blinder1973wage}. No coefficient or feature importance score is read as a measure of consumer taste. The main price index uses a single fitted surface for all months, with time entering as a feature of $\mathbf{h}_j$. The Oaxaca-Blinder decomposition in \cref{subsec:ob_decomposition} uses two period-specific fits, one for each of $T_1$ and $T_2$, because the decomposition explicitly compares how implicit prices of features differ between the two windows. The pooled setup for the main index follows \citet{bajari2025hedonic}, who fit a single multi-task network across all months of their sample and read period-specific predictions off a shared representation. \citet{diewert2007imputation} argues that period-specific imputation is preferred when characteristic prices drift across time. An explicit stability test on our data rejects stable coefficients ($p < 10^{-4}$), so the pooled model is a practical compromise. \Cref{subsec:alt_hedonic} re-runs the index under a per-period imputation surface and a pooled time-dummy surface and reports the range of answers. This paper does not use the Rosen interpretation. Identifying consumer preferences is done elsewhere in the paper through a separate choice model, not through this surface.

\subsection{Price Indices and Conformal Inference}
\label{subsec:hedonic_indices}

The Jevons index is the geometric mean of raw price relatives across surviving articles $\mathcal{M}_t$,
\begin{equation}
J_{t,t+1} = \prod_{j \in \mathcal{M}_t} \left(\frac{p_{j,t+1}}{p_{j,t}}\right)^{1/|\mathcal{M}_t|}.
\label{eq:jevons}
\end{equation}
The Fisher ideal index uses predicted prices from the pooled hedonic surface, $\hat{p}_{jt} = \exp(\hat f_t(\mathbf{v}_j, \mathbf{h}_j))$, evaluated on the set of articles common to adjacent months, $\mathcal{M}_{t,t+1} = \mathcal{N}_t \cap \mathcal{N}_{t+1}$, together with observed quantities $q_{jt}$,
\begin{equation}
F_{t,t+1} = \sqrt{L_{t,t+1} \cdot P_{t,t+1}}, \quad
L = \frac{\sum_{j \in \mathcal{M}_{t,t+1}} \hat{p}_{j,t+1} q_{jt}}{\sum_{j \in \mathcal{M}_{t,t+1}} \hat{p}_{jt} q_{jt}}, \quad
P = \frac{\sum_{j \in \mathcal{M}_{t,t+1}} \hat{p}_{j,t+1} q_{j,t+1}}{\sum_{j \in \mathcal{M}_{t,t+1}} \hat{p}_{jt} q_{j,t+1}}.
\label{eq:fisher}
\end{equation}
The chained index is $F_T = \prod_{t=1}^{T} F_{t-1,t}$. Because predicted prices are estimated, the index is estimated with error too. I quantify index uncertainty through split conformal prediction \citep{vovk2005algorithmic, lei2018distribution}.\footnote{The idea is to set aside a fresh slice of the data, record how large the prediction errors are on that slice, and use the empirical quantile of those errors as an honest error bar on new predictions, without assuming anything about the shape of the error distribution.} A Monte Carlo stress test with $\hedMCDraws$ draws from out-of-fold residuals establishes a $\confBandWidthPP$ percentage point band on the aggregate index.

\begin{figure}[H]
\centering
\includegraphics[width=0.8\textwidth]{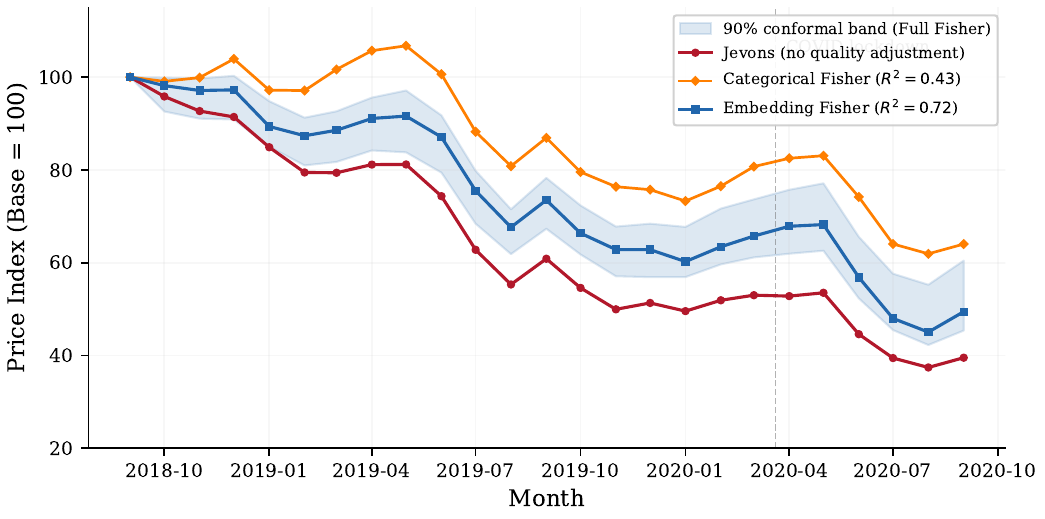}
\caption{Chained price indices for dresses over the two-year window, with the $\confConfidence\%$ conformal band on the embedding Fisher. The matched-item Jevons overstates the true deflation by the gap between the red and blue lines, which widens as catalogue turnover accumulates.}
\label{fig:index_comparison}
\end{figure}

\Cref{fig:index_comparison} plots the three indices alongside the conformal band. Quality-adjusted indices show less deflation than naive ones, and the gap widens over time. The Jevons index declines $\hedJevonsDecline\%$ over two years, while the embedding Fisher declines $\hedFisherDecline\%$, yielding a terminal gap of $\hedGap$ percentage points. The gap stays positive at the 5th percentile of the conformal stress test, so compositional shifts dwarf the model's own prediction uncertainty. \Cref{subsec:alt_hedonic} re-runs the same index under three alternative hedonic methods from the literature; the deep multi-task network recovers a similar Fisher decline and gap (both inside the conformal band), while two classical five-characteristic surfaces return much larger gaps that sit outside the band.

This pattern is not unique to dresses. \Cref{tab:index_cross_category} reports the same gap computed per category. Every entry is positive. In every product category I looked at, the matched-model approach overstates the true deflation, with the gap ranging from about 4 percentage points for blouses to almost 30 for trousers.

\begin{table}[H]
\centering
\caption{Jevons minus Fisher gap by product category}
\label{tab:index_cross_category}
\begin{tabular}{lr@{\hspace{1.5em}}lr}
\toprule
Category & Gap (pp) & Category & Gap (pp) \\
\midrule
Trousers & $\hedGapTrousers$ & Shorts   & $\hedGapShorts$ \\
T-shirt  & $\hedGapTShirt$   & Blouse   & $\hedGapBlouse$ \\
Top      & $\hedGapTop$      & Skirt    & $\hedGapSkirt$ \\
Shirt    & $\hedGapShirt$    & Vest top & $\hedGapVestTop$ \\
Sweater  & $\hedGapSweater$  & Jacket   & $\hedGapJacket$ \\
Dress    & $\hedGapDress$    &          &  \\
\bottomrule
\end{tabular}
\end{table}

\begin{figure}[H]
\centering
\includegraphics[width=\textwidth]{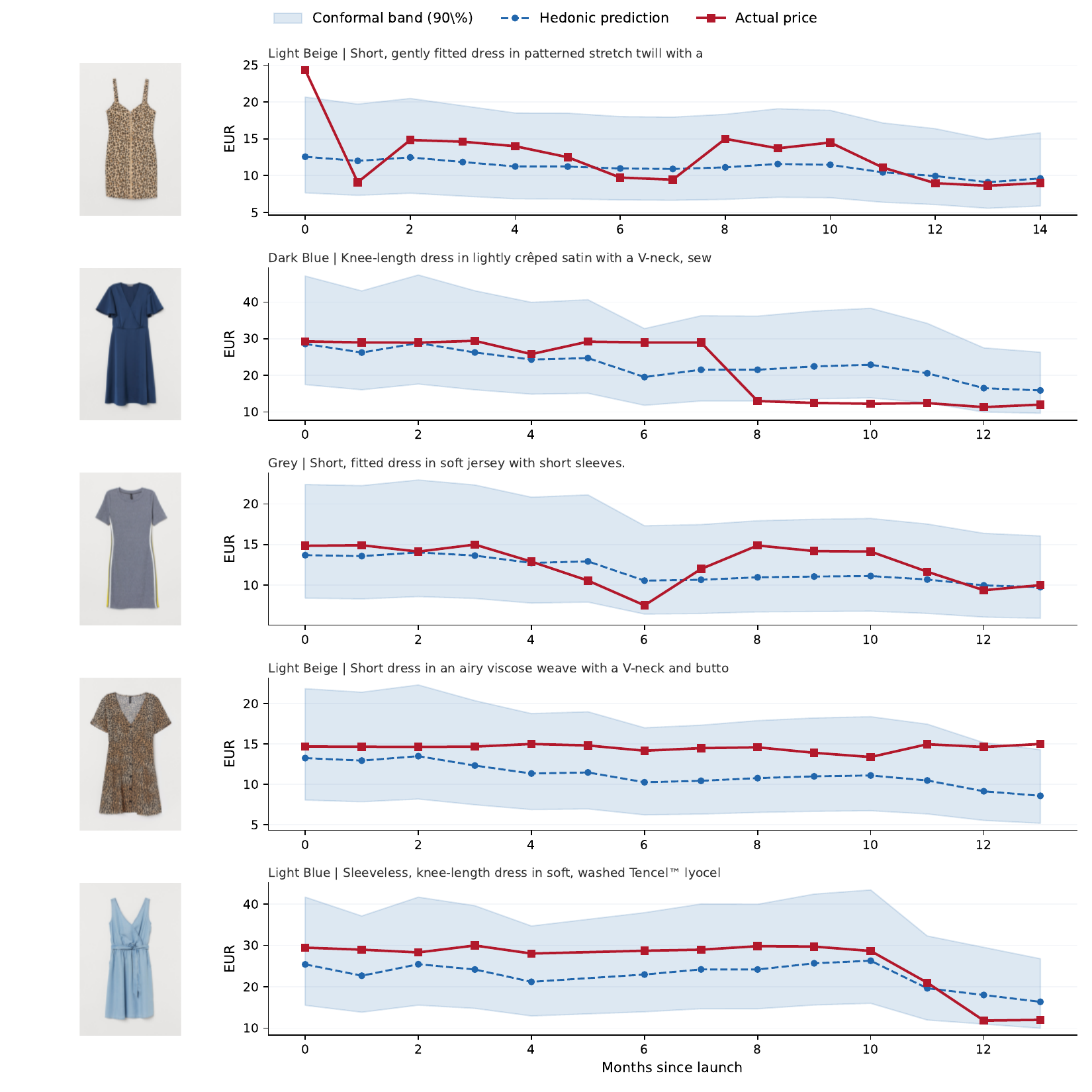}
\caption{Fair-value tracking for five new dresses introduced only in the $T_2$ window. Each panel shows one dress: the actual selling price each month (solid line), the model's fair-value estimate (dashed line), and the $\confConfidence\%$ conformal prediction band (shaded ribbon). The band is about $\confBandFigPP$ percentage points wide. All five dresses settle inside the band within three months of launch, consistent with the $\confMonthThreePct\%$ aggregate coverage reported in the text.}
\label{fig:conformal_settlement}
\end{figure}

Beyond quantifying index uncertainty, the conformal band doubles as a fair-value reference for cold-start pricing. During the $T_2$ window, $\confNewDressN$ dress articles appear that did not exist in $T_1$. For each new article I check whether the actual selling price sits inside the $\confConfidence\%$ conformal prediction band around the model's fair-value estimate each month. At launch, $\confInitialPct\%$ are already inside the band; coverage climbs with age, dips briefly at month six under the summer markdown cycle, and reaches $\confMonthTwelvePct\%$ by month twelve. \Cref{fig:conformal_settlement} traces the trajectory for five representative new dresses. The hedonic surface therefore prices products it has never seen using only features observed at launch, and it predicts those prices well against eventual market outcomes.

\subsection{Decomposition}
\label{subsec:ob_decomposition}

The nonlinear Oaxaca-Blinder extension\footnote{The Oaxaca-Blinder decomposition \citep{oaxaca1973male, blinder1973wage} splits a difference in mean outcomes between two groups into two parts. The first part is driven by differences in observed characteristics and is called composition. The second part is driven by differences in how those characteristics are valued and is called valuation. The nonlinear extension of \citet{bauer2008extension} replaces the linear regressions on each side with arbitrary fitted functions. Here those functions are the period-specific gradient-boosted hedonic surfaces $f_1$ and $f_2$.} \citep{bauer2008extension} decomposes the price decline into pure repricing (valuation) versus assortment composition. With $f_1$ and $f_2$ denoting the fitted surfaces for early ($T_1$) and late ($T_2$) time windows:
\begin{equation}
\underbrace{\overline{\log p}_2 - \overline{\log p}_1}_{\text{total}} = \underbrace{\tfrac{1}{2}\!\left[(\overline{f_1}^{T_2} - \overline{f_1}^{T_1}) + (\overline{f_2}^{T_2} - \overline{f_2}^{T_1})\right]}_{\text{composition}} + \underbrace{\tfrac{1}{2}\!\left[(\overline{f_2}^{T_2} - \overline{f_1}^{T_2}) + (\overline{f_2}^{T_1} - \overline{f_1}^{T_1})\right]}_{\text{valuation}} + \xi,
\label{eq:ob_decomposition}
\end{equation}
where $\overline{f_k}^{T_\ell} = \mathbb{E}_{T_\ell}[f_k(\mathbf{v},\mathbf{h})]$ is the average prediction of surface $f_k$ on the articles in window $T_\ell$, and $\xi$ is a small residual that absorbs the difference between observed mean log-prices and their hedonic predictions. This averages the two standard reference-window decompositions. It is the Cotton average of the $T_1$-reference and $T_2$-reference splits, which reduces sensitivity to the choice of reference window.

\begin{figure}[H]
\centering
\includegraphics[width=0.75\textwidth]{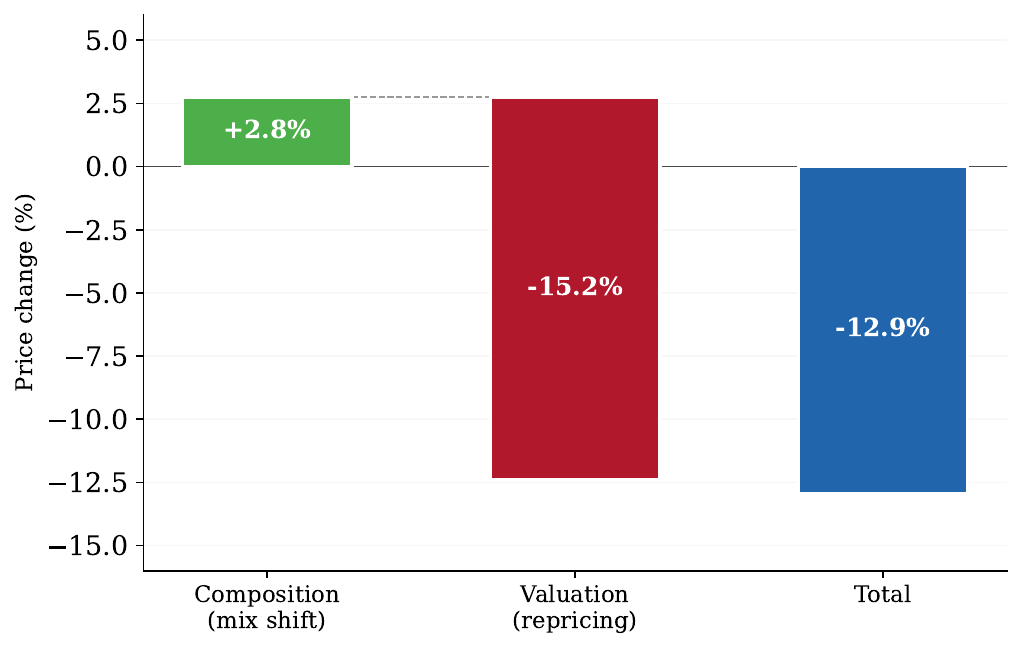}
\caption{Oaxaca-Blinder waterfall decomposition of the dress log-price change between $T_1$ and $T_2$. The total decline of $-\obTotalPct\%$ breaks into composition ($+\obCompPct\%$) and valuation ($-\obValPct\%$), which means the catalogue's mix improved slightly, but H\&M priced the same features meaningfully lower in the second window.}
\label{fig:ob_waterfall}
\end{figure}

\begin{figure}[H]
\centering
\includegraphics[width=0.8\textwidth]{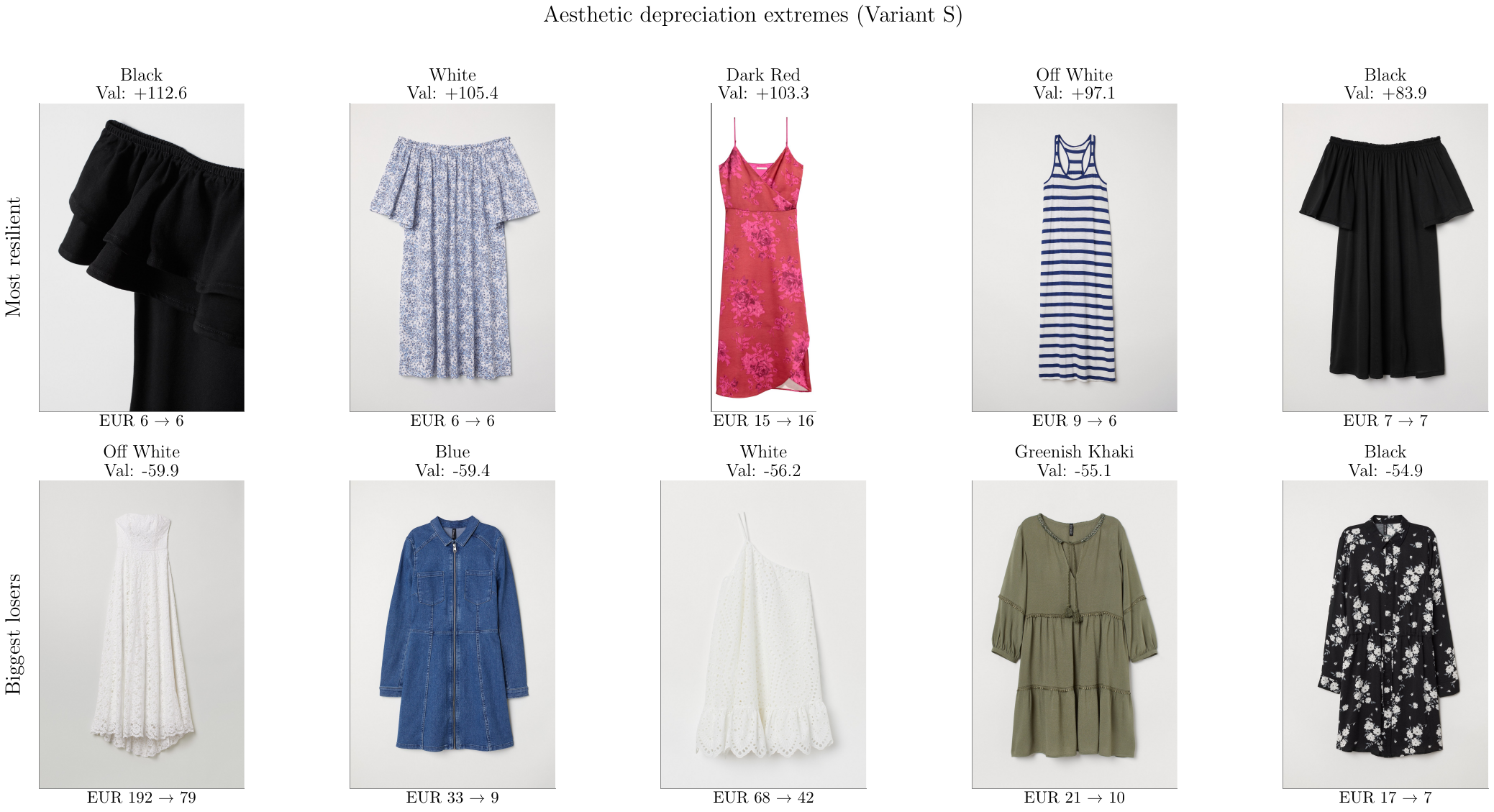}
\caption{Dresses ranked by the per-product valuation effect from the Oaxaca-Blinder decomposition. Top row, dresses whose hedonic value rose the most between $T_1$ and $T_2$. Bottom row, those whose hedonic value fell the most. The pricing surface rewards novelty, consistent with trend alignment driving value in a fashion-forward category.}
\label{fig:ob_ranked_examples}
\end{figure}

\Cref{tab:ob_results} reports the deseasonalized decomposition. The dress price decline is almost entirely repricing, slightly offset by an improving mix. The valuation effect accounts for a $\obValPct\%$ drop in prices, while the assortment composition shifted slightly ($+\obCompPct\%$) toward features the model values higher. \Cref{fig:ob_waterfall} visualizes the same decomposition as a waterfall, and \cref{fig:ob_ranked_examples} shows the dresses at the top and bottom of the valuation ranking. Trendy, fashion-forward items such as pink satin, ruffled off-shoulder, and patterned summer dresses gained hedonic value from $T_1$ to $T_2$. Basics such as plain cami, simple pleated, and jersey dresses lost the most value.

\begin{table}[H]
\centering
\caption{Oaxaca-Blinder decomposition (dresses, deseasonalized)}
\label{tab:ob_results}
\begin{threeparttable}
\begin{tabular}{lccc}
\toprule
& Total & Composition & Valuation \\
\midrule
Point estimate & $-$$\obTotalPct$\% & $+$$\obCompPct$\% & $-$$\obValPct$\% \\
Bootstrap $\obConfLevel\%$ CI & [$-$$\obTotalCILo$, $-$$\obTotalCIHi$] & [$-$$\obCompCILo$, $+$$\obCompCIHi$] & [$-$$\obValCILo$, $-$$\obValCIHi$] \\
\bottomrule
\end{tabular}
\begin{tablenotes}
\small
\item \textit{Notes:} Valuation negative in $\obValNegPct\%$ of $\obBootstrapN$ bootstrap replicates. Composition positive in $\obCompPosPct\%$.
\end{tablenotes}
\end{threeparttable}
\end{table}

%% file: 09_event_study.tex
\section{Event Study: COVID-19 Lockdown}\label{sec:event_study}

\subsection{Empirical Framework}

Index $t$ runs over calendar dates. Index $i$ denotes the unit of observation: the single aggregate series in M0, and one of twelve item clusters or ten user clusters in M1. Let $Y_{it}$ denote daily transaction counts for unit $i$ on date $t$, and let $\mathbf{X}_t$ collect daily covariates common to all units, namely day-of-week dummies and a deep-discount indicator. The coefficient $\kappa_p$ is the absolute period-$p$ effect in a single-series fit. The regression coefficients in §9 ($\kappa_p$, $\kappa_{\text{trend}}$, $\nu_m$, $\boldsymbol{\zeta}$) use locally scoped Greek letters that do not collide with the demand-side networks $\alpha_c$, $\beta$-net, $\gamma_c$, or $\boldsymbol{\delta}_m$ of \cref{sec:theory}. Counts are fit by Poisson quasi-maximum likelihood \citep{gourieroux1984pseudo, wooldridge1999distribution}. Two specifications are estimated, M0 (aggregate) and M1 (per cluster); both share the Poisson likelihood and the common covariate block $\mathbf{X}_t$ and differ only in the unit of observation.

The COVID-19 lockdown hit every product and every store on March 20, 2020. Every unit was treated on the same day, so the sample contains no untreated 2020 observations and no control group inside the treated year.\footnote{When every unit is treated on the same day and no never-treated unit is available, the aggregate treatment effect is not identified without an outside assumption \citep{miller2023introductory}. The 2019 seasonal pattern stands in as the counterfactual for 2020. All lockdown coefficients reported below are absolute effects under that outside assumption.} The lockdown effect is therefore measured by comparing 2020 to 2019 under the assumption that the 2019 seasonal pattern would have continued in 2020 absent the shock. The 2020 lockdown coefficient is the deviation from that baseline. The assumption is not testable from within the sample. All estimates that follow, including differences across item and user clusters, are conditional on it. The item and user embeddings were trained only on pre-COVID transactions, so the embedding space itself does not absorb the shock.\footnote{Clustering first and regressing afterwards treats the partition as fixed, so first-stage uncertainty is not reflected in standard errors. Two design choices limit this: the $K$-means partition is fit on pre-COVID data only, and $K$ is chosen by pre-COVID silhouette rather than by lockdown outcomes. A seed-stability check in \cref{subsec:appendix_post_clustering} refits M1 under twenty random $K$-means initialisations and finds the heterogeneity range within ten percentage points of the main-text estimate on nineteen of twenty seeds.}

\subsubsection{Aggregate and per-cluster single-series: M0 and M1}

M0 uses total daily dress transactions; the index $i$ is trivial (one aggregate series) and is carried in the notation only to match M1. M1 uses per-cluster daily transactions, re-fitting the same equation once for each of twelve item clusters and once for each of ten user clusters, so $i$ identifies the cluster and every coefficient below is cluster-specific. Both share the same single-series specification,\footnote{The structural-break form with a linear time trend, calendar-month dummies, day-of-week controls, and one indicator per post-shock period is the standard fit for a single time series with a known intervention date, documented in the applied literature under the names interrupted time series, segmented Poisson regression, and structural break regression \citep{linden2015conducting}.}
\begin{equation}
    \log \mathbb{E}[Y_{it} \mid X_t] = \mu_0 + \kappa_{\text{trend}} \cdot t + \sum_{m=2}^{12} \nu_m \, \text{Month}_{mt} + \sum_{p \in \mathcal{P}} \kappa_p \, \mathbf{1}_{pt} + \boldsymbol{\zeta}' \mathbf{X}_t,
    \label{eq:poisson_structural_break}
\end{equation}
where $\mu_0$ is the intercept and $t \in [0, 1]$ is the date index rescaled to the unit interval, so $\kappa_{\text{trend}}$ is the log-scale change across the whole sample. The month block $\text{Month}_{mt}$ is a set of eleven calendar-month indicators (January omitted) with coefficients $\nu_m$. The period block $\mathbf{1}_{pt}$ gives one indicator per period in $\mathcal{P} = \{\text{WHO}, \text{Lockdown}, \text{Reopening}, \text{Post-Recovery}\}$, with coefficients $\kappa_p$. The control block $\boldsymbol{\zeta}' \mathbf{X}_t$ loads on six day-of-week indicators and the deep-discount indicator. The coefficients have a semi-elasticity interpretation. In particular, $\exp(\kappa_p) - 1$ is the proportional change in expected transactions during period $p$ relative to the pre-shock seasonal norm. M0 delivers the category-level lockdown estimate, and M1 returns one lockdown estimate per cluster by re-running the same specification on each cluster's daily count.

\subsection{Heterogeneous Demand Responses}

The aggregate model (M0) estimates a lockdown effect of $-\evtAggregatePct\%$ on daily dress transactions ($p < 0.01$), and the WHO announcement period shows a $-\evtWHOPct\%$ decline ($p < 0.001$), confirming that consumer responses began before formal lockdown measures.

The shock is not specific to dresses. \Cref{tab:cross_category_lockdown} reports the aggregate Poisson estimate for each of the eleven fashion categories in the H\&M catalog. Occasion-wear categories (sweaters, blouses, shirts, jackets, skirts) fall sharply. Loungewear-adjacent categories (t-shirts, shorts, vest tops) rise. Dresses fall in between. The same within-category heterogeneity holds across all six main product groups. Full per-category item and user cluster breakdowns appear in \cref{sec:appendix_cross_category}.

\begin{table}[H]
\centering
\caption{Aggregate lockdown effect by product category, from M0 (\cref{eq:poisson_structural_break} on total daily counts). Column `Lockdown \%' reports $100 \cdot [\exp(\hat{\kappa}_{\text{Lockdown}}) - 1]$.}
\label{tab:cross_category_lockdown}
\begin{tabular}{lr@{\hspace{2em}}lr}
\toprule
Category & Lockdown \% & Category & Lockdown \% \\
\midrule
Sweater  & $-\evtCatSweater$  & T-shirt  & $+\evtCatTShirt$ \\
Blouse   & $-\evtCatBlouse$   & Shorts   & $+\evtCatShorts$ \\
Jacket   & $-\evtCatJacket$   & Vest top & $+\evtCatVestTop$ \\
Shirt    & $-\evtCatShirt$    &          &  \\
Skirt    & $-\evtCatSkirt$    &          &  \\
Dress    & $-\evtCatDress$    &          &  \\
Top      & $-\evtCatTop$      &          &  \\
Trousers & $-\evtCatTrousers$ &          &  \\
\bottomrule
\end{tabular}
\end{table}

A second striking aggregate fact is the collapse of in-store transactions during lockdown. Before lockdown the store-to-online split of H\&M transactions was about $\evtPrelockStorePct$/$\evtPrelockOnlinePct$, close to the long-run share in the full sample. During the lockdown weeks, physical store transactions fell to zero because every brick-and-mortar location was closed, so all activity moved online. After reopening the split moved to $\evtPostlockStorePct$/$\evtPostlockOnlinePct$, with a slight permanent lift in store share relative to the pre-lockdown baseline. The event compresses the usual retail channel mix into a pure online channel for roughly seven weeks and then only partly reverses. The cluster-level Poisson models below combine purchases across both channels into a single daily count per cluster, so channel substitution enters the estimated lockdown coefficient rather than a separate control. The store-closure mechanism is therefore part of the measured lockdown effect.

The per-cluster models (M1) reveal substantial heterogeneity across the twelve item embedding clusters \citep{lloyd1982least}. These clusters cut across the retailer's product taxonomy rather than aligning with it, so different kinds of dresses were hit very differently and the embedding partition shows where the drop was largest.\footnote{The number of clusters, $K=\evtKItem$ for items and $K=\evtKUser$ for users, is chosen by maximizing silhouette score across $K \in \{4, 6, 8, 10, 12, 15, 20\}$ for items and $K \in \{4, 6, 8, 10, 12\}$ for users, among values where every cluster averages at least $\evtClusterMinTxnsPerDay$ daily transactions in a pre-COVID window of matched length (24 January to 11 March 2020). The silhouette score for a point $i$ is $s_i = (b_i - a_i)/\max(a_i, b_i)$, where $a_i$ is the mean distance from $i$ to other points in its cluster and $b_i$ is the mean distance from $i$ to points in the nearest other cluster; it is bounded in $[-1, 1]$ and higher is better. The pre-COVID viability filter keeps $K$-selection independent of the shock and rejects $K=15$ and $K=20$ because their smallest clusters fall below the $\evtClusterMinTxnsPerDay$-per-day threshold.}

\begin{table}[H]
    \centering
    \caption{Lockdown impact by product cluster from M1 (per-cluster single-series, \cref{eq:poisson_structural_break} fit once per cluster with $i$ indexing the cluster). Column `Lockdown (\%)' reports $100 \cdot [\exp(\hat{\kappa}_{\text{Lockdown}}) - 1]$ for each cluster.}
    \label{tab:item_lockdown_poisson}
    \begin{threeparttable}
    \begin{tabular}{l r r r r}
        \toprule
        Segment & Items & Daily Txn & Lockdown (\%) & 95\% CI \\
        \midrule
        Item Cluster 0  & $\evtICzeroItems$   & $\evtICzeroTxns$   & $-\evtICzeroEffect^{***}$ & $[-\evtICzeroCILo, -\evtICzeroCIHi]$ \\
        Item Cluster 1  & $\evtIConeItems$    & $\evtIConeTxns$    & $-\evtIConeEffect$        & $[-\evtIConeCILo, +\evtIConeCIHi]$ \\
        Item Cluster 2  & $\evtICtwoItems$    & $\evtICtwoTxns$    & $-\evtICtwoEffect^{***}$  & $[-\evtICtwoCILo, -\evtICtwoCIHi]$ \\
        Item Cluster 3  & $\evtICthreeItems$  & $\evtICthreeTxns$  & $-\evtICthreeEffect^{**}$ & $[-\evtICthreeCILo, -\evtICthreeCIHi]$ \\
        Item Cluster 4  & $\evtICfourItems$   & $\evtICfourTxns$   & $-\evtICfourEffect^{***}$ & $[-\evtICfourCILo, -\evtICfourCIHi]$ \\
        Item Cluster 5  & $\evtICfiveItems$   & $\evtICfiveTxns$   & $-\evtICfiveEffect^{**}$  & $[-\evtICfiveCILo, -\evtICfiveCIHi]$ \\
        Item Cluster 6  & $\evtICsixItems$    & $\evtICsixTxns$    & $+\evtICsixEffect$        & $[-\evtICsixCILo, +\evtICsixCIHi]$ \\
        Item Cluster 7  & $\evtICsevenItems$  & $\evtICsevenTxns$  & $-\evtICsevenEffect^{**}$ & $[-\evtICsevenCILo, -\evtICsevenCIHi]$ \\
        Item Cluster 8  & $\evtICeightItems$  & $\evtICeightTxns$  & $-\evtICeightEffect^{**}$ & $[-\evtICeightCILo, -\evtICeightCIHi]$ \\
        Item Cluster 9  & $\evtICnineItems$   & $\evtICnineTxns$   & $-\evtICnineEffect^{***}$ & $[-\evtICnineCILo, -\evtICnineCIHi]$ \\
        Item Cluster 10 & $\evtICtenItems$    & $\evtICtenTxns$    & $-\evtICtenEffect^{***}$  & $[-\evtICtenCILo, -\evtICtenCIHi]$ \\
        Item Cluster 11 & $\evtICelevenItems$ & $\evtICelevenTxns$ & $-\evtICelevenEffect$     & $[-\evtICelevenCILo, +\evtICelevenCIHi]$ \\
        \midrule
        \multicolumn{5}{l}{Range: $\evtItemRangePP$ percentage points} \\
        \bottomrule
    \end{tabular}
    \begin{tablenotes}
    \small
    \item \textit{Notes}: Poisson QMLE with seven-day Newey-West standard errors \citep{newey1987simple}. Daily Txn is the mean daily transaction count during lockdown.
    \item $^{***}$, $^{**}$, $^{*}$ denote significance at 0.1\%, 1\%, 5\% after Benjamini-Hochberg FDR correction \citep{benjamini1995controlling}.
    \end{tablenotes}
    \end{threeparttable}
\end{table}

\Cref{tab:item_lockdown_poisson} reports a $\evtItemRangePP$ percentage point range across twelve embedding clusters. Womens Tailoring falls the most; jersey-knit everyday basics carry the only positive point estimate. Matched-$K$ TF-IDF text clustering and H\&M's section taxonomy produce comparable raw ranges (see \cref{tab:ari_comparison} for the partition-agreement comparison). On raw variance the three segmentations are comparable. On partition structure they are not.

\Cref{tab:ari_comparison} reports pairwise Adjusted Rand Index\footnote{The Adjusted Rand Index compares two partitions of the same items. Let $n_{ij}$ denote the number of items assigned to cluster $i$ in the first partition and cluster $j$ in the second, with row sums $a_i = \sum_j n_{ij}$ and column sums $b_j = \sum_i n_{ij}$, and total $n$. The index is
$\text{ARI} = \frac{\sum_{ij} \binom{n_{ij}}{2} - [\sum_i \binom{a_i}{2} \sum_j \binom{b_j}{2}] / \binom{n}{2}}{\frac{1}{2}[\sum_i \binom{a_i}{2} + \sum_j \binom{b_j}{2}] - [\sum_i \binom{a_i}{2} \sum_j \binom{b_j}{2}] / \binom{n}{2}}$.
It equals one for identical partitions and zero when the overlap matches chance \citep{hubert1985comparing}.} and Normalized Mutual Information\footnote{Normalized Mutual Information rescales the mutual information between two partitions by their average entropy: $\text{NMI}(U, V) = 2\, I(U; V) / [H(U) + H(V)]$, where $I(U; V) = \sum_{ij} (n_{ij}/n) \log\!\big(n\, n_{ij} / (a_i b_j)\big)$ is the mutual information and $H(U) = -\sum_i (a_i/n) \log(a_i/n)$ is the partition entropy (with $H(V)$ defined symmetrically). It is bounded in $[0, 1]$, equals one for identical partitions and zero when the partitions are statistically independent.} between the three partitions. The embedding partition is almost orthogonal to the TF-IDF partition and only partially aligned with H\&M section labels (\cref{tab:ari_comparison}). Several embedding clusters cross-cut the section taxonomy. Cluster 6 draws roughly a quarter of its items each from Womens Everyday Collection, Everyday Basics, and Womens Casual, a grouping no text-only method reproduces.

\begin{table}[H]
    \centering
    \caption{Pairwise agreement between item-side segmentations on the Dress catalog.}
    \label{tab:ari_comparison}
    \begin{tabular}{l r r}
        \toprule
        Pair of segmentations & ARI & NMI \\
        \midrule
        Embedding vs TF-IDF       & $\evtARIEmbTFIDF$  & $\evtNMIEmbTFIDF$  \\
        Embedding vs H\&M section & $\evtARIEmbHM$     & $\evtNMIEmbHM$     \\
        TF-IDF vs H\&M section    & $\evtARITFIDFHM$   & $\evtNMITFIDFHM$   \\
        \bottomrule
    \end{tabular}
\end{table} 

\begin{figure}[H]
    \centering
    \includegraphics[width=\textwidth]{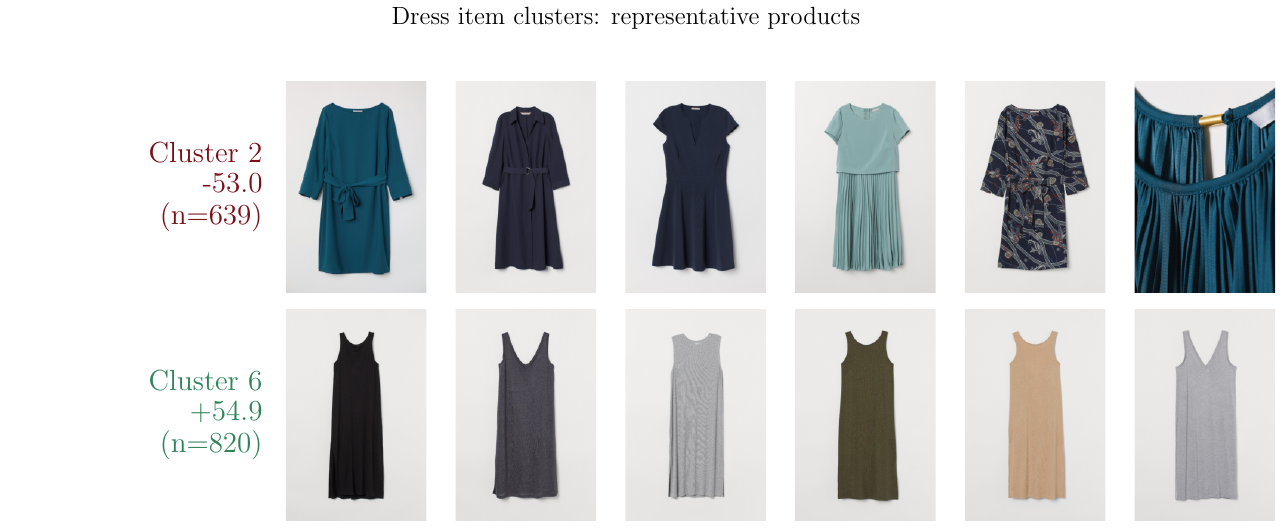}
    \caption{Item-cluster view. Representative products from the two most extreme dress item embedding clusters, ranked by the M1 estimate $\exp(\hat{\kappa}_{\text{Lockdown}}) - 1$ (\cref{eq:poisson_structural_break}). Item Cluster~6 (jersey-knit everyday basics, $+\evtItemICsixMostPos\%$) is the only cluster with a positive point estimate. Item Cluster~2 (Womens Tailoring officewear, $-\evtItemICtwoMostNeg\%$) is the hardest hit.}
    \label{fig:cluster_product_grid}
\end{figure}

\Cref{fig:cluster_product_grid} shows representative products from the two extreme clusters. Cluster 2 is almost entirely Womens Tailoring, the structured officewear most exposed to office closures. Cluster 6 cross-cuts Womens Everyday Collection, Everyday Basics, and Womens Casual, and contains jersey-knit solid-colour dresses of the kind worn at home.

\begin{figure}[H]
    \centering
    \includegraphics[width=0.7\textwidth]{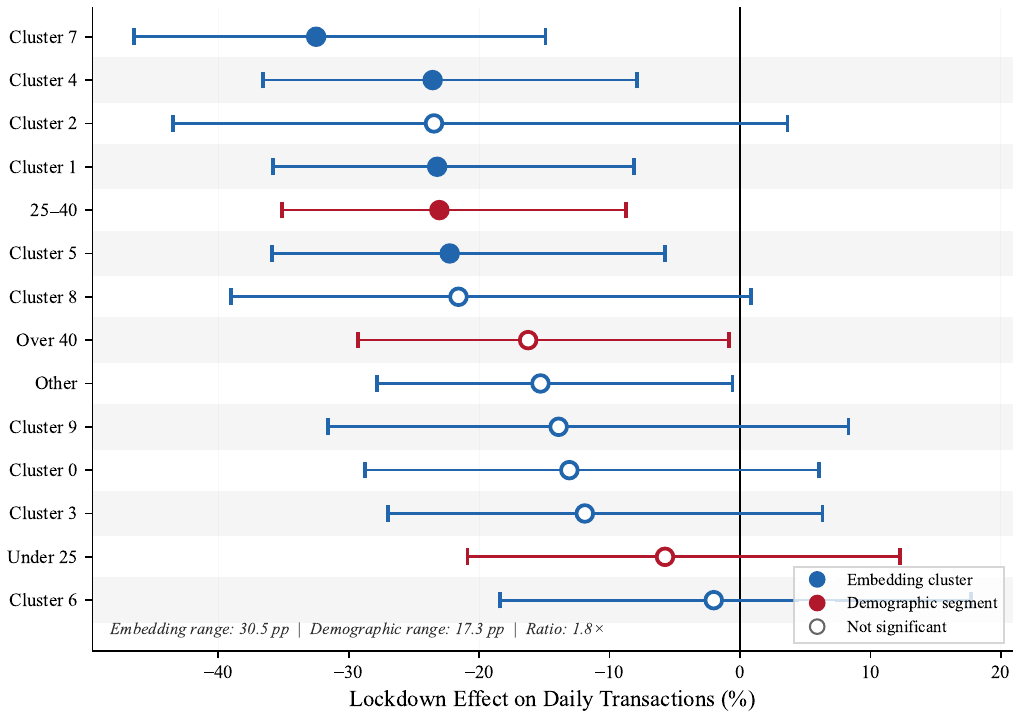}
    \caption{User-cluster view. Lockdown effect per user embedding cluster (blue) and per age-based demographic segment (red), from M1 (\cref{eq:poisson_structural_break} fit once per segment). Each point is $100 \cdot [\exp(\hat{\kappa}_{\text{Lockdown}}) - 1]$. The user embedding clusters span $\evtUserVsDemo\times$ the range of the demographic segments.}
    \label{fig:user_heterogeneity}
\end{figure}

Turning to consumer heterogeneity, embedding-based user clusters are estimated from the highest-activity customer tier, whose transaction histories are sufficiently dense for reliable continuous representation, and these clusters cleanly cross-cut traditional demographic bins. The per-segment M1 fits give a $\evtUserRangePP$ percentage point range across embedding clusters, $\evtUserVsDemo\times$ wider than the range across three age-based demographic segments (\cref{fig:user_heterogeneity}). The most and least affected embedding clusters both fall outside the demographic range, despite sharing nearly identical age distributions.

\begin{figure}[H]
    \centering
    \includegraphics[width=\textwidth]{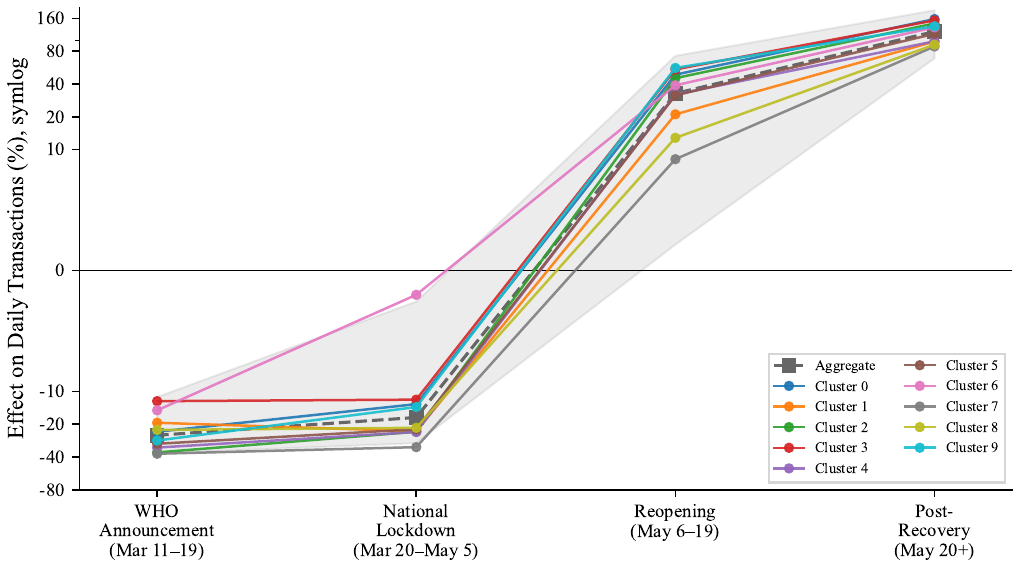}
    \caption{User-cluster view. Recovery trajectories by user embedding cluster across the four event periods, from M1 (\cref{eq:poisson_structural_break} fit once per user cluster). Each line plots $\exp(\hat\kappa_p) - 1$ for one user cluster across $p \in \mathcal{P}$. Shaded bands are 95\% CIs.}
    \label{fig:user_recovery_trajectories}
\end{figure}

\Cref{fig:user_recovery_trajectories} plots the M1 recovery path for each user cluster across the four event periods. The hardest-hit and most resilient user clusters diverge during lockdown and remain partially separated through the post-recovery window, so the user-side heterogeneity is not just a one-period anomaly.

\begin{figure}[H]
    \centering
    \includegraphics[width=0.62\textwidth]{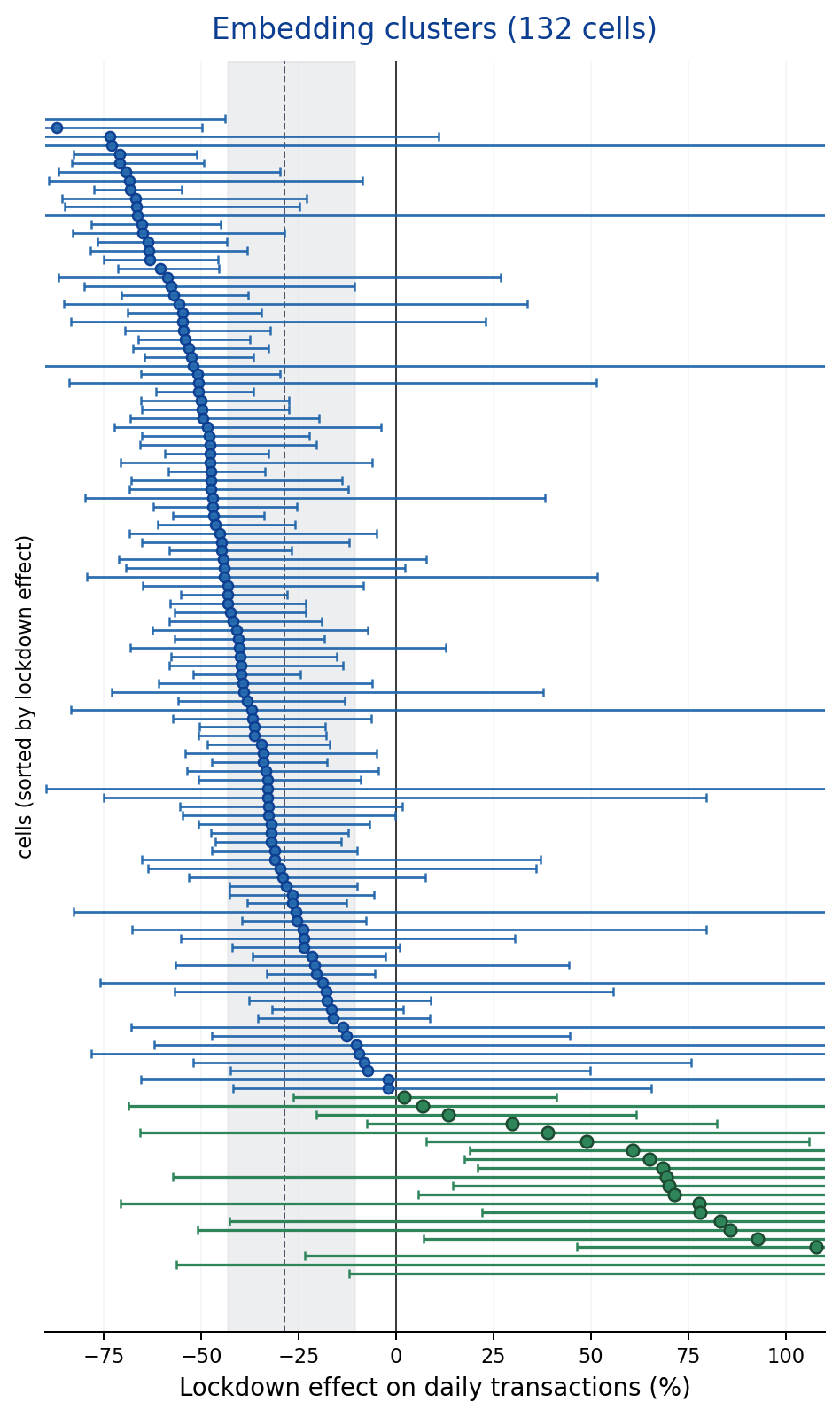}
    \caption{Item $\times$ user-cell view. Lockdown effect on daily dress transactions at the item-cluster by user-cluster cell level, from per-cell M1 fits (\cref{eq:poisson_structural_break} fit once per $(i, j)$ cell on that cell's daily count). Each point is $100 \cdot [\exp(\hat{\kappa}_{\text{Lockdown}}) - 1]$ for one of the $\evtCellN$ cells; horizontal bars are 95\% confidence intervals. Green markers flag cells with positive point estimates. The dashed vertical line and light grey band mark the aggregate M0 estimate ($-\evtAggregatePct\%$) and its 95\% CI. The cell-level spread is wider than either the item-only ($\evtItemRangePP$pp) or user-only ($\evtUserRangePP$pp) range in \cref{tab:item_lockdown_poisson,fig:user_heterogeneity}.}
    \label{fig:cross_cell_embed}
\end{figure}

A per-cell fit of M1 (\cref{eq:poisson_structural_break}) on each item-cluster by user-cluster cell yields a cell-level lockdown range of $\evtCellRangePP$ percentage points across the sufficient-volume cells. This range exceeds both the item-only and user-only margins reported in \cref{tab:item_lockdown_poisson,fig:user_heterogeneity}. \Cref{fig:cross_cell_embed} plots all cells. The cell ordering does not reduce to a rescaling of either marginal cut.

\begin{figure}[H]
    \centering
    \includegraphics[width=0.95\textwidth]{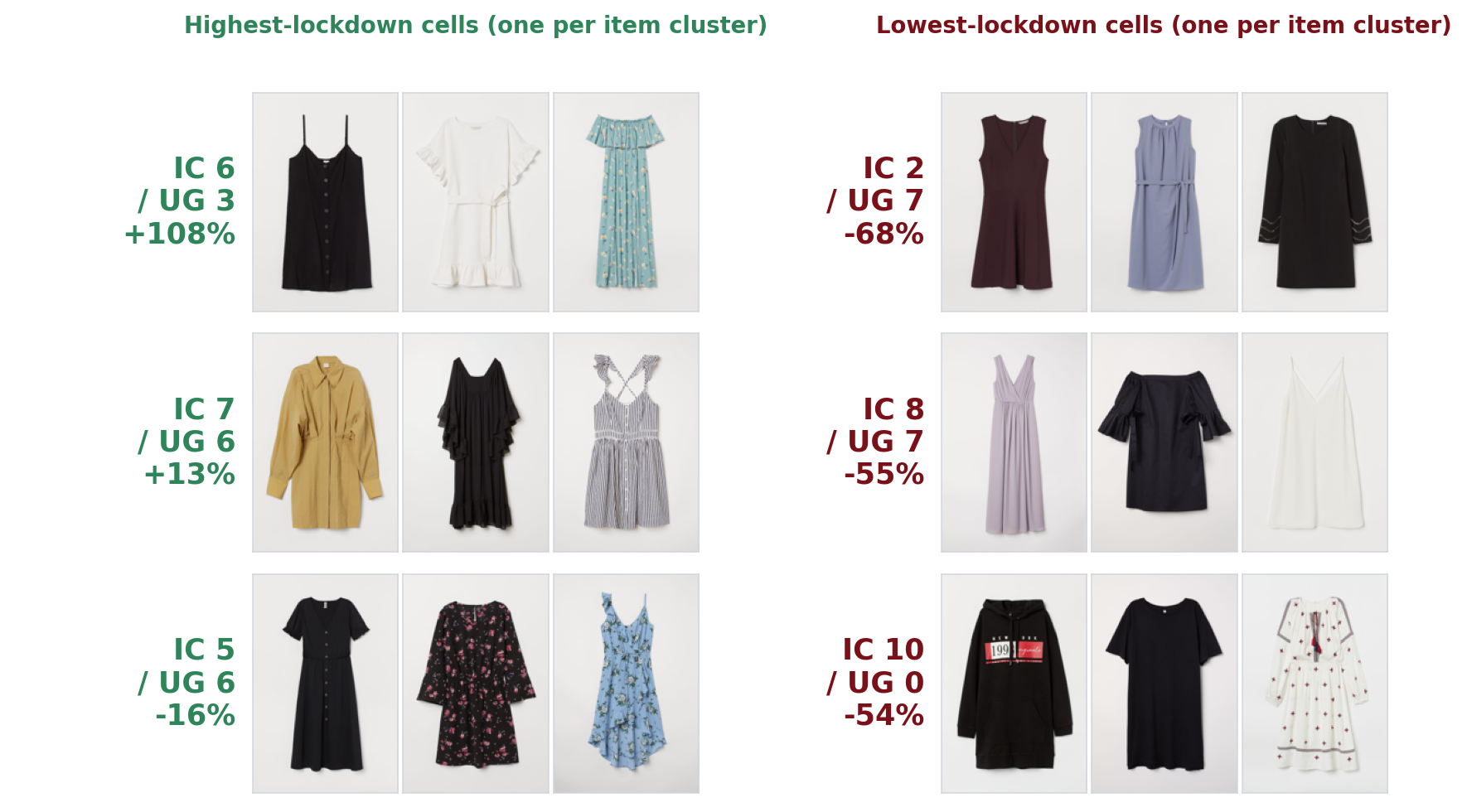}
    \caption{Item $\times$ user-cell view. Tail cross-cells from the per-cell M1 distribution in \cref{fig:cross_cell_embed}, shown as product images. Left panel: three distinct item clusters whose best sufficient-volume cell is most positive. Right panel: three distinct item clusters whose best sufficient-volume cell is most negative. Each row shows three items from the named cluster; the label above each row gives the item cluster, the user cluster, and $100 \cdot [\exp(\hat{\kappa}_{\text{Lockdown}}) - 1]$ from the cell-level M1 fit.}
    \label{fig:cross_cell_tails}
\end{figure}

\Cref{fig:cross_cell_tails} puts product images on the two tails of the cell-level M1 distribution in \cref{fig:cross_cell_embed}. For each tail, three distinct item clusters are shown, each represented by its best sufficient-volume cell and three items from that cluster. The display keeps the item clusters distinct so the span of the distribution is visible as variation in product type.

User Cluster 6 has the least-negative row average in the cross-cell panel; the remaining rows sit deeply negative. UC6 is a young cohort skewed toward ladies' dresses. Within this row, the jersey-knit everyday basics cluster shows a positive point estimate while the Womens Tailoring cluster is deeply negative. The pattern is not visible from the item-only or user-only cut alone. Several cross-cell estimates rest on fewer than fifteen daily transactions with confidence intervals that cross zero, and are not interpreted individually.

%% file: 11_conclusion.tex
\section{Discussion and Conclusion}
\label{sec:conclusion}

\subsection{Implications}

For retailers, the value of the representation learning framework is aesthetic segmentation, which identifies which products appeal to which consumer classes and enables targeted assortment decisions. The two latent classes differ meaningfully in price sensitivity, with the minority class substantially more responsive to markdowns than the majority. The classes also separate along taste dimensions, so the retailer can potentially gain from both class-targeted assortment and class-differentiated pricing. Seasonal timing dominates seasonal pricing. The low-rank monthly taste rotation $\boldsymbol{\delta}_m$ shows the aesthetic direction of demand shifting noticeably during the spring collection turnover, so assortment timing, not price timing, is what moves with the calendar.\footnote{The seasonal demand extension uses a low-rank monthly taste rotation and does not add a seasonal component to the price coefficient; the master latent-class model holds $\alpha_c$ fixed across months.} The hedonic composition bias implies that retailers calibrating markdown timing to matched-item price indices will systematically mistime their interventions.

For industrial organization theory, the logit specification imposes the Independence of Irrelevant Alternatives conditional on class membership, but the two-class latent formulation generates substantial market-level departures. Substitution patterns depend on class shares, so a price increase drops demand onto products favoured by the same class rather than dispersing symmetrically. Encoding visual similarity in the choice probabilities turns latent taste heterogeneity into an aggregate-level relaxation of the IIA restriction. On the measurement side, contrastive pre-training on purchase records recovers economic structure, and the three-tower embeddings act as sufficient statistics for unobservable preferences across several disjoint applications. The near-zero Adjusted Rand Index between embedding-based behavioural clusters and traditional demographic segments confirms that the learned representations capture dimensions orthogonal to age and income.

\subsection{Limitations}

The analysis uses data from a single retailer in one market segment over two years, so the specific estimates are not portable without replication across retailers, categories, and market structures. Price is the dominant predictor of demand but not a purely causal estimate, since identification rests on a hedonic control function rather than an exogenous cost-shock instrument. The outside option is calibrated by bisection rather than jointly estimated, because session-level logs recording non-purchases are not in the data. Choice-model parameters come with consumer-bootstrap standard errors (\cref{tab:bootstrap_ses}); a full two-stage bootstrap that also propagates first-stage embedding noise is left to future work. Rival retailers are absorbed into the outside option rather than modelled as best-response players.

The data does not record fit or per-size availability, and survey evidence ranks fit as the most valued attribute in fast-fashion purchases, ahead of price and trend \citep{deKok2022, Ruigrok2009, Leinenga2019}. A non-purchase cannot be separated into a taste rejection, a price rejection, or a stock-out in the customer's size. The taste sensitivity network therefore encodes a bundle of aesthetic preference and item availability rather than pure aesthetic taste. Session-level logs linked to real-time size-level inventory would be needed to separate these channels cleanly.

\subsection{Future Research}

Several extensions would address these limitations and expand the framework. The most direct path forward is a richer supply-side specification. The current static Bertrand-Nash assumption recovers a multi-product Lerner index modestly above H\&M's reported gross margin, qualitatively consistent with the monopolist conduct assumption. The static model overstates realized margins because it treats every price as an instantaneous profit maximum, while in practice markdown calendars and seasonal clearance push realized margins below that benchmark. A dynamic pricing extension that models inventory clearance and markdown timing could close this gap by accounting for intertemporal substitution.

Extending the framework also requires a deeper understanding of cross-category substitution and budgeting behaviour. The current choice set is restricted to dresses, which implicitly assumes out-of-category substitution behaves exactly like the outside good. Accounting for complementary purchases (such as buying a dress and matching shoes) and broader wallet constraints would naturally lower aggregate elasticity, expand margins, and better match observed retail behaviour.

%% file: B_robustness.tex
\section{Choice Model Robustness}
\label{sec:appendix_robustness}

\subsection{Control Function}
\label{subsec:control_function}

As a first-stage check for price endogeneity I follow the control-function approach of \citet{petrin_train_2010} and add the per-product hedonic residual $\hat\lambda_j = \overline{\log p_j} - \hat g(\mathbf{v}_j)$ to the utility specification, where $\overline{\log p_j}$ is the mean log price of product $j$ over all weeks it sells and $\hat g$ is a five-fold cross-validated LightGBM regression of mean log price on the three-tower item embedding $\mathbf{v}_j$. The residual captures the cross-sectional component of price the item embedding cannot explain and is constant within each product. Two variants are tested, a linear control and a nonparametric MLP. The price coefficient $\hat\alpha$ is stable across both, and the control-function coefficient is economically small.

\subsection{Consumer Bootstrap}
\label{sec:appendix_bootstrap}

To check whether the estimates in \cref{sec:choice_model} are stable at the sample size used, I refit the demand model on ten consumer bootstrap draws of the estimation panel. \Cref{tab:bootstrap_ses} reports the bootstrap mean and standard error for the class-mean price coefficients, the mixture weight of the more price-sensitive class, and the validation fit metrics. All ten refits reach early-stopping convergence under EM without class collapse. The standard errors are small on every row, and the gap between the two class-mean price coefficients sits many standard errors above zero. The full-sample numbers cited in the main text are therefore stable at this sample size rather than an artefact of a single optimisation draw.

\input{tables/bootstrap_ses}

\subsection{Outside-Good Calibration Sensitivity}
\label{sec:appendix_tau_robustness}

The demand model calibrates the outside-good utility $V_0$ by bisection to a target inside share $\hat\tau = \dataTau$ (\cref{sec:choice_model}). The robustness sweep covers $\hat\tau \in [\dataTauLo, \dataTauHi]$, bracketing the plausible single-category online share range. Across this range mean own-price elasticity moves by less than $\lcEpsTauVarPct\%$, mean Lerner shifts by roughly $\robLernerSwingPP$ percentage points, and the fraction of positive marginal costs falls from $\robMcPosHi\%$ to $\robMcPosLo\%$. Only the level of the recovered markup is sensitive to the outside-good calibration. \Cref{tab:supply-sensitivity} reports the full sweep on the demand-side primitive.

\begin{table}[H]
  \centering
  \small
  \caption{Sensitivity of recovered marginal costs to the calibrated outside share $\hat\tau$ under the two-class latent deep logit. All specifications use multi-product monopolist conduct ($\boldsymbol{\Omega} = \mathbf{1}$). The outside good utility $V_0$ is calibrated via bisection.}
  \label{tab:supply-sensitivity}
  \begin{tabular}{cccccc}
    \toprule
    $\hat{\tau}$ & $V_0$ & Mean $|\varepsilon|$ & Mean $\widehat{mc}$ (\texteuro{}) & Mean $L$ & $\widehat{mc} > 0$ \\
    \midrule
    0.02 & $\robTauTwoVZero$     & $\robTauTwoEps$     & $\robTauTwoMc$     & $\robTauTwoLerner$     & $\robTauTwoMcPos$/$\dataJ$ \\
    0.04 & $\robTauFourVZero$    & $\robTauFourEps$    & $\robTauFourMc$    & $\robTauFourLerner$    & $\robTauFourMcPos$/$\dataJ$ \\
    0.06 & $\robTauSixVZero$     & $\robTauSixEps$     & $\robTauSixMc$     & $\robTauSixLerner$     & $\robTauSixMcPos$/$\dataJ$ \\
    0.08 & $\robTauEightVZero$   & $\robTauEightEps$   & $\robTauEightMc$   & $\robTauEightLerner$   & $\robTauEightMcPos$/$\dataJ$ \\
    0.10 & $\robTauTenVZero$     & $\robTauTenEps$     & $\robTauTenMc$     & $\robTauTenLerner$     & $\robTauTenMcPos$/$\dataJ$ \\
    0.15 & $\robTauFifteenVZero$ & $\robTauFifteenEps$ & $\robTauFifteenMc$ & $\robTauFifteenLerner$ & $\robTauFifteenMcPos$/$\dataJ$ \\
    0.20 & $\robTauTwentyVZero$  & $\robTauTwentyEps$  & $\robTauTwentyMc$  & $\robTauTwentyLerner$  & $\robTauTwentyMcPos$/$\dataJ$ \\
    0.30 & $\robTauThirtyVZero$  & $\robTauThirtyEps$  & $\robTauThirtyMc$  & $\robTauThirtyLerner$  & $\robTauThirtyMcPos$/$\dataJ$ \\
    \bottomrule
  \end{tabular}
\end{table}

\Cref{tab:tau_welfare} turns the sweep into welfare deltas from the aesthetic-collapse counterfactual. The profit drop stays between $\robProfitDropLo\%$ and $\robProfitDropHi\%$ across the five calibrations, so this within-model result does not hinge on the outside-good choice.

\begin{table}[H]
  \centering
  \small
  \caption{Sensitivity of the aesthetic-collapse counterfactual to the outside-good calibration $\hat\tau$, under monopolist ownership. Class-specific item projections are pulled one standard deviation toward the class centroid. $\Delta$Profit is percent of baseline firm profit; $\Delta$CS per consumer is percent of baseline mean consumer surplus. Baseline PS grows with $\hat\tau$ because a larger inside share scales the weekly profit pool. The $\hat\tau = 0.04$ row is the paper's main-text baseline.}
  \label{tab:tau_welfare}
  \begin{tabular}{crrr}
    \toprule
    $\hat\tau$ & Baseline PS (\texteuro) & $\Delta$Profit & $\Delta$CS/$i$ \\
    \midrule
    0.02 & \phantom{0}$\robTauTwoPS$   & $-\robTauTwoProfitDelta$   & $-\robTauTwoCSDelta$ \\
    0.04 & $\robTauFourPS$             & $-\robTauFourProfitDelta$  & $-\robTauFourCSDelta$ \\
    0.06 & $\robTauSixPS$              & $-\robTauSixProfitDelta$   & $-\robTauSixCSDelta$ \\
    0.08 & $\robTauEightPS$            & $-\robTauEightProfitDelta$ & $-\robTauEightCSDelta$ \\
    0.10 & $\robTauTenPS$              & $-\robTauTenProfitDelta$   & $-\robTauTenCSDelta$ \\
    \bottomrule
  \end{tabular}
\end{table}

\subsection{Alternative Pre-trained Item Features}
\label{subsec:item_embedding_robustness}

We assess whether the demand and markup estimates for the Dress category depend on the choice of pre-trained item feature. The main analysis uses the $\embConcatDim$-dimensional concatenation of our three-tower $\embTowerDim$D item vector and the $\embCLIPDim$-dimensional raw FashionCLIP vector (\cref{sec:choice_model} footnote). We re-estimate the same two-class mixture demand model, with the same sample, hyperparameters, and random seed, under two ablations that drop back to one channel at a time. The first replaces the three-tower vector with FashionCLIP, a 512-dimensional image-and-text matching model fine-tuned on around 800{,}000 products from the Farfetch retail site, fed directly into the demand model. The second uses the $\embConcatDim$-dimensional concatenation of the two vectors. \Cref{tab:item_embedding_robustness} reports the main fit, demand, and supply quantities, together with correlations of mixture-weighted item-taste-network scores across the three variants. The class structure and the supply-side markup summary are close across specifications. Item-level taste scores correlate less tightly, so the three models produce distinct item-by-item rankings while agreeing at the aggregate level. The paper's main Dress conclusions rest on aggregate elasticities and markups, not on item-by-item rankings, so they do not depend on the three-tower pre-training choice in particular.

\begin{table}[H]
\centering
\caption{Dress demand and supply estimates under three pre-trained item features.}
\label{tab:item_embedding_robustness}
\begin{threeparttable}
\begin{tabular}{l rrr}
\toprule
 & Three-tower & FashionCLIP & Concatenation \\
 & 64D (legacy) & 512D direct & $64{+}512 = 576$D \textbf{(main)} \\
\midrule
\multicolumn{4}{l}{\textit{Demand fit}} \\
Validation negative log-likelihood               & $\phantom{-}\robIEBlegacyValNLL$       & $\phantom{-}\robIEBfcValNLL$           & $\phantom{-}\robIEBmainValNLL$ \\
Validation McFadden $R^{2}$                      & $+\robIEBlegacyValRSqrPct\%$           & $+\robIEBfcValRSqrPct\%$               & $+\robIEBmainValRSqrPct\%$ \\
Parameters in the item taste network             & $\phantom{-}\robIEBlegacyItemParams$   & $\phantom{-}\robIEBfcItemParams$       & $\phantom{-}\robIEBmainItemParams$ \\
\midrule
\multicolumn{4}{l}{\textit{Demand class structure}} \\
Mean price coefficient, class 1 ($\bar\alpha_{1}$) & $\robIEBlegacyAlphaOne$              & $\robIEBfcAlphaOne$                    & $\robIEBmainAlphaOne$ \\
Mean price coefficient, class 2 ($\bar\alpha_{2}$) & $\robIEBlegacyAlphaTwo$              & $\robIEBfcAlphaTwo$                    & $\robIEBmainAlphaTwo$ \\
Share of class 1 consumers ($\pi_{1}$)             & $\phantom{-}\robIEBlegacyPiOne$      & $\phantom{-}\robIEBfcPiOne$            & $\phantom{-}\robIEBmainPiOne$ \\
\midrule
\multicolumn{4}{l}{\textit{Supply (Bertrand-Nash, $\hat\tau=\dataTau$, monopolist)}} \\
Mean absolute own-price elasticity               & $\phantom{-}\robIEBlegacyEps$          & $\phantom{-}\robIEBfcEps$              & $\phantom{-}\robIEBmainEps$ \\
Mean Lerner index                                & $\phantom{-}\robIEBlegacyLerner$       & $\phantom{-}\robIEBfcLerner$           & $\phantom{-}\robIEBmainLerner$ \\
Share of products with $\mathrm{mc} > 0$         & $\phantom{-}\robIEBlegacyMcPos$        & $\phantom{-}\robIEBfcMcPos$            & $\phantom{-}\robIEBmainMcPos$ \\
Cross-elasticity departure from IIA              & $\phantom{-}\robIEBlegacyIIA$          & $\phantom{-}\robIEBfcIIA$              & $\phantom{-}\robIEBmainIIA$ \\
\midrule
\multicolumn{4}{l}{\textit{Agreement with 64D reference}} \\
Per-item taste-score correlation, Pearson        & $\phantom{-}\robIEBlegacyItemRho$      & $\phantom{-}\robIEBfcItemRho$          & $\phantom{-}\robIEBmainItemRho$ \\
Per-user taste-score correlation, Pearson        & $\phantom{-}\robIEBlegacyUserRho$      & $\phantom{-}\robIEBfcUserRho$          & $\phantom{-}\robIEBmainUserRho$ \\
Per-item taste-score correlation, Spearman       & $\phantom{-}\robIEBlegacyItemRhoSpearman$ & $\phantom{-}\robIEBfcItemRhoSpearman$ & $\phantom{-}\robIEBmainItemRhoSpearman$ \\
\bottomrule
\end{tabular}
\begin{tablenotes}[flushleft]
\footnotesize
\item Column 3 (576D concatenation) is the main specification for Dress; column 1 (three-tower 64D) is retained as a legacy robustness comparison. All three columns report post-V0-fix supply quantities under their respective LC artifacts. Agreement rows correlate each variant's mixture-weighted item-taste-network score against the three-tower 64D reference, aggregated per item and per user.
\end{tablenotes}
\end{threeparttable}
\end{table}

%% file: tables/bootstrap_ses.tex
\begin{table}[H]
\centering
\caption{Consumer-bootstrap standard errors for the Dress demand model, $B=10$ draws of the 38{,}918 consumers sampled with replacement. Inside each replicate the two classes are reordered so the more price-sensitive one is labelled class 1.}
\label{tab:bootstrap_ses}
\begin{tabular}{l rr}
\toprule
 & Bootstrap mean & Bootstrap SE \\
\midrule
$\bar\alpha_1$ (minority, more price-sensitive) & -0.216 & 0.004 \\
$\bar\alpha_2$ (majority, less price-sensitive) & -0.124 & 0.003 \\
$\lvert\bar\alpha_1 - \bar\alpha_2\rvert$ & 0.093 & 0.003 \\
$\pi_1$ (mixture weight of more price-sensitive class) & 0.402 & 0.010 \\
Val NLL & 6.10 & 0.02 \\
Val $R^{2}$ & +6.10\% & +0.34\% \\
\bottomrule
\end{tabular}
\end{table}

%% file: D_blp.tex
\section{Alternative Estimators}
\label{sec:demand_comparison}

\subsection{Demand}
\label{subsec:alt_demand}

\Cref{tab:blp_comparison} benchmarks the two-class deep latent class against four classical alternatives on the same $J = \bnchJ$ top-selling-dress, $\dataT$-week pre-COVID panel.\footnote{The deep-logit row in this benchmark uses the 64-dimensional three-tower item features to match the PyBLP comparison panel. The main-paper Dress LC master in \cref{sec:choice_model} uses the 576-dimensional concatenation of three-tower and FashionCLIP as the $\beta$-net item input; the BLP comparison is a separate J=500 panel kept architecturally aligned with the classical estimators.} The alternatives are a plain aggregate logit, a random-coefficient logit by simulated maximum likelihood, the \citet{berry1995automobile} GMM estimator with differentiation instruments following \citet{gandhi2020measuring} and the \citet{conlon2020best} checklist, and an individual mixed logit by simulated maximum likelihood. Rows one and three collapse the price coefficient toward zero because per-product shares of order $10^{-4}$ dominate the share inversion, and the single-firm Bertrand-Nash inversion returns negative marginal costs on every product. The random-coefficient SML estimator recovers an economically meaningful $\hat\alpha$ with positive marginal costs on 90\% of products. The individual mixed logit has positive costs on 17\% of products. The deep latent class recovers a mean price coefficient roughly two orders of magnitude larger than the classical aggregate rows and positive marginal costs on essentially every dress. The classical specifications cannot support the individual-level counterfactuals in \cref{sec:counterfactuals} because they do not map an observed consumer to a specific $(\alpha_i, \mathbf{r}_i)$ pair.

\begin{table}[H]
\centering
\input{tables/blp_comparison_6row.tex}
\caption{Five demand models on the dress category. All rows use the same $J = \bnchJ$ top-selling-dress panel over $\dataT$ pre-COVID weekly markets. Rows 1 to 4 use $\bnchPCADim$ principal-component features; row 5 uses the full $\embTowerDim$-dimensional item embedding with individual purchase records. $\hat\alpha$ is the mean price coefficient per euro. $|\varepsilon|$ is the mean-$\alpha$ own-price elasticity at mean prices. mc${>}0$ is the fraction of products with positive inferred marginal cost under the single-firm Bertrand-Nash inversion. Lerner is the mean implied markup as a share of price. NLL/obs is per-inside-purchase conditional log loss on the held-out validation set.}
\label{tab:blp_comparison}
\end{table}

\subsection{Hedonic}
\label{subsec:alt_hedonic}

The composition-gap headline in \cref{sec:hedonic} uses a LightGBM hedonic surface. To check that the choice of surface is not driving the result I re-estimate the same monthly-chained Fisher index on the same Dress sample under three alternatives: the pooled time-dummy hedonic of \citet{court1939, griliches1961}, the per-period imputation of \citet{diewert2007imputation}, and a deep multi-period network following \citet{bajari2025hedonic}. \Cref{tab:alt_hedonic} reports the comparison. LightGBM and the deep network agree that the quality-adjusted Fisher decline sits between $-50\%$ and $-60\%$ and that the Jevons-minus-Fisher composition gap is in the single digits. The disagreement between the two stays inside the $\confBandWidthPP$ percentage-point conformal band reported in \cref{sec:hedonic}. The two classical five-characteristic rows explain under $40\%$ of price variation and show gaps roughly four times larger; they stand as a thin-feature reference rather than a competing estimate.

\begin{table}[H]
\centering
\caption{Alternative hedonic methods for dresses, reported for the Dress category only; the cross-category gap table in \cref{tab:index_cross_category} is not re-estimated. Same sample and quantity weights as \cref{sec:hedonic}; the Jevons index declines $\hedJevonsDecline\%$ in every row by construction.}
\label{tab:alt_hedonic}
\begin{threeparttable}
\small
\begin{tabular}{l c c c}
\toprule
Method & OOS $R^{2}$ & Index decline & Gap vs Jevons (pp) \\
\midrule
Pooled time dummy \citep{court1939, griliches1961}      & $\bnchHedPooledR$    & $-\bnchHedPooledDecline\%$    & $\hedPooledGap$ \\
Per-period imputation \citep{diewert2007imputation}     & $\bnchHedPerPeriodR$ & $-\bnchHedPerPeriodDecline\%$ & $\hedPerPeriodGap$ \\
Multi-price network \citep{bajari2025hedonic}           & $\bnchHedDeepR$      & $-\hedDeepFisherDecline\%$    & $\phantom{0}\hedDeepGap$ \\
Paper LightGBM (\cref{sec:hedonic})                     & $\hedRSqrInFull$     & $-\hedFisherDecline\%$        & $\phantom{0}\hedGap$ \\
\bottomrule
\end{tabular}
\begin{tablenotes}[para]
\footnotesize
\item \textit{Notes:} OOS $R^{2}$ is from 5-fold article-level cross-validation, quantity-weighted, for all four rows (unseen articles, not out-of-time; not comparable to the in-time and forward $R^{2}$ columns in \cref{tab:hedonic_results}). The LightGBM row's Fisher decline and gap are transcribed from \cref{sec:hedonic}; its $R^{2}$ is re-scored under the same protocol as the other three rows and happens to coincide with the in-time figure in \cref{tab:hedonic_results} to two decimal places. The pooled time-dummy row is a fixed-base index; the other three are monthly-chained Fisher indices. The deep-network row is the mean across three random seeds (standard deviation $\bnchHedDeepStdPP$ pp on the Fisher decline and on the gap).
\end{tablenotes}
\end{threeparttable}
\end{table}

%% file: tables/blp_comparison_6row.tex
\begin{tabular}{lrrrrr}
\toprule
\textbf{Model} & $\hat\alpha$ & $|\varepsilon|$ & $\text{mc}>0$ & Lerner & NLL/obs \\
\midrule
Plain aggregate logit & -0.01 & 0.11 & 0\% & +11.85 & 6.50 \\
RC aggregate logit & -0.12 & 2.16 & 90\% & +0.60 & 7.77 \\
BLP with PCA features & -0.02 & 0.40 & 1\% & +3.27 & 6.50 \\
\midrule
Individual mixed logit & -0.04 & 0.73 & 17\% & +1.79 & 6.50 \\
Deep latent class & -0.18 & 3.13 & 98\% & +0.42 & 7.11 \\
\bottomrule
\end{tabular}

%% file: E_cross_category.tex
\section{Cross-Category Comparison}
\label{sec:appendix_cross_category}

All six product categories (Dress, Trousers, Sweater, T-shirt, Shirt, and Shorts) are estimated using the same three-tower embedding architecture and estimation pipeline.\footnote{For Dress, the $\beta$-net item input in the latent-class demand master concatenates the 64-dimensional three-tower item vector with the 512-dimensional raw FashionCLIP vector (see \cref{sec:choice_model} footnote). The other five categories use the 64-dimensional three-tower item vector only. All six use the same three-tower user embedding and price tower, and the same R3.2 $K=2$ unconditional-gating specification.} The comparisons test whether the structural findings from the preceding sections, which focus on Dress, generalize across product types with different price levels, seasonal patterns, and demand structures.

\subsection{Demand Estimation}

Across the six categories, the $K=2$ latent-class master recovers class-mean price sensitivities between $-\xcatAlphaMinAbs$ and $-\xcatAlphaMaxAbs$ per euro, with class-separation gaps ranging from $\xcatGapMin$ (T-shirt, collapsed on price, split by taste only) to $\xcatGapMax$ (Shirt). Mean own-price elasticity runs from $\xcatEpsMin$ for T-shirts to $\xcatEpsMax$ for Trousers. The within-tier aesthetic percentile exceeds $50$ for four of five non-Dress categories, significant at $p < 0.001$. \Cref{tab:cross_choice} reports the per-category estimates.

\begin{table}[H]
\centering
\caption{Demand model estimates across product categories.}
\label{tab:cross_choice}
\small
\begin{threeparttable}
\begin{tabular}{l rr rrr r rr rr r}
\toprule
& \multicolumn{2}{c}{Scale} & \multicolumn{3}{c}{Price ($K=2$ LC master)} & & \multicolumn{2}{c}{Val NLL} & \multicolumn{2}{c}{Decomposition} & \\
\cmidrule(lr){2-3} \cmidrule(lr){4-6} \cmidrule(lr){8-9} \cmidrule(lr){10-11}
Category & $I$ & $J$ & $\bar{\alpha}_0$ & $\bar{\alpha}_1$ & $\pi_0$ & $|\bar{\varepsilon}|$ & Static & Time & $R^2(p)$ & $R^2(\text{aes})$ & WT\% \\
\midrule
Dress$^\ddagger$ & $\dataI$             & $\dataJ$             & $\lcAlphaOne$            & $\lcAlphaTwo$            & $\lcPiOne$             & $\lcEpsBar$  & --                            & --                          & $\xcatDressRSqrP$    & $\xcatDressRSqrAes$    & -- \\
Trousers        & $\xcatTrousersI$     & $\xcatTrousersJ$     & $\xcatTrousersAlphaZero$ & $\xcatTrousersAlphaOne$  & $\xcatTrousersPiZero$  & $\xcatTrousersEps$  & $\xcatTrousersValNLLStatic$   & $\xcatTrousersValNLLTime$   & $\xcatTrousersRSqrP$ & $\xcatTrousersRSqrAes$ & $\xcatTrousersWT^{***}$ \\
Sweater         & $\xcatSweaterI$      & $\xcatSweaterJ$      & $\xcatSweaterAlphaZero$  & $\xcatSweaterAlphaOne$   & $\xcatSweaterPiZero$   & $\xcatSweaterEps$   & $\xcatSweaterValNLLStatic$    & $\xcatSweaterValNLLTime$    & $\xcatSweaterRSqrP$  & $\xcatSweaterRSqrAes$  & $\xcatSweaterWT$ \\
T-shirt         & $\xcatTShirtI$       & $\xcatTShirtJ$       & $\xcatTShirtAlphaZero$   & $\xcatTShirtAlphaOne$    & $\xcatTShirtPiZero$    & $\xcatTShirtEps$    & $\xcatTShirtValNLLStatic$     & $\xcatTShirtValNLLTime$     & $\xcatTShirtRSqrP$   & $\xcatTShirtRSqrAes$   & $\xcatTShirtWT^{***}$ \\
Shirt           & $\xcatShirtI$        & $\xcatShirtJ$        & $\xcatShirtAlphaZero$    & $\xcatShirtAlphaOne$     & $\xcatShirtPiZero$     & $\xcatShirtEps$     & $\xcatShirtValNLLStatic$      & $\xcatShirtValNLLTime$      & $\xcatShirtRSqrP$    & $\xcatShirtRSqrAes$    & $\xcatShirtWT^{***}$ \\
Shorts          & $\xcatShortsI$       & $\xcatShortsJ$       & $\xcatShortsAlphaZero$   & $\xcatShortsAlphaOne$    & $\xcatShortsPiZero$    & $\xcatShortsEps$    & $\xcatShortsValNLLStatic$     & $\xcatShortsValNLLTime$     & $\xcatShortsRSqrP$   & $\xcatShortsRSqrAes$   & $\xcatShortsWT^{***}$ \\
\bottomrule
\end{tabular}
\begin{tablenotes}
\small
\item \textit{Notes:} $I$ = consumers, $J$ = products in the $K=2$ LC estimation sample (prep\_phase3, pre-COVID window 2019-07-01 to 2020-02-02, $T=31$ weeks). $\bar{\alpha}_0$, $\bar{\alpha}_1$ = class-mean price sensitivities (EUR$^{-1}$) from the $K=2$ LC master; $\pi_0$ = unconditional class-$0$ mixture weight ($\pi_1 = 1 - \pi_0$). $|\bar{\varepsilon}|$ = mean own-price elasticity from the $K=2$ LC master. Val NLL Static and Time are retained three-tower MNL diagnostics on the full sample and are not $K=2$ LC numbers. $R^2(p)$ and $R^2(\text{aes})$ = share of predicted market share variance explained by price and aesthetic components respectively. WT\% = within-tier aesthetic percentile (50 = no discrimination). $^{***}$ $p < 0.001$.
\end{tablenotes}
\end{threeparttable}
\end{table}

\subsection{Hedonic Pricing}

The gap between the Jevons and hedonic Fisher indices runs from $\hedGap$ percentage points for Dress to $\xcatTrousersHedGap$ percentage points for Trousers. The hedonic fit stays within a narrow band, $R^2$ between $\xcatHedRMin$ and $\xcatHedRMax$. The gap is positive for all six categories. \Cref{tab:cross_hedonic} reports the per-category values.

\begin{table}[H]
\centering
\caption{Hedonic price indices across product categories.}
\label{tab:cross_hedonic}
\begin{threeparttable}
\begin{tabular}{l rrrr}
\toprule
Category & $R^2$ & Jevons $\downarrow$ & Fisher $\downarrow$ & Gap (pp) \\
\midrule
Dress    & $\xcatDressHedR$    & $\hedJevonsDecline\%$  & $\hedFisherDecline\%$  & $\hedGap$ \\
Trousers & $\xcatTrousersHedR$ & $\xcatTrousersJevons\%$ & $\xcatTrousersFisher\%$ & $\xcatTrousersHedGap$ \\
Sweater  & $\xcatSweaterHedR$  & $\xcatSweaterJevons\%$  & $\xcatSweaterFisher\%$  & $\xcatSweaterHedGap$ \\
T-shirt  & $\xcatTShirtHedR$   & $\xcatTShirtJevons\%$   & $\xcatTShirtFisher\%$   & $\xcatTShirtHedGap$ \\
Shirt    & $\xcatShirtHedR$    & $\xcatShirtJevons\%$    & $\xcatShirtFisher\%$    & $\xcatShirtHedGap$ \\
Shorts   & $\xcatShortsHedR$   & $\xcatShortsJevons\%$   & $\xcatShortsFisher\%$   & $\xcatShortsHedGap$ \\
\bottomrule
\end{tabular}
\begin{tablenotes}
\small
\item \textit{Notes:} $R^2$ = hedonic regression fit. Jevons and Fisher columns report cumulative price decline over the sample period. Gap = composition bias (Jevons $-$ Fisher). Positive gap means the na\"ive index overstates true price decline.
\end{tablenotes}
\end{threeparttable}
\end{table}

\subsection{Lockdown Event Study}

Aggregate lockdown effects run from $-\xcatSweaterAggregate\%$ for Sweaters to $+\xcatTShirtAggregate\%$ for T-shirts. Item-cluster ranges span $\xcatItemRangeMin$ to $\xcatItemRangeMax$ percentage points, user-cluster ranges $\xcatUserRangeMin$ to $\xcatUserRangeMax$ percentage points. The ratio of user-cluster range to demographic-group range runs from $\xcatUserDemoRatioMin\times$ to $\xcatUserDemoRatioMax\times$. The Adjusted Rand Index between embedding clusters and age groups sits near $\xcatARIEmbAgeTypical$ for every category. \Cref{tab:cross_lockdown} reports the full set.

\begin{table}[H]
\centering
\caption{COVID-19 lockdown effects across product categories.}
\label{tab:cross_lockdown}
\begin{threeparttable}
\begin{tabular}{l r rr rr r}
\toprule
& & \multicolumn{2}{c}{Item clusters} & \multicolumn{2}{c}{User clusters} & \\
\cmidrule(lr){3-4} \cmidrule(lr){5-6}
Category & Aggregate & $K$ & Range (pp) & $K$ & Range (pp) & User/Demo \\
\midrule
Dress    & $-\xcatDressAggregate$\%$^{***}$    & $\xcatDressKItem$    & $\xcatDressRangeItem$    & $\xcatDressKUser$    & $\xcatDressRangeUser$    & $\xcatDressUserDemo\times$ \\
Trousers & $-\xcatTrousersAggregate$\% n.s.    & $\xcatTrousersKItem$ & $\xcatTrousersRangeItem$ & $\xcatTrousersKUser$ & $\xcatTrousersRangeUser$ & $\xcatTrousersUserDemo\times$ \\
Sweater  & $-\xcatSweaterAggregate$\%$^{***}$  & $\xcatSweaterKItem$  & $\xcatSweaterRangeItem$  & $\xcatSweaterKUser$  & $\xcatSweaterRangeUser$  & $\xcatSweaterUserDemo\times$ \\
T-shirt  & $+\xcatTShirtAggregate$\%$^{*}$     & $\xcatTShirtKItem$   & $\xcatTShirtRangeItem$   & $\xcatTShirtKUser$   & $\xcatTShirtRangeUser$   & $\xcatTShirtUserDemo\times$ \\
Shirt    & $-\xcatShirtAggregate$\%$^{***}$    & $\xcatShirtKItem$    & $\xcatShirtRangeItem$    & $\xcatShirtKUser$    & $\xcatShirtRangeUser$    & $\xcatShirtUserDemo\times$ \\
Shorts   & $+\xcatShortsAggregate$\% n.s.      & $\xcatShortsKItem$   & $\xcatShortsRangeItem$   & $\xcatShortsKUser$   & $\xcatShortsRangeUser$   & $\xcatShortsUserDemo\times$ \\
\bottomrule
\end{tabular}
\begin{tablenotes}
\small
\item \textit{Notes:} Aggregate = Poisson QMLE average treatment effect. $K$ = number of embedding clusters. Range = spread between most positive and most negative cluster effects (percentage points). User/Demo = ratio of embedding-cluster heterogeneity range to demographic-group heterogeneity range. $^{*}$ $p < 0.05$; $^{***}$ $p < 0.001$; n.s.\ = not significant.
\end{tablenotes}
\end{threeparttable}
\end{table}

\subsection{Post-Clustering Robustness for the Heterogeneity Range}
\label{subsec:appendix_post_clustering}

The per-cluster range of lockdown coefficients reported in \cref{sec:event_study} depends on one K-means partition of the Dress item embeddings at $K=12$. A standard concern with any two-step estimate that first clusters and then tests is that the partition is itself estimated, and the second-stage regression treats it as fixed. To check whether the heterogeneity finding is an artifact of a single favorable seed, I re-estimate the partition under twenty different K-means initialisations, refit M1 (per-cluster Poisson QMLE with Newey-West standard errors) on each resulting assignment, and record the range of lockdown coefficients across the twelve clusters for each seed. The main-text value of $\evtItemRangePP$ percentage points is the seed-42 estimate.

\begin{figure}[H]
    \centering
    \includegraphics[width=0.7\textwidth]{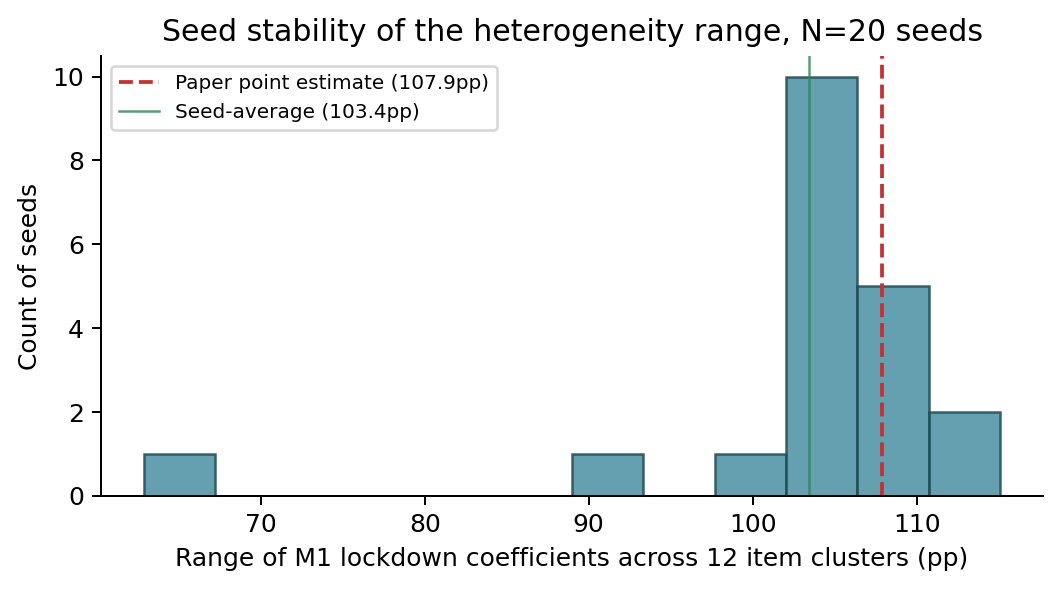}
    \caption{Distribution of the M1 heterogeneity range across $\evtKmeansSeedsN$ random K-means initialisations on the pre-COVID item embeddings at $K=\evtKItem$. The dashed red line is the main-text point estimate; the green line is the seed-average.}
    \label{fig:kmeans_seed_stability}
\end{figure}

\Cref{fig:kmeans_seed_stability} plots the distribution of the range across the $\evtKmeansSeedsN$ seeds. Mean range is $\evtKmeansSeedMeanPP$ percentage points with standard deviation $\evtKmeansSeedSDPP$. The 5th-to-95th percentile interval is $[\evtKmeansSeedPctFifthLo, \evtKmeansSeedPctFifthHi]$. The main-text point estimate sits inside this interval. Nineteen of twenty seeds give ranges between $\evtKmeansSeedRangeLo$ and $\evtKmeansSeedRangeHi$ percentage points; the one outlier seed (range $\evtKmeansSeedOutlierRange$) corresponds to a K-means local minimum that merges two of the occasion-wear clusters. The heterogeneity finding is not an artifact of a single favorable initialisation.

The check varies the K-means initialisation but not the embedding stage or the item sample. A fuller treatment would bootstrap the item sample, re-train the embeddings, and re-estimate the partition on each draw before refitting M1. The narrower seed check rules out one of the two first-stage concerns.

%% file: H_embedding_validation.tex
\section{Embedding Validation}
\label{sec:embedding_validation}

\begin{figure}[H]
\centering
\includegraphics[width=0.78\textwidth]{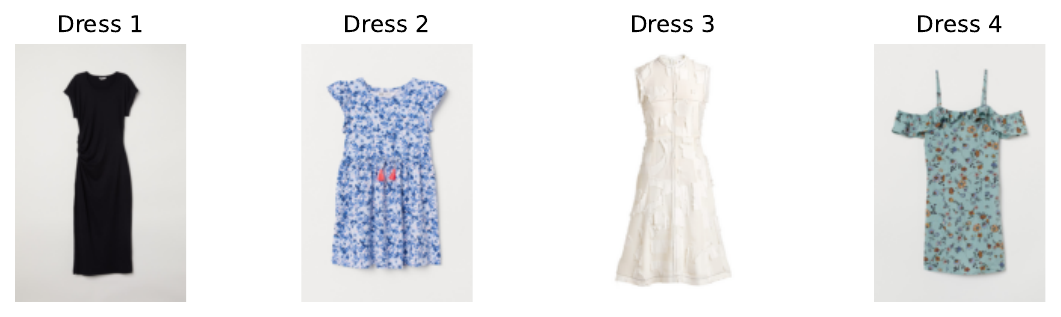}

\vspace{0.3cm}
\begin{minipage}[t]{0.54\textwidth}
\centering
\textit{Hand-coded attributes}\\[0.2cm]
\footnotesize
\setlength{\tabcolsep}{4pt}
\begin{tabular}{llll}
\toprule
       & Pattern           & Color      & Group         \\
\midrule
Dress 1 & Solid             & Black      & Jersey Basic  \\
Dress 2 & All over pattern  & White      & Jersey Fancy  \\
Dress 3 & Jacquard          & Off White  & Studio        \\
Dress 4 & All over pattern  & Turquoise  & Ladies        \\
\bottomrule
\end{tabular}
\end{minipage}\hfill
\begin{minipage}[t]{0.42\textwidth}
\centering
\textit{Learned embedding $\mathbf{v}_j \in \mathbb{R}^{64}$}\\[0.2cm]
\includegraphics[width=\textwidth]{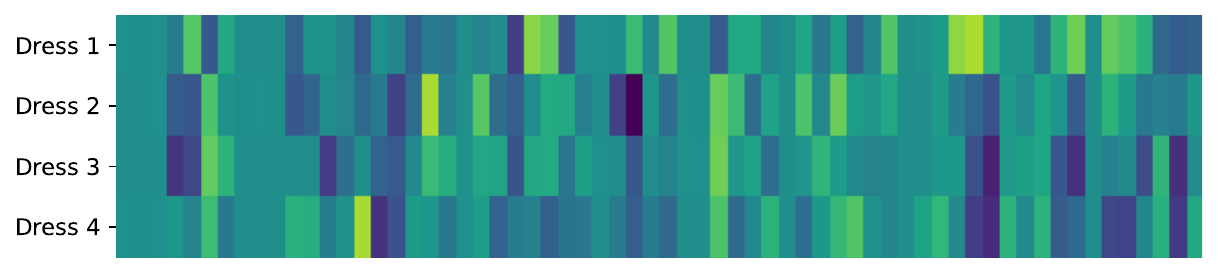}
\end{minipage}
\caption{What the 64-dimensional item embedding looks like compared with the hand-coded product attributes available in the raw data. Four representative dresses are shown at the top. The bottom-left panel lists their three recorded attribute fields, which is the kind of input a hedonic regression or a demand model with categorical fixed effects would use. The bottom-right panel shows the same four dresses as rows of a heatmap, with each column being one of the 64 learned dimensions. The embedding spreads fine aesthetic distinctions across many dimensions instead of forcing them into a short list of named categories.}
\label{fig:attribute_vs_embedding}
\end{figure}

Before running any quantitative diagnostics, \cref{fig:attribute_vs_embedding} shows what a single item embedding looks like side by side with the raw categorical information the dataset records for each product. Each dress has only three recorded attribute fields in the raw data, pattern, color, and product group, which is the kind of input a standard hedonic regression or a demand model with categorical fixed effects would use. The learned 64-number embedding spreads the differences between dresses across many dimensions instead of collapsing them into a small set of named categories.

\begin{figure}[H]
    \centering
    \includegraphics[width=\textwidth]{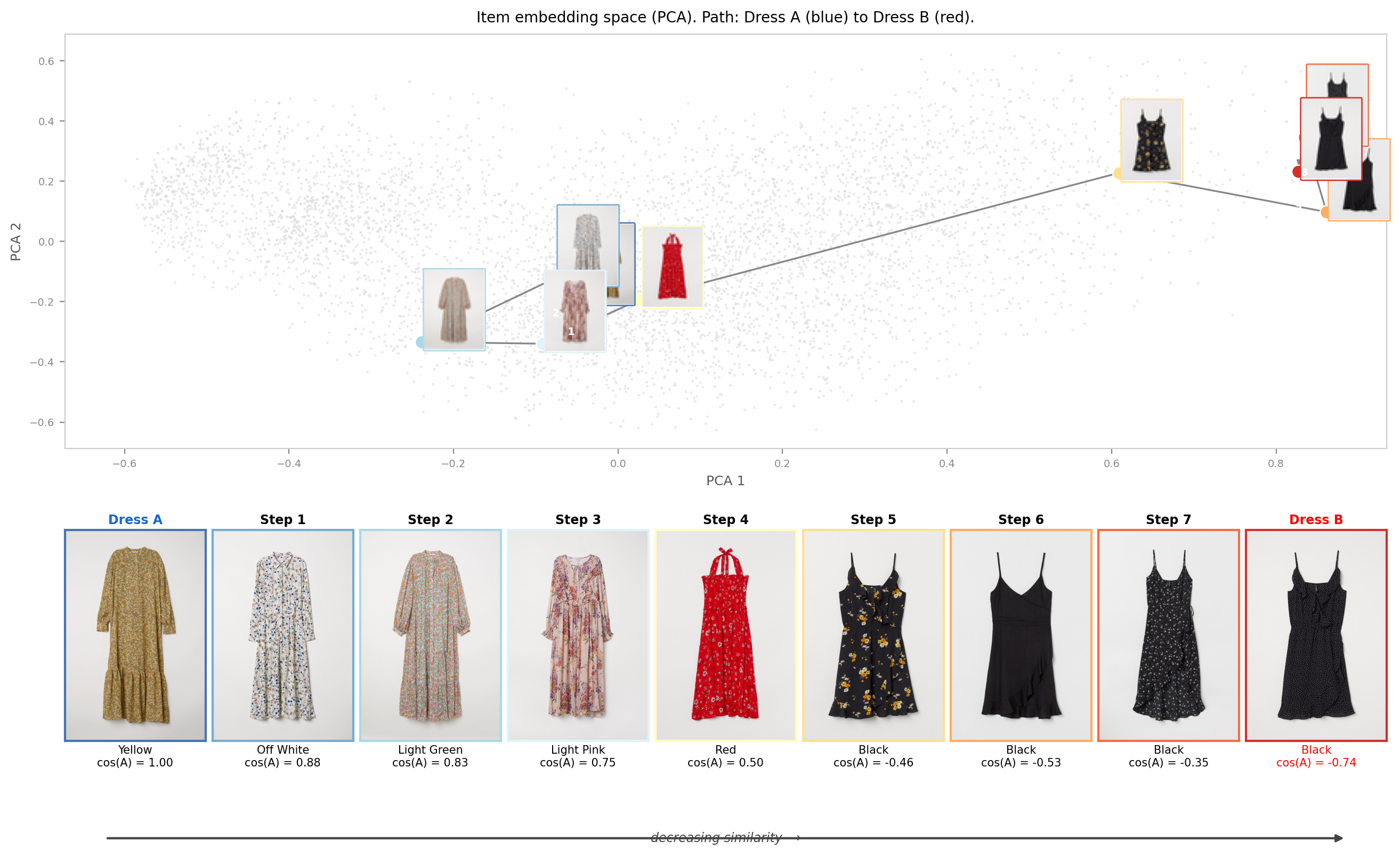}
    \caption{A path through the $\embTowerDim$-dimensional dress embedding space. \textit{Top:} A PCA projection\protect\footnote{The scatter uses a two-component PCA projection. PCA is chosen over t-SNE here because linear interpolation in $\embTowerDim$-dimensional space remains a straight line under any linear projection. Under t-SNE, which optimizes for local neighborhood structure rather than global variance, the two globally extreme dresses cluster near the same region and the path does not trace a visible trajectory.} of all $\embDressN$ dress embeddings (gray dots). The nine interpolation dresses are highlighted as a colored path from blue (Dress A) to red (Dress B), with thumbnail images at each stop. \textit{Bottom:} Starting from Dress A (a yellow printed dress, cosine $= \embCosineStart$), each column shows the actual dress nearest to an evenly-spaced interpolation point along the straight line toward Dress B (a black strap dress, cosine $= \embCosineEnd$). The cosine similarity to Dress A falls from $\embCosineStart$ to $\embCosineEnd$ across the nine columns, tracing how the model's notion of visual similarity shifts continuously from one aesthetic extreme to the other.}
    \label{fig:cosine_similarity}
\end{figure}

For the embeddings to be useful downstream, they must encode economically meaningful spatial structure. \Cref{fig:cosine_similarity} illustrates this by tracing a straight path between the two most dissimilar dresses in the space.

\begin{figure}[H]
    \centering
    \begin{minipage}{0.48\textwidth}
        \centering
        \includegraphics[width=\textwidth]{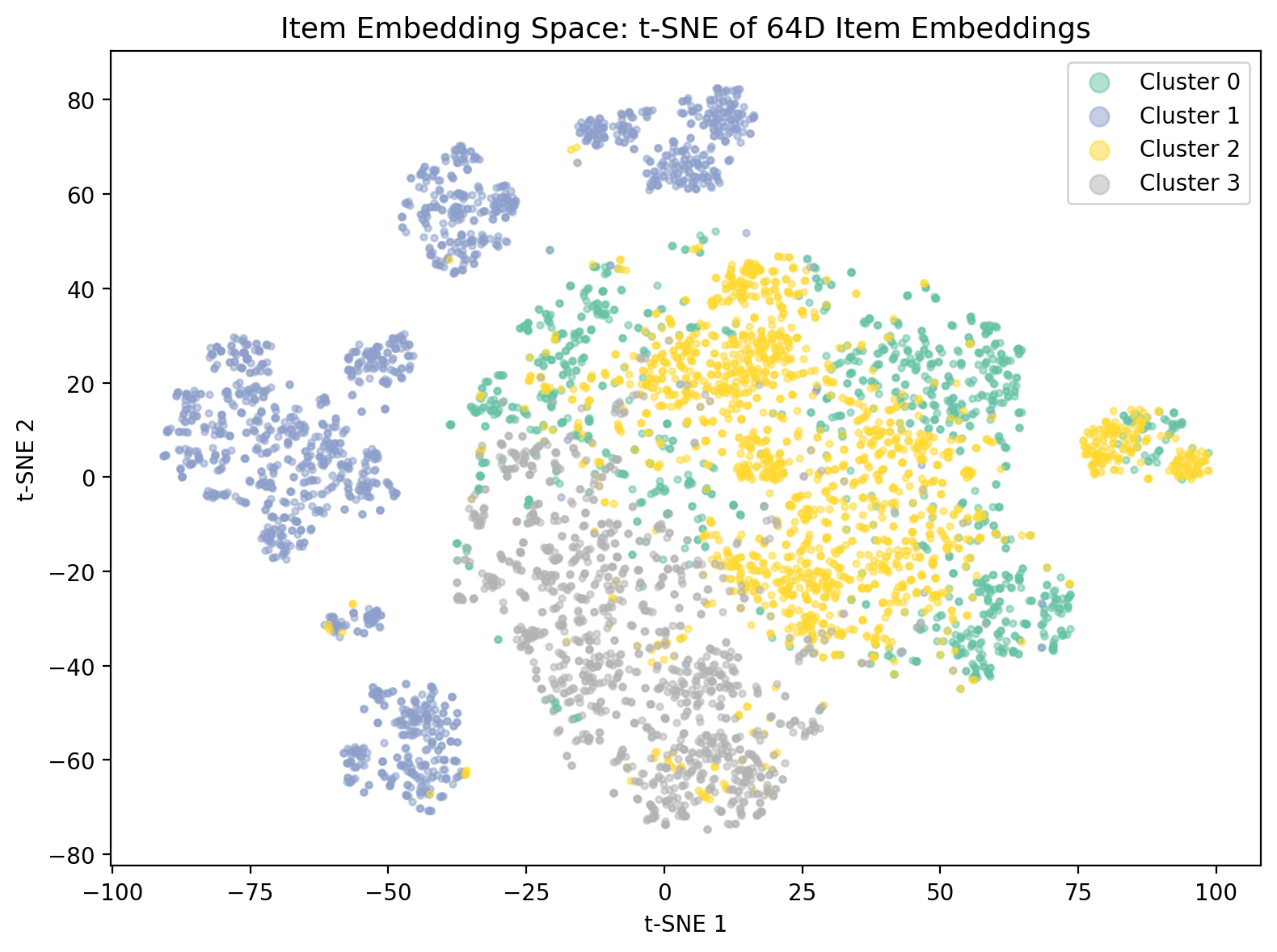}
        \caption{t-SNE projection of item embeddings with K-Means ($k=4$).}
        \label{fig:item_tsne}
    \end{minipage}\hfill
    \begin{minipage}{0.48\textwidth}
        \centering
        \includegraphics[width=\textwidth]{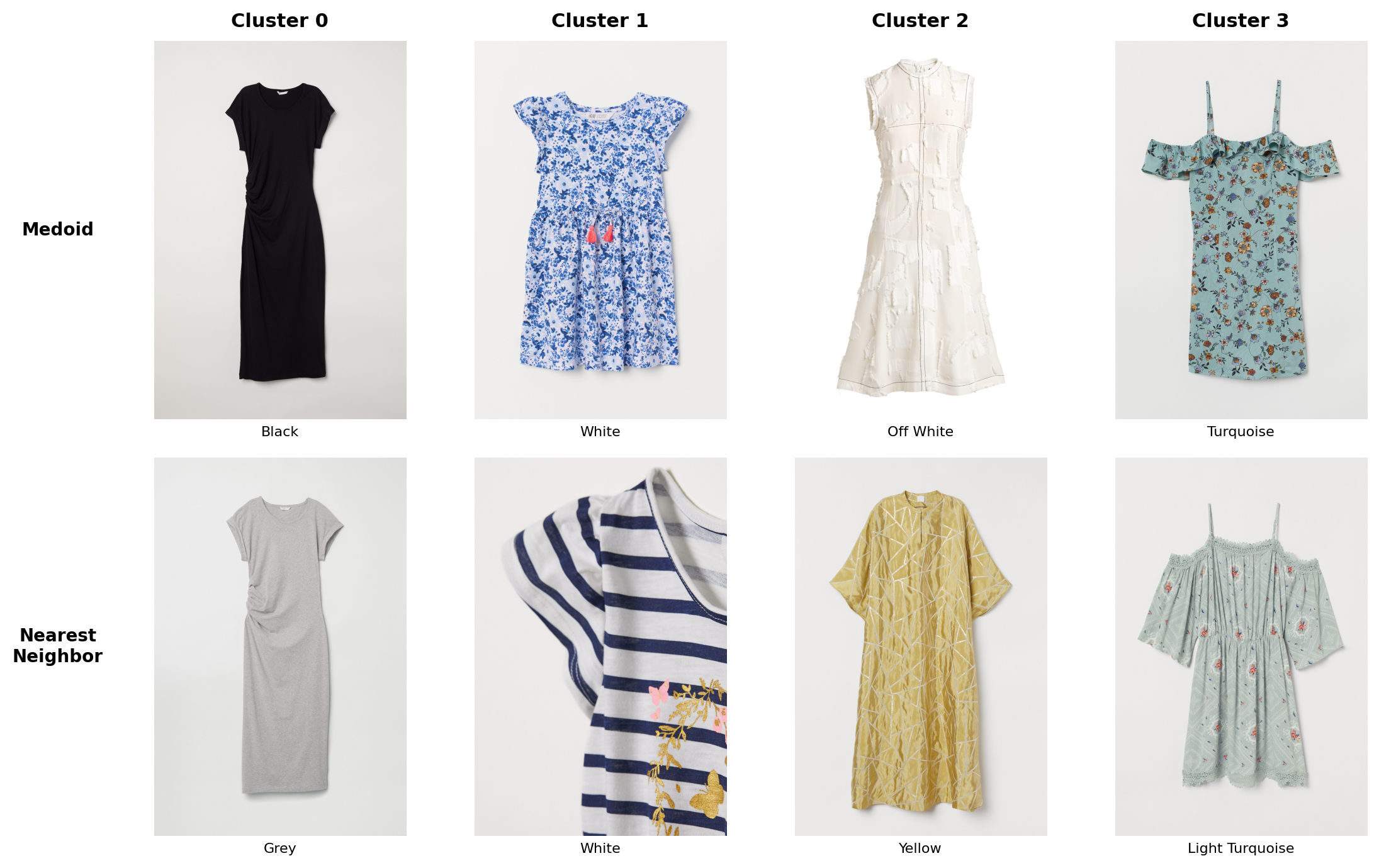}
        \caption{Representative product images from the four clusters.}
        \label{fig:item_cluster_examples}
    \end{minipage}
\end{figure}

K-Means clustering\footnote{K-Means partitions items into $k$ groups by iteratively assigning each item to its nearest group center and recomputing each center as the mean of its assigned items. The algorithm repeats until assignments stop changing.} \citep{lloyd1982least} with $k=\embClusterK$ on $\embDressN$ dress items confirms this structure. The learned embeddings group dresses by how they actually look rather than by their product codes. \Cref{fig:item_tsne} plots a t-SNE projection\footnote{t-SNE (t-Distributed Stochastic Neighbor Embedding) projects high-dimensional data into two dimensions while preserving local neighborhood relationships. Items that are close together in the original 64-dimensional space tend to appear close together in the 2D picture.} \citep{vandermaaten2008visualizing} of the embeddings with cluster assignments, and \cref{fig:item_cluster_examples} shows representative items. The clusters group visual styles (e.g., solids versus all-over patterns) that traverse product categories.

The user tower encodes purchase behaviour orthogonal to observed demographics. The Adjusted Rand Index between embedding-based user clusters and age-based demographic segments is near zero ($\embARIUserAge$).\footnote{The Adjusted Rand Index measures agreement between two groupings of the same items. It equals 1 when both groupings are identical and is near 0 when overlap is no better than chance.} Standard demand models attribute whatever variation demographics fail to capture to unobserved taste shocks. The user embeddings capture this behaviour systematically, mapping consumers with diverging aesthetic tastes into distinct spatial segments even when their observed demographics are identical.

\Cref{tab:quantitative_validation} reports quantitative diagnostics for the embeddings, focusing on cluster concentration and out-of-sample predictive power. Panel A details the cluster profiles and shows the extreme skew in demand, with Cluster 0 dominating transactions ($\embClusterSalesZero\%$) despite containing only $\embClusterZeroItemPct\%$ of the products. Panel B validates the value of contrastive training over zero-shot representations. In regressions predicting article-level log market shares, the $\embTowerDim$-dimensional three-tower embeddings outperform the raw $\embCLIPDim$-dimensional CLIP baseline, achieving an OLS $R^2$ of $\embRSqrSharesTTOLS$ versus $\embRSqrSharesCLIPOLS$.

\begin{table}[H]
\centering
\small
\caption{Quantitative Validation of Item Embeddings}
\label{tab:quantitative_validation}
\textbf{Panel A: Cluster Profiles for Dresses ($N=\embDressN$)} \vspace{0.1cm}\\
\begin{tabular}{lrllrr}
\toprule
Cluster & $N$ & Modal Color & Modal Pattern & Mean Price & Sales \% \\
\midrule
0 & \embClusterNZero  & Black     & Solid             & \embClusterPriceZero  & \embClusterSalesZero  \\
1 & \embClusterNOne   & Dark Blue & All over pattern  & \embClusterPriceOne   & \embClusterSalesOne   \\
2 & \embClusterNTwo   & Black     & Solid             & \embClusterPriceTwo   & \embClusterSalesTwo   \\
3 & \embClusterNThree & Black     & All over pattern  & \embClusterPriceThree & \embClusterSalesThree \\
\bottomrule
\end{tabular}

\vspace{0.4cm}
\textbf{Panel B: Predictive Performance on Downstream Outcomes} \vspace{0.1cm}\\
\begin{tabular}{lcccc}
\toprule
Feature Set & \multicolumn{2}{c}{$R^2$ for Log Shares} & \multicolumn{2}{c}{$R^2$ for Log Prices} \\
\cmidrule(lr){2-3} \cmidrule(lr){4-5}
& OLS & LightGBM & OLS & LightGBM \\
\midrule
Raw CLIP ($\embCLIPDim$D)             & \embRSqrSharesCLIPOLS & \embRSqrSharesCLIPLGB & \embRSqrPricesCLIPOLS & \embRSqrPricesCLIPLGB \\
Three-Tower Embeddings ($\embTowerDim$D) & \embRSqrSharesTTOLS   & \embRSqrSharesTTLGB   & \embRSqrPricesTTOLS   & \embRSqrPricesTTLGB   \\
\bottomrule
\end{tabular}
\begin{tablenotes}
\small
\item Note: Panel A uses K-Means ($k=4$). Mean Price is normalized transaction price; Sales \% is the cluster's share of total category volume. Panel B tests article-level predictions on an 80/20 train-test split, with both linear OLS and a gradient-boosted decision tree fit via LightGBM \citep{ke2017lightgbm}; a gradient-boosted tree builds many small decision trees in sequence, each correcting the residuals of the ones before it. The 64D Three-Tower embeddings achieve superior share prediction, while structural design appropriately suppresses their price predictivity. The Dress LC demand model in \cref{sec:choice_model} consumes both representations jointly as a $\embConcatDim$D concatenation; Panel B here documents their individual share-prediction power, which motivates retaining three-tower as the distilled signal.
\end{tablenotes}
\end{table}

The log price regressions in Panel B show the three-tower item embeddings still predict price about as well as the raw image features ($R^2 = \embRSqrPricesTTOLS$ versus $\embRSqrPricesCLIPOLS$ under OLS). The point of the three-tower design is not that the item embedding loses all correlation with price. It is that the item tower never sees price as an input during training, so any predictive power comes from genuine visual and categorical features, not from a learned copy of the price signal. That architectural separation is what the downstream demand model needs, because price enters the utility function as its own term with its own coefficient.

%% file: I_glossary.tex
\section{Glossary of Machine Learning Terms}
\label{sec:glossary}

This thesis uses a small set of terms from the machine learning literature that may be unfamiliar to economists. The table below gives a plain-English explanation of each term and the canonical reference. Terms that are already standard in econometrics (softmax, logit, cross-validation, principal components, K-means, simulated maximum likelihood, Halton draws) are not included.

\begingroup
\footnotesize
\renewcommand{\arraystretch}{1.25}
\begin{longtable}{p{3.3cm} p{9.7cm} p{2.2cm}}
\caption{Machine learning terms used in this thesis.}\label{tab:glossary} \\
\toprule
\textbf{Term} & \textbf{Explanation} & \textbf{Reference} \\
\midrule
\endfirsthead
\toprule
\textbf{Term} & \textbf{Explanation} & \textbf{Reference} \\
\midrule
\endhead
\bottomrule
\endlastfoot

Embedding &
A learned low-dimensional vector that represents a product or a consumer. Proximity in the vector space corresponds to similarity in the task the model was trained on. &
\citet{mikolov2013distributed} \\

Pre-training &
Training a model on a large generic corpus so that it learns broadly useful representations. The item tower in this thesis starts from a pre-trained CLIP checkpoint. &
\citet{radford2021learning} \\

Fine-tuning &
Continuing to train a pre-trained model on the target sample so that the generic representations specialise to the task at hand. &
\citet{goodfellow2016deep} \\

Three-tower / dual-encoder architecture &
A neural network with two or three parallel sub-networks whose outputs are scored against each other by an inner product, rather than being chained in series. CLIP is the prototype dual-encoder; this thesis uses a three-tower variant. &
\citet{radford2021learning} \\

Contrastive learning &
A training objective that pulls the representation of a matched pair (for example, a consumer and a product they bought) close together and pushes unmatched pairs apart. &
\citet{oord2018representation} \\

InfoNCE loss &
The specific contrastive loss used in this thesis. Algebraically equivalent to a categorical log-likelihood over the product category, and therefore to a multinomial logit likelihood with within-category negatives. &
\citet{oord2018representation} \\

CLIP &
Contrastive Language-Image Pre-training. A vision-language model that maps images and short text descriptions into a shared vector space, trained contrastively on internet-scale image-caption pairs. &
\citet{radford2021learning} \\

FashionCLIP &
CLIP fine-tuned on fashion product catalogues. Used as an additional 512-dimensional feature bundle concatenated with the three-tower item vector in the dress demand master, alongside the CLIP-based three-tower embedding. &
\citet{chia2022contrastive} \\

BERT, ResNet &
Pre-trained text and image models used by \citet{bajari2025hedonic} as drop-in feature extractors for hedonic regressions. BERT returns a vector for any short string; ResNet returns a vector for any image. &
\citet{devlin2019bert}, \citet{he2016deep} \\

Multi-layer perceptron (MLP) &
A feedforward neural network built from a stack of linear layers separated by elementwise nonlinearities. The $\alpha$-net and the taste projections in this thesis are MLPs. &
\citet{goodfellow2016deep} \\

Softplus activation &
The function $\log(1 + e^x)$, a smooth non-negative approximation to $\max(0, x)$. The paper negates its output to enforce $\alpha_c < 0$ in the price-sensitivity network. &
\citet{goodfellow2016deep} \\

LightGBM, gradient-boosted trees &
An ensemble of shallow decision trees, fit sequentially so that each new tree targets the residuals of the previous ones. Used here as the hedonic pricing surface. &
\citet{ke2017lightgbm} \\

t-SNE &
t-Distributed Stochastic Neighbor Embedding. A non-linear two-dimensional projection that preserves local neighbourhood structure. Used only for visualising the item embedding in figures, never as an input to estimation. &
\citet{vandermaaten2008visualizing} \\

Adjusted Rand Index &
A measure of agreement between two clusterings of the same items, corrected for chance. Equals zero when the two clusterings agree no better than chance and one when they are identical. &
\citet{hubert1985comparing} \\

Hit@K, Recall@K &
Retrieval accuracy metrics widely used in information retrieval and recommender systems. Hit@$K$ is one if the held-out item appears in the model's top-$K$ list and zero otherwise, averaged across queries. Recall@$K$ is the fraction of relevant items that appear in the top-$K$ list. &
-- \\

Adam &
Stochastic-gradient optimizer that adapts the step size per parameter. Used throughout this paper for neural network training. &
-- \\

Epoch &
One full pass through the training dataset. Training stops when validation error stops improving. &
-- \\

Validation set &
A held-out slice of the data, not used for fitting, used to decide when to stop training. Distinct from the test set reserved for final reporting. &
-- \\

Mini-batch &
A small random subset of the training data used for one gradient update. Replaces full-dataset gradients when data is large. &
-- \\

Checkpoint &
A saved snapshot of a neural network's weights at a point during training. Used for resuming, reuse, or freezing an encoder for downstream work. &
-- \\

L2 normalisation &
Rescaling a vector to unit length. After this step, the inner product of two vectors equals the cosine of the angle between them, so magnitudes no longer influence scoring. &
-- \\

Negative sampling &
In contrastive training, the denominator of the loss is summed over a random subsample of non-matching items rather than the full set, which keeps each gradient step cheap on a large catalogue. &
\citet{mikolov2013distributed} \\

Overfitting &
A model that fits training-sample noise loses accuracy on new data. Common counters are regularisation, low-rank structure, and validating on a held-out sample. &
\citet{goodfellow2016deep} \\

Low-rank factorisation &
Writing a matrix as the product of two narrower matrices, which lowers the parameter count and acts as a regulariser. Used here for the monthly taste shift $\boldsymbol{\delta}_m = W \mathbf{z}_m$. &
-- \\

EM algorithm &
Expectation-Maximization. An iterative procedure for fitting mixture models: the E step computes posterior class probabilities given current parameters, and the M step re-estimates parameters weighted by those probabilities. &
-- \\

AdamW &
A variant of Adam (see above) that separates weight decay from the adaptive learning-rate step. The decoupling typically improves generalisation compared to plain Adam and is the optimiser used in this paper. &
-- \\

Validation NLL &
The negative log-likelihood of the fitted model evaluated on the validation set (see above). Lower is better; the demand model in this paper reports validation NLL as its main held-out fit metric. &
-- \\

TF-IDF &
Term frequency-inverse document frequency. A text feature that counts words in one document and down-weights words that appear across many documents, so distinctive vocabulary receives more weight. &
-- \\

Conformal prediction &
A distribution-free way to build prediction intervals with a target coverage level, by recording prediction errors on a fresh slice and reading off their empirical quantile. &
\citet{vovk2005algorithmic} \\

Silhouette score &
A goodness-of-clustering metric in $[-1,1]$. For each point it compares distance to own-cluster mean against distance to the nearest other cluster; higher values indicate tighter, better-separated groups. &
-- \\

Normalized Mutual Information &
A measure of agreement between two clusterings that rescales their mutual information by the average entropy of the two partitions. Bounded in $[0,1]$; equals one for identical partitions. &
-- \\

Variational inference &
An approximate method for fitting probabilistic models with unobserved variables. The exact posterior is replaced with the closest member of a tractable family, which is optimised instead. &
-- \\

Poisson QMLE &
Poisson quasi-maximum likelihood. Consistent for the conditional mean of a count outcome even when the variance does not equal the mean, with sandwich standard errors handling overdispersion. &
\citet{gourieroux1984pseudo} \\

Oaxaca-Blinder decomposition &
Splits a mean outcome gap between two groups into a composition part (differences in observed characteristics) and a valuation part (differences in coefficients on those characteristics). The nonlinear extension replaces linear regressions with any fitted function. &
\citet{oaxaca1973male, blinder1973wage, bauer2008extension} \\

\end{longtable}
\endgroup